\documentclass[
reprint,
nofootinbib,
 amsmath,amssymb,
 aps, 
]{revtex4-1}

\newcommand{\grafe}[1]{\left\{ #1 \right\}}
\newcommand{\tonde}[1]{\left( #1 \right)}
\newcommand{\quadre}[1]{\left[ #1 \right]}
\usepackage{graphicx,subcaption}
\usepackage{xcolor}
\usepackage{amssymb}
\usepackage{amsmath}
\usepackage{dcolumn}
\usepackage{dsfont}
\usepackage{amsthm, amsfonts}
\usepackage{bbm}
\usepackage{eucal}
\usepackage{tikz}
\usepackage{tikz-feynman}
\usepackage{empheq}
\usepackage{mathtools}

\makeatletter
\newcommand*\bigcdot{\mathpalette\bigcdot@{.5}}
\newcommand*\bigcdot@[2]{\mathbin{\vcenter{\hbox{\scalebox{#2}{$\m@th#1\bullet$}}}}}
\makeatother

\DeclareMathOperator{\Tr}{Tr}

\begin{document}

\preprint{APS/123-QED}
\title{Triplets of local minima in a high-dimensional random landscape:\\  Correlations, clustering, and memoryless  activated jumps}

\author{Alessandro Pacco}
\email{alessandro.pacco@etu-upsaclay.fr}
\author{Alberto Rosso}
\author{Valentina Ros}%
\affiliation{Université Paris-Saclay, CNRS, LPTMS, 91405, Orsay, France}

\date{\today}

\begin{abstract}
We compute the distribution of triplets of stationary points in the energy landscape of the spherical $p$-spin model, by evaluating the quenched three-point complexity by means of the Kac-Rice formalism. 
We show the occurrence of transitions in the organization of stationary points in the landscape, identifying regions where local minima and saddles accumulate and cluster around other stationary points, thus displaying the presence of correlations in the landscape. We discuss the implications of these findings for the dynamical exploration of the energy landscape in the activated regime, specifying conditions under which transitions between local minima are expected to exhibit correlated rates and when, conversely, activated jumps are likely to be memoryless. 
\end{abstract}

\maketitle
Understanding low-temperature dynamics in disordered systems is a longstanding challenge that has been tackled through complementary approaches over the years. In low dimension, new algorithms and ideas have provided significant insights into the low-temperature dynamics of interacting particles and disordered elastic interfaces~\cite{ninarello2017models, ferrero2017spatiotemporal,liu2018creep}. One of the most striking observations made in recent years, both in experiments~\cite{durin2024earthquakelike, korchinski2024microscopic} and in numerical simulations~\cite{ferrero2021creep, scalliet2022thirty,tahaei2023scaling, de2024dynamical}, are {\em thermal avalanches}, i.e. the occurrence of a cascade of smaller activations following a slow activated nucleation. 
In parallel, high-dimensional models have been a fundamental playground to develop a powerful theoretical framework, able to capture glassy phenomena and out-of-equilibrium dynamics~\cite{parisi2020theory, charbonneau2017glass, agoritsas2018out, agoritsas2019out}. Significant insight has been gained by studying mean-field dynamical equations, which are exact for fully connected models in the limit of infinite system size~\cite{sompolinsky1981dynamic}.
Starting from the seminal works~\cite{crisanti1993spherical, cugliandolo93analytical}, the pure spherical $p$-spin model, a simple model of random landscapes with Gaussian statistics, has been the prototypical toy model used to gain an analytical understanding of \emph{relaxational} glassy dynamics~\cite{cugliandolo93analytical,kirkpatrick1989scaling, cugliandolo2011effective,franz2013quasi, bouchaud1998out}. Developing a theory for \emph{activated} dynamics in complex energy landscapes remains however an open challenge, even in the context of simple models such as the pure $p$-spin. Indeed, in such fully-connected model the energy barriers diverge in the limit of infinite system size (mean-field limit); thus, activated dynamical processes, where systems jump between local minima of the energy landscape by crossing these energy barriers, cannot be captured directly. For systems with large but finite size, two complementary approaches can be considered: one can attempt to describe these rare dynamical processes by examining large deviations in the mean-field dynamical equations~\cite{lopatin1999instantons, rizzo2021path}; in parallel, insight into activated dynamics can also be sought by studying the structure of the energy landscape~\cite{ros2021dynamical, ros2019complexity, ros2020distribution}.

\begin{figure}[ht!]
\centering
\captionsetup{justification=justified}
\includegraphics[width=0.41\textwidth, trim= 3 3 3 3, clip]{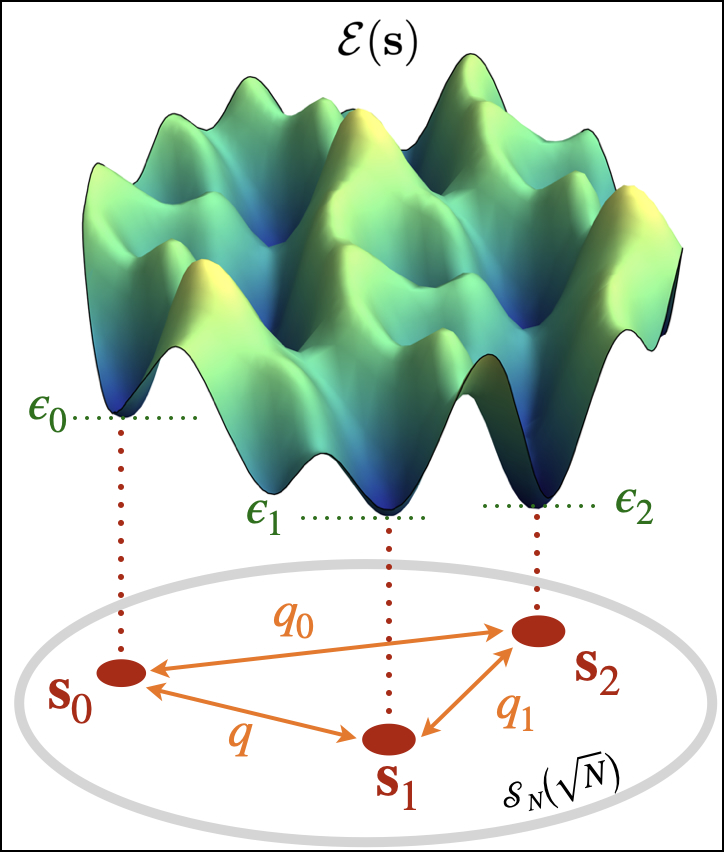}
\caption{Pictorial representation of the energy landscape defined on the configuration space $\mathcal{S}_N(\sqrt{N})$, with three stationary points ${\bf s}_0$, ${\bf s}_1$ and ${\bf s}_2$ at overlaps $q, q_0$ and $q_1$. 
}
\label{fig:landscape}
\end{figure}

In this work, we consider this second approach: motivated by the problem of activated dynamics, we investigate the distribution of triplets of stationary points in the energy landscape of the spherical $p$-spin model, see Fig.~\ref{fig:landscape}. More precisely, we compute a three-point complexity, which counts the number of stationary points ${\bf s}_2$ at given energy density $\epsilon_2$, as a function of their overlaps (i.e., proximity) in configuration space from a pair of stationary points ${\bf s}_1,{\bf s}_0$ at energy densities $\epsilon_1, \epsilon_0$. We use the distribution of ${\bf s}_2$ as a probe of the structure of the landscape in the vicinity of ${\bf s}_0$, ${\bf s}_1$. 
We identify two distinct scenarios: when the energy densities are small enough, the maximal three-point complexity reduces to the two-point complexity previously computed in \cite{ros2019complexity, cavagna1997investigation}. Hence, the majority of the configurations ${\bf s}_2$ are arranged in the landscape in a way that is independent of ${\bf s}_0$ (or of  ${\bf s}_1$). Instead, when the energy densities $\epsilon_1, \epsilon_2$ are large enough compared to $\epsilon_0$, the landscape displays different regimes when tuning the overlap $q$ between ${\bf s}_0$ and ${\bf s}_1$: (i) for small overlap $q$ (when ${\bf s}_0$ and  ${\bf s}_1$ are well separated in configuration space), the presence of ${\bf s}_0$ generates a \emph{depletion} of the stationary points ${\bf s}_2$ around ${\bf s}_1$; (ii) for intermediate $q$ we observe an anomalous \emph{accumulation} of the stationary points ${\bf s}_2$ near ${\bf s}_0$; (iii) for large $q$, the points ${\bf s}_2$ \emph{cluster} close to ${\bf s}_1$. For these larger values of $q$, close to a deep minimum, higher-energy stationary points are more densely packed than they are in typical regions of configuration space, far away from the deep minimum. This is a signature of strong local correlations in the energy landscape. 

We derive some implications of our results for the activated dynamics of the pure $p$-spin model. In principle, to characterize activated processes one should have access to the distribution of the energy barriers between minima, a problem that is notoriously challenging (see \cite{ros2023high} for a brief discussion of recent advancements). It is however reasonable to assume that the optimal energy barrier between minima grows with the distance between them in configuration space, since the larger is the distance, the larger is the amount of local rearrangements to connect them. Within this assumption, we consider an effective dynamics where the system jumps among closest minima at a given (possibly, the same) energy. 
Making use of our landscape's results we show
that jumps between local minima at \emph{small enough} energy density are ‘‘memoryless": they are characterized by a typical energy barrier that does not depend on configurations previously visited by the system. Based on this, we argue that thermal avalanches do not play a role in the $p$-spin activated dynamics, at least when the system visits low energy configurations. However, we also find a precursor of the thermal avalanches: subsequent jumps \emph{from a deep minimum to high-energy minima} are not memoryless but display correlations, and large energy barriers are systematically followed by smaller ones.  

The work is structured as follows: in Sec.~\ref{sec:the_model} we introduce the model. In Sec.~\ref{sec:2and3compelxity} we recap known results on the two-point complexity, we define the three-point complexity, which is the focus of this work. In Sec.~\ref{subsec:defs_clustering} we define the notions of \textit{local accumulation} and \textit{clustering} of stationary points, which we use as an indication of the local correlations in the distribution of the stationary points. In Sec.~\ref{sec:3point_results} we present in detail the results of our landscape analysis. In Sec.~\ref{sec:Dynamics} we discuss the implications for the activated dynamics, and we draw our conclusions in Sec.~\ref{sec:conclusion}. In the Appendices we present details on the calculations of the three-point complexity within an approximation scheme (the so called annealed approximation), and we discuss the analysis of the stability of the stationary points counted by the complexity. Ref.~\cite{PaperLungo} will be devoted to reporting details of the calculation of the three-point complexity in the quenched setting.

\section{The pure spherical $p$-spin model}
\label{sec:the_model}

The pure spherical $p$-spin model is one of the simplest models exhibiting glassiness \cite{gross1984simplest, crisanti1992sphericalp, crisanti1993spherical}. 
The degrees of freedom are soft spins ${\mathbf{s}} =( s_1,\ldots,  s_N)\in\mathbb{R}^N$ lying on the surface of an hypersphere in dimension $N$, i.e., ${\mathbf{s}}^2=\sum_{i=1}^N  s_i^2=N$. We denote such hypersphere with $\mathcal{S}_N(\sqrt N)$ (in short, $\mathcal{S}_N$) in the following.  The energy associated to each configuration is a random function with Gaussian statistics, 
\begin{equation}
\label{eq:landscape}
\mathcal{E}({\bf s})= \sqrt{\frac{ p!}{2 N^{p-1}}}\sum_{i_1<\ldots <i_p}a_{i_1\ldots i_p} s_{i_1}\ldots  s_{i_p},
\end{equation}
where $p \geq 3$, the quenched random couplings $a_{i_1\ldots i_p}$ are i.i.d Gaussian random variables with zero mean and unit variance, and the prefactor in \eqref{eq:landscape} guarantees the extensive scaling of the energy. We denote with $q({\bf s}_0, {\bf s}_1)= N^{-1} \tonde{{\bf s}_0\cdot {\bf s}_1}$ the overlap between two configurations, which measures how close they are in configuration space, and with  $\epsilon({\bf s}) = N^{-1}\,\mathcal{E}({\bf s})$ the energy density associated to a configuration. The function \eqref{eq:landscape} has zero average and  covariance
\begin{equation}\label{eq:Enp}
\mathbb{E}\left[ \mathcal{E}({\bf s}_0)  \mathcal{E}({\bf s}_1)\right] = \frac{N}{2} \,  \quadre{q({\bf s}_0, {\bf s}_1)}^p,
\end{equation}
where $\mathbb{E}\left[  \cdot  \right]$ denotes the average with respect to fluctuations of the couplings $a_{i_1\ldots i_p}$. As it appears from this formula, the values taken by the energy landscape at configurations at zero overlap are independent random variables. Stationary points of the landscape are configurations ${\bf s}$ which satisfy the constraint $\nabla_\perp \mathcal{E}({ \bf s})=0$, where $\nabla_\perp$ denotes the gradient of the function $\mathcal{E}({\bf s})$ restricted to the surface of the sphere $\mathcal{S}_N(\sqrt{N})$ (meaning that the variations of the function in the radial direction of the sphere, parallel to the vector ${\bf s}$, are not considered). 

The statistical properties of the stationary points of the Gaussian landscape \eqref{eq:landscape} have been object of interest since the early works on mean-field glassy models \cite{cavagna1997investigation, cavagna1998stationary, franz1995recipes, cavagna1997structure}; in recent years, this interest has revived in connections to problems of high-dimensional inference and  topology \cite{ros2019complex, arous2019landscape, auffinger2013random, kent2021complex}. Gaussian landscapes obtained by taking linear combinations of \eqref{eq:landscape} for different values of $p$, known as mixed spherical $p$-spin models \cite{MixedModelNieuwenhuizen95,crisanti2006spherical}, have also garnered renewed attention recently~\cite{barbier2020constrained, kent2024arrangement, kent2024conditioning}, motivated by intriguing results on the associated Langevin dynamics \cite{folena2020rethinking, EquilMixedLeuzzi07}. For the model \eqref{eq:landscape}, it is by now mathematically proven~\cite{auffinger2013random, subag2017complexity} that each energy level $\epsilon$ above the ground state energy $\epsilon_{\rm gs}$ hosts an exponentially-large number $\mathcal{N}(\epsilon)$ of stationary points, measured by the landscape's complexity
\begin{equation}
    \Sigma(\epsilon)= \lim_{N \to \infty} \frac{1}{N}\mathbb{E}[ \log \mathcal{N}(\epsilon)].
\end{equation}
At the threshold energy density $\epsilon_{\rm th}=-\sqrt{2 (p-1)/p}$ a stability transition occurs: for $\epsilon > \epsilon_{\rm th}$ the stationary points are typically saddles\footnote{In this context, typically means that the exponential majority (in $N$) of stationary points at the given energy density is a saddle.} with a large index $\propto N$ (the index is the number of directions in configuration space along which the landscape has negative curvature). For $\epsilon < \epsilon_{\rm th}$, instead,  the exponential majority of stationary points are local minima (even though a smaller, but still exponentially large number of saddles of low-index is present  \cite{cavagna1998stationary, auffinger2013random, auffinger2020number}). This low-energy portion of the landscape corresponds to equilibrium energies at low temperature. Equilibration at these temperatures requires the exploration of regions of the energy landscape dominated by local minima separated by high energy barriers, that scale extensively with $N$. Such regions are inaccessible to the mean-field dynamics ($N \to \infty$) starting from random initial conditions, which instead relaxes to the threshold energy $\epsilon_{\rm th}$ at asymptotically large times. For $N$ large but finite, the low-energy portion of the landscape is eventually explored at exponentially-large (in $N$) timescales by means of rare activated jumps between local minima. In contrast to relaxational dynamics, this activated regime of the dynamics is much less well understood.

\begin{figure}[t!]
\includegraphics[width=0.46
\textwidth, trim=6 6 6 6,clip]{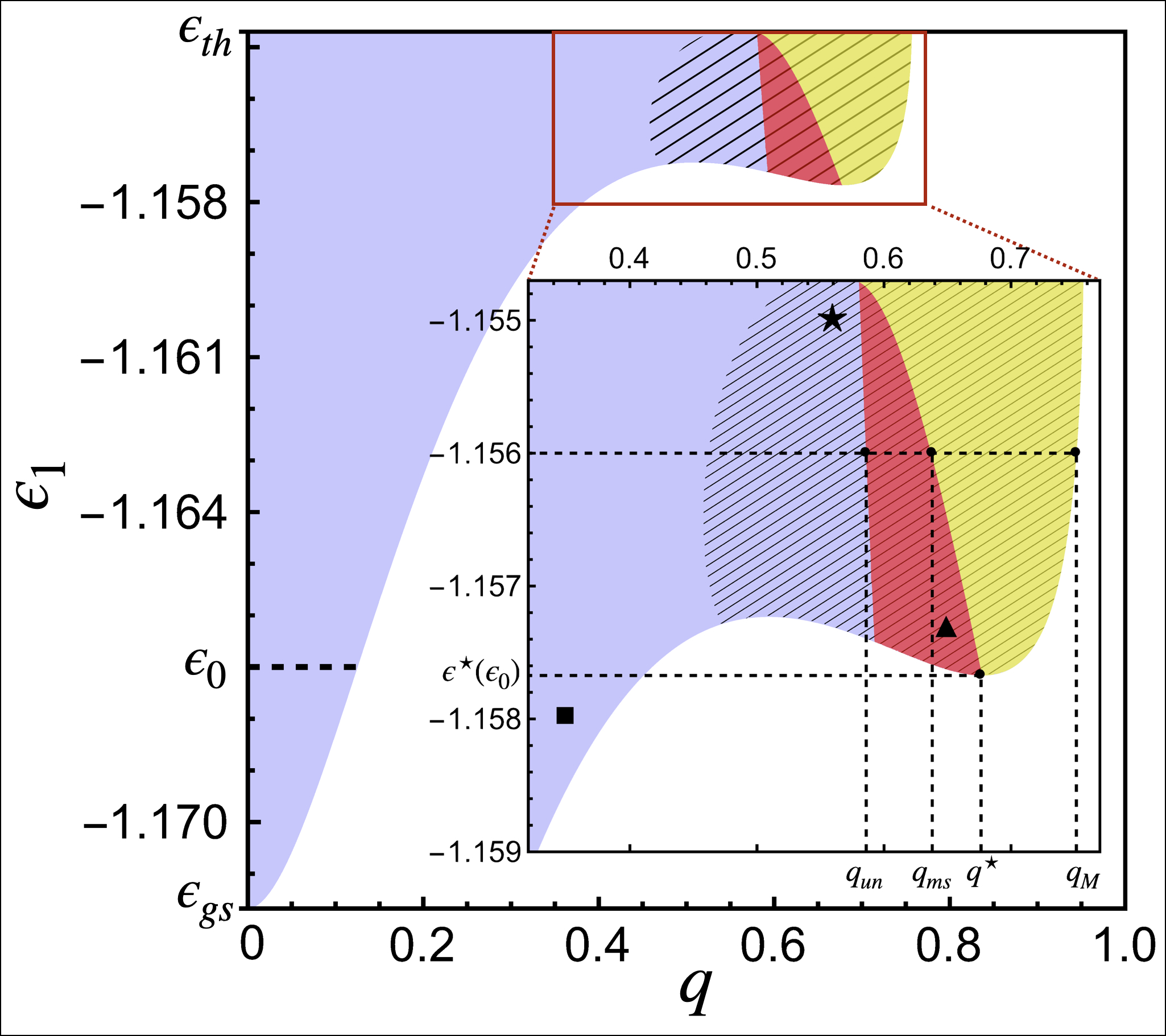}\\
\includegraphics[width=0.46\textwidth, trim= 7 6 6 6, clip]{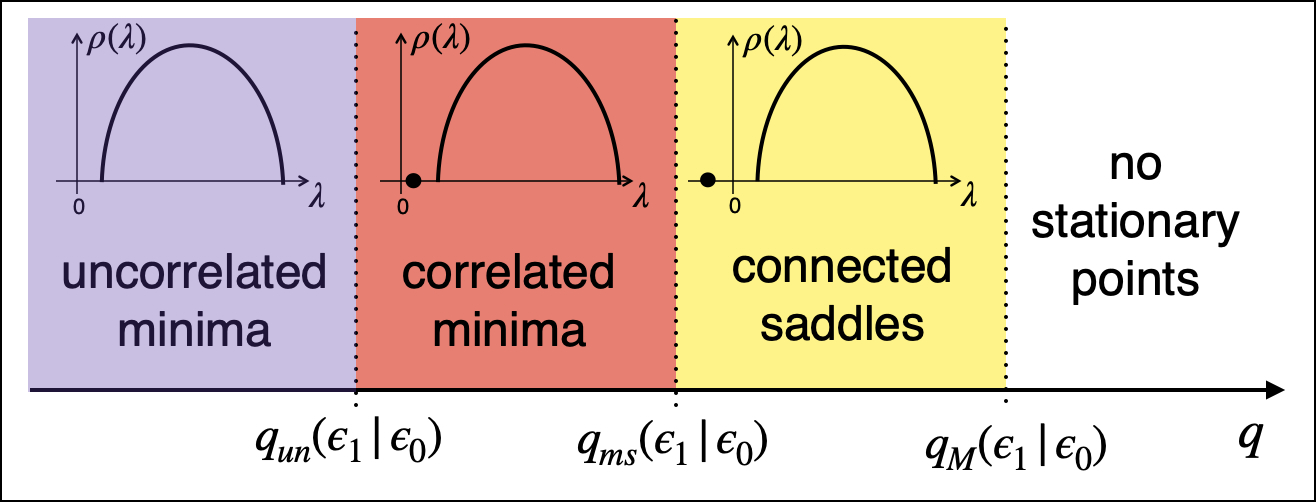}
\caption{ \textit{Top.} The colored area shows the range of $\epsilon_1$ and $q$ where $\Sigma^{(2)}(\epsilon_1,q|\epsilon_0) \geq 0$ for $\epsilon_0 = -1.167$. Blue indicates minima, red indicates correlated minima, and yellow indicates index-1 saddles (see \textit{Bottom} picture). The lowest energy of index-1 saddles is $\epsilon^\star(\epsilon_0)$, with overlap $q^\star(\epsilon_0)$. \textit{Inset}. The black hatched region marks the values of $q$ for which clustering occurs with $\epsilon_2 = \epsilon_1$, and $\epsilon_0 = -1.167$. We refer to Sec.~\ref{subsec:defs_clustering} for a definition of clustering. The symbols $\blacksquare, \blacktriangle, \bigstar$ mark specific values of parameters listed in Table \ref{fig:table} and considered in  Fig.~\ref{Fig:Plots3D}. \textit{Bottom.} Eigenvalue distribution of the Hessian at the stationary points ${\bf s}_1$ for $\epsilon_1 > \epsilon^\star(\epsilon_0)$. For $q < q_{\rm un}(\epsilon_1|\epsilon_0)$, the eigenvalues are distributed according to a semicircle law and ${\bf s}_1$ is an uncorrelated minimum. For $q_{\rm un}(\epsilon_1|\epsilon_0) < q < q_{\rm ms}(\epsilon_1|\epsilon_0)$, the Hessian has a positive isolated eigenvalue, and  ${\bf s}_1$ is a correlated minimum. For $q_{\rm ms}(\epsilon_1|\epsilon_0) < q < q_M(\epsilon_1|\epsilon_0)$, ${\bf s}_1$ is an index-1 saddle. No stationary points exist for $q > q_M(\epsilon_1|\epsilon_0)$.}
\label{fig:2D_plot_clustering}
\end{figure}

\section{The two-point and three-point complexity}\label{sec:2and3compelxity}

\subsection{Previous results: the two-points complexity}\label{sec:TwoPoint}
The geometry of the landscape in the vicinity of a local minimum can be described~\cite{ros2019complexity,ros2020distribution} by computing a two-point complexity $\Sigma^{(2)}(\epsilon_1, q |\epsilon_0)$, defined as
\begin{align}
\label{eq:quenchedcomp2}
\Sigma^{(2)}(\epsilon_1,q|\epsilon_0)=\lim_{N\to\infty}\frac{1}{N}\mathbb{E}\left[\log\mathcal{N}_{{\bf s}_0}(\epsilon_1,q )\right]_0.
\end{align}
Here $\mathcal{N}_{{\bf s}_0}(\epsilon_1,q)$ 
is the number of stationary points  
 ${\bf s}_1$ of energy $\epsilon_1$, that are at overlap $q$ with another stationary point ${\bf s}_0$. The average $\mathbb{E} \left[\cdot \right]_0$ denotes both a flat average over all stationary points ${\bf s}_0$ with energy $\epsilon_0$ at fixed realization of the landscape, and over the realizations of the landscape. More precisely:
 \begin{equation}
     \mathbb{E} \left[\cdot \right]_0= \mathbb{E}\left[ \frac{1}{\mathcal{N}(\epsilon_0)} \int_{\mathcal{S}_N} d{\bf s}_0 \, \omega_{\epsilon_0}({\bf s}_0) \, \cdot \; \right],
 \end{equation}
 where 
  \begin{equation}\label{eq:Measure0}
 \omega_{\epsilon_0}({\bf s}_0)=|\det \nabla^2_\perp \mathcal{E}({\bf s}_0)|\delta(\nabla_\perp \mathcal{E}({\bf s}_0))\delta(\mathcal{E}({\bf s}_0)-N\epsilon_0)
\end{equation}
 is the measure selecting configurations ${\bf s}_0$ that are stationary points of energy density $\epsilon_0$. In Eq.~\eqref{eq:Measure0}, $\nabla^2_\perp \mathcal{E}$ denotes the Hessian matrix of the random landscape $\mathcal{E}$ restricted to the surface of the hypersphere. By normalization, we have
\begin{equation}\label{eq:Norma1}
    \begin{split}
&\mathcal{N}(\epsilon_0)= \int_{\mathcal{S}_N} d {\bf s}_0  \,\omega_{\epsilon_0}({\bf s}_0).
    \end{split}
\end{equation}
The explicit expression of the two-point complexity has been derived in \cite{ros2019complexity} (see also \cite{cavagna1997investigation, subag2017complexity}), and it reads: 
\begin{equation}\label{eq:FinalComplexityPostSaddle}
\Sigma^{(2)}(\epsilon_1,q|\epsilon_0) =   \frac{Q(q)}{2}-f(\epsilon_0,\epsilon_1, q)+ I\tonde{\epsilon_1 \sqrt{\frac{p}{p-1}}} 
 \end{equation}
 where: 
\begin{equation*}
 \begin{split}
           &Q(q)=  1+\log \tonde{\frac{2(p-1)(1-q^2)}{1-q^{2p-2}}},\\
           &f(\epsilon_0,\epsilon_1, q)= \epsilon_0^2 U_0(q)+ \epsilon_0 \epsilon_1 U(q)+ \epsilon^2_1 U_1(q)
            \end{split}
\end{equation*}
 with 
\begin{equation}\label{eq:UC}
 \begin{split}
  U_0(q)&=\frac{q^{2 p} [p q^2-q^4(p-1)-q^{2 p}]}{\mathcal{A}(q)},\\
  U(q)&=\frac{2 q^{3 p} \left(p \left(q^2-1\right)+1\right)-2 q^{p+4}}{\mathcal{A}(q)},\\
  U_1(q)&=\frac{q^4-q^{2 p} - p q^{2 p}  [(p-1) q^4+(3-2 p) q^2+p-2]}{\mathcal{A}(q)},\\
  \mathcal{A}(q)&=q^{4 p}-q^{2 p}[(p-1)^2 (1+q^4)-2 (p-2) p q^2] +q^4,
 \end{split}
\end{equation}
and where 
the symmetric function $I(y)$ reads explicitly
\begin{align}\label{eq:IDef}
    I(y)=\begin{cases}I_-(y)
        \quad \text{if } y\leq -\sqrt{2}\\
        I_+(y)\quad \text{if }-\sqrt{2}\leq y\leq 0
    \end{cases}
\end{align}
with 
\begin{equation}
\begin{split}
&I_-(y)=\frac{y^2-1}{2}+\frac{y}{2}\sqrt{y^2-2}+\log\left(\frac{-y+\sqrt{y^2-2}}{2}\right)\\
& I_+(y)=\frac{1}{2}y^2-\frac{1}{2}(1+\log 2).
 \end{split}
\end{equation}
The results of this calculation are summarized in Fig.~\ref{fig:2D_plot_clustering} for a representative value of $\epsilon_0 < \epsilon_{\rm th}$. The colored region in the figure identifies the values of $q, \epsilon_1$ for which the function \eqref{eq:quenchedcomp2} is positive (the plot is cutoff at $\epsilon_1=\epsilon_{\rm th}$, since at $\epsilon_1>\epsilon_{\rm th}$ the landscape is dominated by saddles with large index; this portion of the landscape is easily explored by relaxational dynamics, and it is therefore not of interest for our analysis). For $q=0$, the range of energy density is maximal, and extends down to the ground state energy $\epsilon_{\rm gs}$: at $q=0$ one has the largest two-points complexity, meaning that most of the stationary points of the landscape are at zero overlap with the reference one at energy $\epsilon_0$. In fact,  one has  
\begin{equation}
    \lim_{q \to 0}\Sigma^{(2)}(\epsilon_1,q|\epsilon_0)= \Sigma(\epsilon_1),
\end{equation}
meaning that one recovers the  expression of the unconstrained complexity~\cite{crisom95, cavagna1998stationary} counting the number of minima irrespective of their location in configurations space.
When $q$ increases, the range of energies at which one finds a positive complexity first decreases, and then increases again at the larger values of $q$, reaching a local maximum at a given $q^*(\epsilon_0)$. The maximal $q$ at which one finds stationary points at energy below the threshold one is $q=q_M(\epsilon_0)$. For each value of $\epsilon_1$, one can define the maximal overlap at which stationary points of that energy density are found: this is denoted with 
\begin{equation}\label{eq:qMAx}
q_M(\epsilon_1|\epsilon_0) \equiv \text{max  } q \text{  such that } \Sigma^{(2)}(\epsilon_1,q|\epsilon_0) \geq 0.
\end{equation}

The different colors in Fig.~\ref{fig:2D_plot_clustering} are related to the linear stability of the stationary points found at those values of $q, \epsilon_1$; the linear stability is described by the spectrum of the Hessian matrices $\nabla^2_\perp \mathcal{E}$ at the stationary points, whose statistical properties are recalled in Sec.~\ref{sec:stability}. The blue and red region correspond to stationary points whose Hessian has all eigenvalues positive: these points are thus local minima of the landscape. The red area corresponds to minima whose Hessian has a single mode that is detached from the rest of the eigenvalues distribution (it is an isolated eigenvalue), that is smaller and whose eigenvector is partially aligned in the direction of the reference minimum of energy $\epsilon_0$; these minima thus display a softest curvature in the direction of the reference minimum, and we call them {\it correlated minima} to emphasize  that their Hessians displays correlations with ${\bf s}_0$. Finally, the yellow area corresponds to rank-1 saddles, with one single Hessian mode that is negative and correlated with the direction of the reference minimum. These saddles are geometrically connected to the minimum, but also dynamically, meaning that the dynamics starting from the saddle relaxes to the local minimum~\cite{ros2021dynamical}. For each $\epsilon_1$, we denote with $q_{\rm un}(\epsilon_1|\epsilon_0)$ (where the subscript  “un" stands for  “uncorrelated") and  $q_{\rm ms}(\epsilon_1|\epsilon_0)$ (where the subscript  “ms" stands for  “minima-to-saddles") the overlaps at which the corresponding transitions occur, see Fig.~\ref{fig:2D_plot_clustering} {\it bottom} for a sketch. In Table~\ref{fig:table} we list the values of these special overlap parameters for a fixed choice of $\epsilon_0$, which is the value that we choose for all the plots presented in this work. 

The overlap $q^*(\epsilon_0)$ is associated to an energy density $\epsilon^*(\epsilon_0)$, that plays a crucial role in our subsequent discussion. Fig.~\ref{fig:2D_plot_clustering} shows that this is also the critical energy above which index-1 saddles and correlated minima appear in the landscape in the vicinity of the reference minimum ${\bf s}_0$. How this critical energy depends on $\epsilon_0$ is illustrated in the inset of Fig.~\ref{fig:2D_plot_clustering_e1e2}.

\begingroup
\renewcommand{\arraystretch}{1.5}
\begin{table}[h!]
    \centering
    \begin{tabular}{|c|c|c|c|c|c|}
        \hline
        \multicolumn{6}{|c|}{$\epsilon_{gs}\approx -1.17167,\quad\epsilon_{th}\approx -1.1547$} \\
        \hline
        \multicolumn{6}{|c|}{$\epsilon_0=-1.167,\quad \epsilon^\star(\epsilon_0)\approx-1.15767$} \\
        \hline
        \text{icon} &$\bm{\epsilon}$ & $\bf{q_{un}}(\epsilon|\epsilon_0)$ & $\bf{q_{ms}}(\epsilon|\epsilon_0)$ & $\bf{q_{M}}(\epsilon|\epsilon_0)$ & $\bf{q}$ \\
        \hline
        $\bigstar$ & -1.155 & 0.5763 & 0.6028 & 0.7564 & 0.56\\
        \hline
        $\blacktriangle$ & -1.1573 & 0.586 & 0.669 & 0.7266 & 0.65\\
        \hline
        $\blacksquare$ & -1.158 &  &  & 0.3829 & 0.35\\
        \hline
    \end{tabular}
    \caption{Values of the parameters associated to the points marked in Fig.~\ref{fig:2D_plot_clustering}.}
    \label{fig:table}
\end{table}
\endgroup

\subsection{The three-points complexity: definition}
\label{subsec:3_pt_def}
In this work we extend the analysis of~\cite{ros2019complexity,ros2020distribution} by computing the asymptotic behavior (for large $N$) of the typical value of the random variable $\mathcal{N}_{{\bf s}_0\, {\bf s}_1}(\epsilon_2, q_0, q_1)$. This is the number of stationary points ${\bf s}_2$ of energy density $\epsilon_2$, that are found at overlap $q_1$ with a stationary point ${\bf s}_1$ and at overlap $q_0$  from another stationary point ${\bf s}_0$, see Fig.~\ref{fig:landscape}. In turn, the stationary points ${\bf s}_1$ and ${\bf s}_0$ are at a given overlap $q$ with each others.
We are interested in the three-point complexity $\Sigma^{(3)}(\epsilon_2, q_0, q_1 | \epsilon_1,\epsilon_0, q)$, defined as 
\begin{align}
\label{eq:quenchedcomp3}
\Sigma^{(3)}=\lim_{N\to\infty} \frac{1}{N}\mathbb{E}\left[\log\mathcal{N}_{{\bf s}_0 \,{\bf s}_1}(\epsilon_2, q_0, q_1)\right]_{0,1}.
\end{align}
 In this case, the average $\mathbb{E} [ \cdot ]_{0,1}$ denotes the flat average over the stationary points ${\bf s}_1$ with energy density $\epsilon_1$, \emph{constrained} to be at overlap $q$ with another stationary point ${\bf s}_0$ with energy density $\epsilon_0$ extracted with a flat measure; additionally, the landscape is averaged over:
\begin{equation}\label{eq:averages}
\mathbb{E} [ \cdot ]_{0,1}=\mathbb{E}\left[
\int_{\mathcal{S}_N} d{\bf s}_0\frac{\omega_{\epsilon_0}({\bf s}_0)}{\mathcal{N}(\epsilon_0)}\int_{\mathcal{S}_N} d{\bf s}_1 \frac{\omega_{\epsilon_1,q}({\bf s}_1|{\bf s}_0)}{\mathcal{N}_{{\bf s}_0}(\epsilon_1,q)} \; \cdot\right],
\end{equation}
where now 
\begin{equation}\label{eq:Measure1}
  \begin{split}
&\omega_{\epsilon_1,q}({\bf s}_1|{\bf s}_0)=|\det \nabla^2_\perp \mathcal{E}({\bf s}_1)|\delta(\nabla_\perp \mathcal{E}({\bf s}_1))\times\\
&\delta(\mathcal{E}({\bf s}_1)-N\epsilon_1)\delta({\bf s}_1 \cdot {\bf s}_0-N q)
\end{split}
\end{equation}
 is the measure that selects configurations ${\bf s}_1$ that are stationary points with given parameters $\epsilon_1, q$.  Similarly as above, we have:
\begin{equation}\label{eq:Norma2}
    \begin{split}
&\mathcal{N}_{{\bf s}_0}(\epsilon_1, q)= \int_{\mathcal{S}_N} d {\bf s}_1\,  \omega_{\epsilon_1,q}({\bf s}_1 | {\bf s}_0).
    \end{split}
\end{equation}
The parameters $\epsilon_1, q$ are chosen in such a way that typically stationary points are found at those values, i.e., $\Sigma^{(2)}(\epsilon_1,q|\epsilon_0) \geq 0$ (the colored region in Fig. \ref{fig:2D_plot_clustering}). The number $\mathcal{N}_{{\bf s}_0\, {\bf s}_1}(\epsilon_2, q_0, q_1)$ is obtained as
\begin{equation}\label{eq:enne}
   \mathcal{N}_{{\bf s}_0\, {\bf s}_1}(\epsilon_2, q_0, q_1)= \int_{\mathcal{S}_N} d {\bf s}_2\,  \omega_{\epsilon_2,q_0, q_1}({\bf s}_2 | {\bf s}_1,{\bf s}_0)
\end{equation}
where now
\begin{equation}\label{eq:Measure2}
  \begin{split}
&\omega_{\epsilon_2,q_0, q_1}({\bf s}_2|{\bf s}_1,{\bf s}_0)=|\det \nabla^2_\perp \mathcal{E}({\bf s}_2)|\delta(\nabla_\perp \mathcal{E}({\bf s}_2))\times\\
&\delta(\mathcal{E}({\bf s}_2)-N\epsilon_2)\delta({\bf s}_2 \cdot {\bf s}_0-N q_0)\delta({\bf s}_2 \cdot {\bf s}_1-N q_1).
\end{split}
\end{equation}
Notice that in the definition of $\Sigma^{(3)}$, the roles of ${\bf s}_1$ and ${\bf s}_0$ are not interchangeable: while ${\bf s}_0$  is selected with no other constraints than its energy, ${\bf s}_1$ is selected with a measure that is \emph{conditioned} to the overlap with ${\bf s}_0$. We also remark that the three-point complexity differs from the zero-temperature limit of the three-replica potential introduced in \cite{cavagna1997structure}, in that the configurations are extracted with a different measures which enforces them to be stationary points of the landscape.\\

\subsection{Quenched \emph{vs} annealed complexity}
\label{subsec:quenched_annealed}
In the language of disordered systems, the three-point complexity~\eqref{eq:quenchedcomp3} is a \emph{quenched} complexity. The terminology indicates the fact that the complexity is obtained averaging the logarithm of the exponentially-scaling quantity $\mathcal{N}_{{\bf s}_0\, {\bf s}_1}(\epsilon_2, q_0, q_1)$; therefore, $\Sigma^{(3)}$ determines the asymptotic scaling of the \emph{typical} number of stationary points ${\bf s}_2$. This has to be compared with the \emph{annealed} complexity, that is defined as: 
\begin{align}
\label{eq:annealedcomp3}
\Sigma^{(3)}_{A}(\epsilon_2, q_0, q_1 | \epsilon_1,\epsilon_0, q)=\lim_{N\to\infty}\frac{\log \mathbb{E}[\mathcal{N}_{{\bf s}_0\,{\bf s}_1}(\epsilon_2, q_0, q_1)]_{0,1}}{N}
\end{align}
with the same average as in \eqref{eq:averages}. Of course, this quantity controls the asymptotic scaling of the \emph{average} number of stationary points ${\bf s}_2$, which may significantly differ from the typical value (this happens whenever the average is dominated by realizations of the random landscape that are exponentially rare, but which present an atypically large number of stationary points). The annealed complexity is an upper bound to the quenched one,
\begin{equation}\label{eq:bound}
\Sigma^{(3)}(\epsilon_2, q_0, q_1 | \epsilon_1,\epsilon_0, q) \leq \Sigma^{(3)}_A(\epsilon_2, q_0, q_1 | \epsilon_1,\epsilon_0, q),
\end{equation}
as it follows from the fact that the logarithm is a concave function. It is often introduced as an approximation to the quenched one, being it significantly simpler to compute (since it does not require to compute the expected value of the logarithm of a random function, which is done exploiting algebraic tricks such as  the replica trick). In the setting we are considering, the annealed complexity \eqref{eq:annealedcomp3} still requires some form of replica trick to be calculated,  due to the presence of the denominators $\mathcal{N}(\epsilon_0)$ and $\mathcal{N}_{{\bf s}_0}(\epsilon_1,q)$ in \eqref{eq:averages}.
 Such denominators are random functions guaranteeing the normalization of the measures with which the two stationary points ${\bf s}_0, {\bf s}_1$ are selected. The standard way to treat them is to make use of the identity $x^{-1}=\lim_{n \to 0} x^{n-1}$, which allows to replace the average of a ratio of random variables with the average of a power, provided that the limit $n \to 0$ is taken afterwards. This is one instance of the replica trick (see \cite{franz1998effective} for a similar calculation). To bypass the use of replicas, one may consider an approximation in which the expectation value of the ratio in \eqref{eq:averages} is factorized into the ratio of expectation values of the numerator and denominator, meaning that the average $ \mathbb{E}[ \cdot ]_{0,1}$ is replaced with:
\begin{equation}\label{eq:averagesAnn}
 \mathbb{E}[ \bigcdot ]_{2A}:= \frac{\mathbb{E}\left[
\int_{\mathcal{S}_N^{\otimes 2}} d{\bf s}_0  d{\bf s}_1 \,  \omega_{\epsilon_0}({\bf s}_0) \omega_{\epsilon_1,q}({\bf s}_1|{\bf s}_0) \; \bigcdot\right]}{\mathbb{E}[\mathcal{N}(\epsilon_0)\,  \mathcal{N}_{{\bf s}_0}(\epsilon_1,q)] }.
\end{equation}
We dub the three-point complexity obtained with this approximation the \emph{doubly-annealed} complexity, and we denote it with
\begin{equation}\label{eq:annealedcomp3doubly}
\Sigma^{(3)}_{2A}(\epsilon_2, q_0, q_1 | \epsilon_1,\epsilon_0, q)=\lim_{N\to\infty} \frac{1}{N}\log \mathbb{E}_{2A}[ \mathcal{N}_{{\bf s}_0\,{\bf s}_1}].
\end{equation}
 Notice that the same approximation can be considered for the two-point complexity defined in Eq.~\eqref{eq:quenchedcomp2}, setting:
\begin{equation}\label{eq:annealedcomp2doubly}
\Sigma^{(2)}_{2A}(\epsilon_1, q| \epsilon_0)=\lim_{N\to\infty} \frac{1}{N}\log \mathbb{E}_{2A}[ \mathcal{N}_{{\bf s}_0}]
\end{equation}
 where in this case:
 \begin{equation}
 \mathbb{E}[ \cdot ]_{2A}:= \frac{\mathbb{E}\left[
\int_{\mathcal{S}_N^{\otimes 2}} d{\bf s}_0  \,  \omega_{\epsilon_0}({\bf s}_0) \cdot\right]}{\mathbb{E}[\mathcal{N}(\epsilon_0)] }.
\end{equation}
For the pure spherical $p$-spin model, it has been shown explicitly in \cite{ros2019complexity} that this approximation is actually exact for the two-point complexity, meaning that for all values of the parameters the quenched two-point complexity \eqref{eq:quenchedcomp2} coincides with the annealed one, which was first computed in \cite{cavagna1997investigation}. As we  discuss below, this is in general not the case for the three-point complexity.

\section{Landscape's geometry: local accumulation and clustering}\label{subsec:defs_clustering}
We now introduce some notions and terminology relevant to the subsequent discussion.
Through the calculation of the three-point complexity \eqref{eq:quenchedcomp3}, we aim at determining to what extent the landscape in the vicinity of a stationary point ${\bf s}_0$ (e.g., a deep local minimum) differs from the landscape in typical regions of configuration space that are not conditioned to be near ${\bf s}_0$. In other words, the knowledge of both the two- and three-point complexity allows us to compare the local structure of the landscape (probed by ${\bf s}_2$) in the vicinity of:\\

\begin{itemize}
    \item[(i)]   \emph{typical} stationary points ${\bf s}_1$ with energy density $\epsilon_1$, i.e., stationary points extracted with the uniform measure over all stationary points with that energy density and no additional constraint;
    \item[(ii)]  \emph{conditioned} stationary points ${\bf s}_1$ with  energy density $\epsilon_1$, at overlap $q$ with another stationary point ${\bf s}_0$ with energy density $\epsilon_0$.
\end{itemize}
The first information is encoded in the two point complexity $\Sigma^{(2)}(\epsilon_2, q_1 |\epsilon_1)$, the second one in the three-point complexity $\Sigma^{(3)}(\epsilon_2, q_0, q_1 | \epsilon_1,\epsilon_0, q)$. \\

When comparing the three-point and two-point complexities, it is straightforward to argue that the following inequality must hold for all values of parameters:
\begin{equation}\label{eq:BoundTrivial}
\Sigma^{(3)}(\epsilon_2, q_0, q_1 | \epsilon_1,\epsilon_0, q) \leq \Sigma^{(2)}(\epsilon_2, q_0 |\epsilon_0).
    \end{equation}
Indeed, the number of stationary points ${\bf s}_2$ conditioned on the properties of ${\bf s}_1$ is smaller than the number of stationary points of the same energy density counted without such conditioning. In both the quantities in \eqref{eq:BoundTrivial}, the stationary points ${\bf s}_2$ are enforced to be at given overlap with  ${\bf s}_0$, and only a fraction of these points also satisfy the constraint on the overlap with ${\bf s}_1$. As we justify below, the bound is saturated  for $q_1=q\cdot q_0$.\\
\noindent On the other hand, one can not assume an analogous bound exchanging the role of ${\bf s}_0$ and ${\bf s}_1$, i.e. comparing $\Sigma^{(2)}(\epsilon_2, q_1 |\epsilon_1)$ with $\Sigma^{(3)}(\epsilon_2, q_0, q_1 | \epsilon_1,\epsilon_0, q)$. Indeed, the properties of ${\bf s}_1$ are not the same in these two quantities: in one case ${\bf s}_1$ is a typical stationary point at that energy density, while in the other case it is conditioned. In fact, we shall show that there are values of parameters for which
\begin{equation}\label{eq:Bound2}
\Sigma^{(3)}(\epsilon_2, q_0, q_1 | \epsilon_1,\epsilon_0, q) > \Sigma^{(2)}(\epsilon_2, q_1 |\epsilon_1).
\end{equation}
This leads us to define the notions of \textit{local accumulation} and \textit{clustering} as specific instances of this phenomenon.\\

 \textbf{Local accumulation.} We say that local accumulation occurs whenever for some fixed values of $\epsilon_0,\epsilon_1,q,q_0$, there exists a region of values of $q_1$ such that Eq.~\eqref{eq:Bound2} holds. In other words, there are regions in the vicinity of ${\bf s}_0$ in which the number of stationary points ${\bf s}_2$ (conditioned to the properties of ${\bf s}_1$) is higher than the value predicted by the two-point complexity, which measures the complexity in absence of the conditioning to ${\bf s}_0$. \\

 \textbf{Clustering.} We use the word clustering to designate a special instance of local accumulation, occurring when $\Sigma^{(3)}(\epsilon_2, q_0, q_1 | \epsilon_1,\epsilon_0, q)>0$ but $\Sigma^{(2)}(\epsilon_2, q_1 |\epsilon_1) = - \infty$: there are exponentially many stationary points ${\bf s}_2$ at an overlap $q_1$ (with ${\bf s}_1$) which is large enough that, typically—i.e., in the absence of ${\bf s}_0$—there would be none. In other words, clustering occurs whenever $\Sigma^{(3)}(\epsilon_2, q_0, q_1 | \epsilon_1,\epsilon_0, q)$ is positive for values of $q_1>q_M(\epsilon_2|\epsilon_1)$, where $q_M(\epsilon_2|\epsilon_1)$ is defined in Eq.~\eqref{eq:qMAx} and it identifies the maximal overlap of  stationary points of energy density $\epsilon_2$ with a stationary point of energy density $\epsilon_1$. The inset of Fig.~\ref{fig:2D_plot_clustering} shows (as the hatched region within the two-point complexity), for fixed $\epsilon_0=-1.167$, the region of energies $\epsilon_1$ (in the particular case where $\epsilon_1=\epsilon_2$) and overlaps $q$ such that clustering exists. This means that within that region there exist values of $q_0$ and $q_1>q_M(\epsilon_2|\epsilon_1)$ such that $\Sigma^{(3)}(\epsilon_2, q_0, q_1 | \epsilon_1,\epsilon_0, q)>0$. \\

It is clear that local accumulation (as well as clustering, which is a special case of it) indicate the fact that the landscape in the vicinity of ${\bf s}_0$ is strongly correlated to ${\bf s}_0$ itself: the distribution of the other stationary points in that region is not the typical one. In Sec.~\ref{eq:SpecialLines} and Sec.~\ref{sec:LandscapeEvolution} we show that such correlations are indeed present in the landscape, through a quantitative analysis of the results of the three-point complexity calculation. These landscape correlations also have an interpretation in the context of activated dynamics: as we elaborate in Sec.~\ref{sec:Dynamics}, they can be seen as a signature of avalanche-like behavior in the dynamics.

\section{The three-points complexity: results}\label{sec:3point_results}
In the following, we discuss the results of the calculation of both the annealed and quenched complexity, and devote \cite{PaperLungo} to presenting in full details the derivation of the quenched one. 
We begin by commenting on the relationship between the different functions (quenched, annealed and doubly-annealed complexity) defined in Sec.~\ref{subsec:quenched_annealed}. First, by computing explicitly the annealed complexity~\eqref{eq:annealedcomp3}, we find that in this model it coincides exactly with the doubly-annealed complexity \eqref{eq:annealedcomp3doubly},
\begin{equation}\label{eq:EqaAnn}
   \Sigma^{(3)}_{2A}(\epsilon_2, q_0, q_1 | \epsilon_1,\epsilon_0, q)= \Sigma^{(3)}_{A}(\epsilon_2, q_0, q_1 | \epsilon_1,\epsilon_0, q).
\end{equation}
Therefore, the factorization of the expectation values \eqref{eq:averagesAnn} is justified within the annealed scheme of the calculation. Due to this coincidence, in Appendix~\ref{app:A_com_2A} we focus on the doubly-annealed complexity only, 
reporting its detailed derivation. 
We find on the other hand that the quenched complexity \eqref{eq:quenchedcomp3} is strictly smaller than the annealed one for generic choices of $q_0, q_1$ (see the Appendix~\ref{app:ann_vs_quench} for a comparison between the two functions). For the values of parameters that we explore, however, the numerical values of quenched and annealed complexity happen to be quite close. Moreover, for special choices of the parameters $q_0, q_1$, the bound \eqref{eq:bound} is saturated and the two functions coincide, as we discuss in Sec.~\ref{eq:SpecialLines}.

\begin{figure*}[t!]
  \begin{subfigure}[b]{1\columnwidth}
\includegraphics[width=0.95
\textwidth, trim= 5 5 5 5, clip]{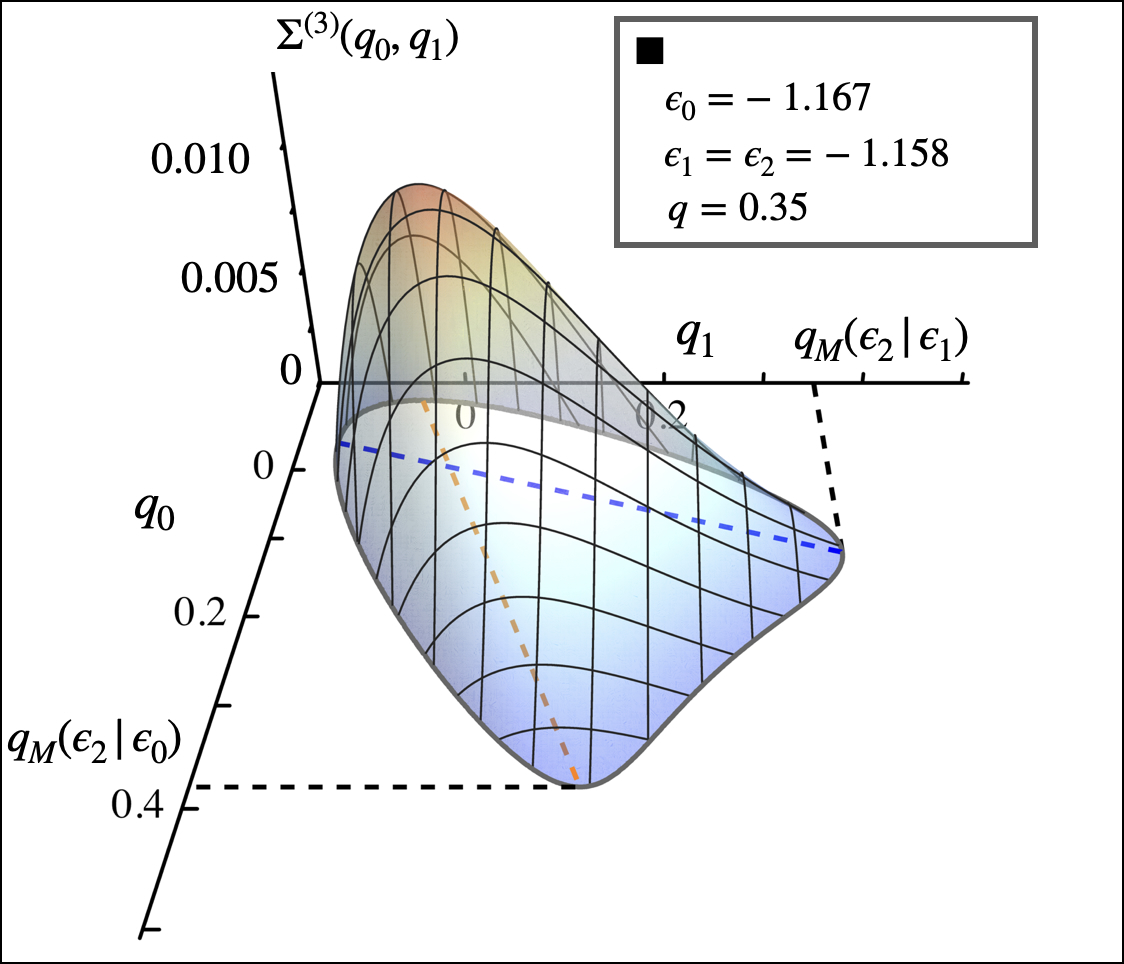}
\caption{}
 \end{subfigure}
 \hfill
  \begin{subfigure}[b]{1\columnwidth}
\includegraphics[width=0.95
\textwidth, trim= 4 4 4 4,clip]{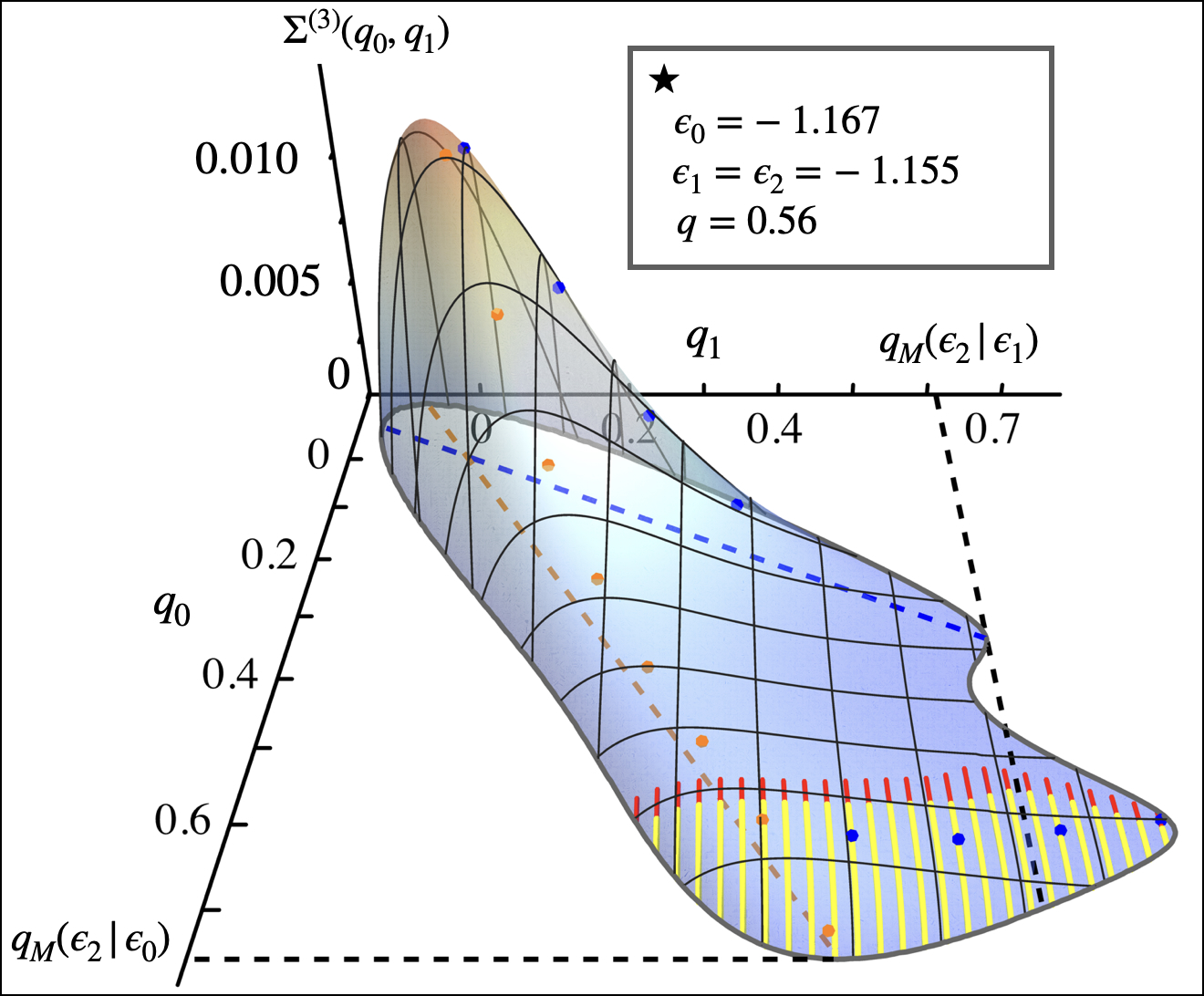}
\caption{}
 \end{subfigure}
 \medskip
   \begin{subfigure}[b]{1\columnwidth}
\includegraphics[width=0.95
\textwidth, trim=4 4 4 4, clip]{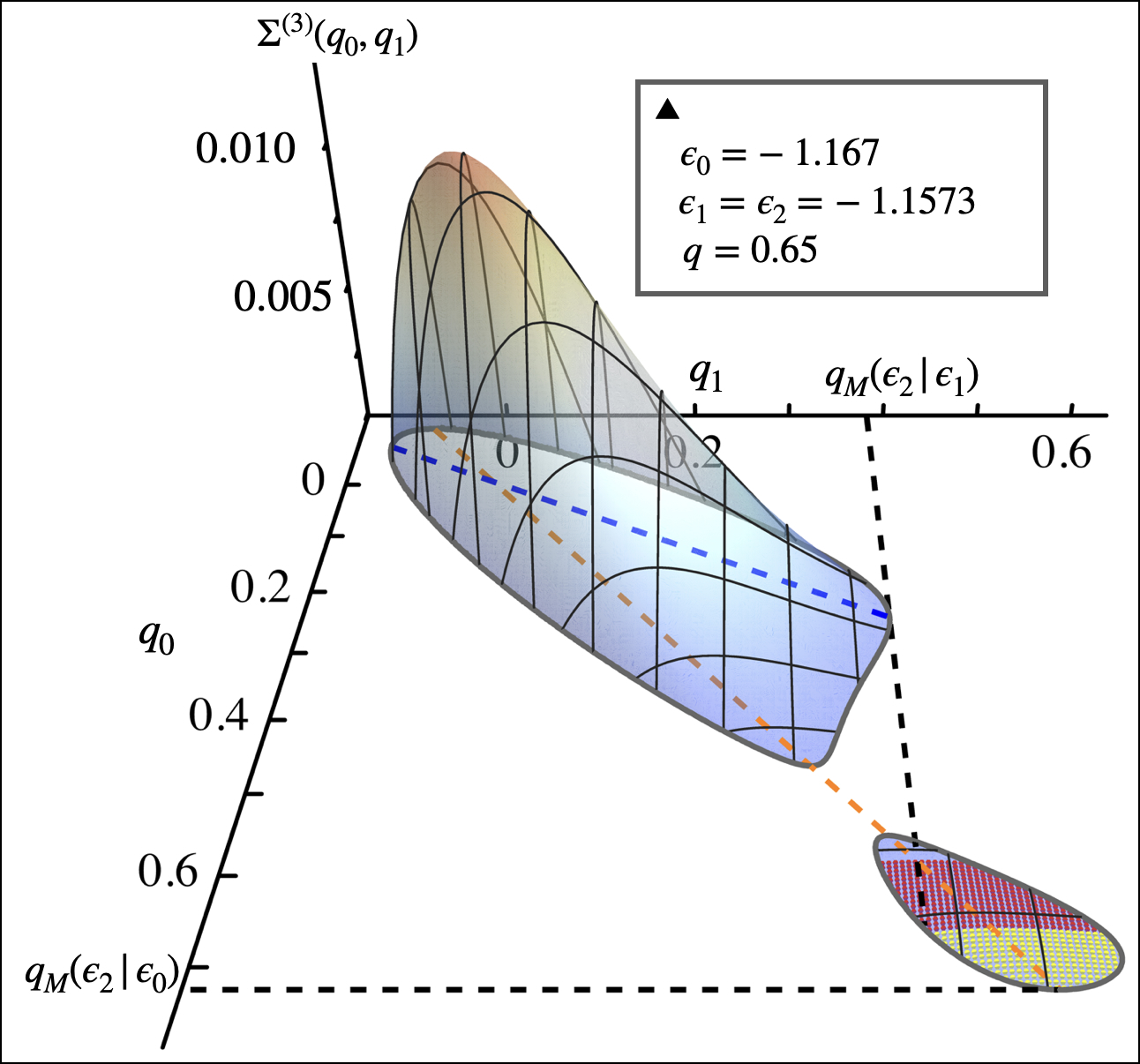}
\caption{}
 \end{subfigure}
 \hfill
 \begin{subfigure}[b]{1\columnwidth}
\includegraphics[width=0.95
\textwidth, trim=4 4 4 4,clip]{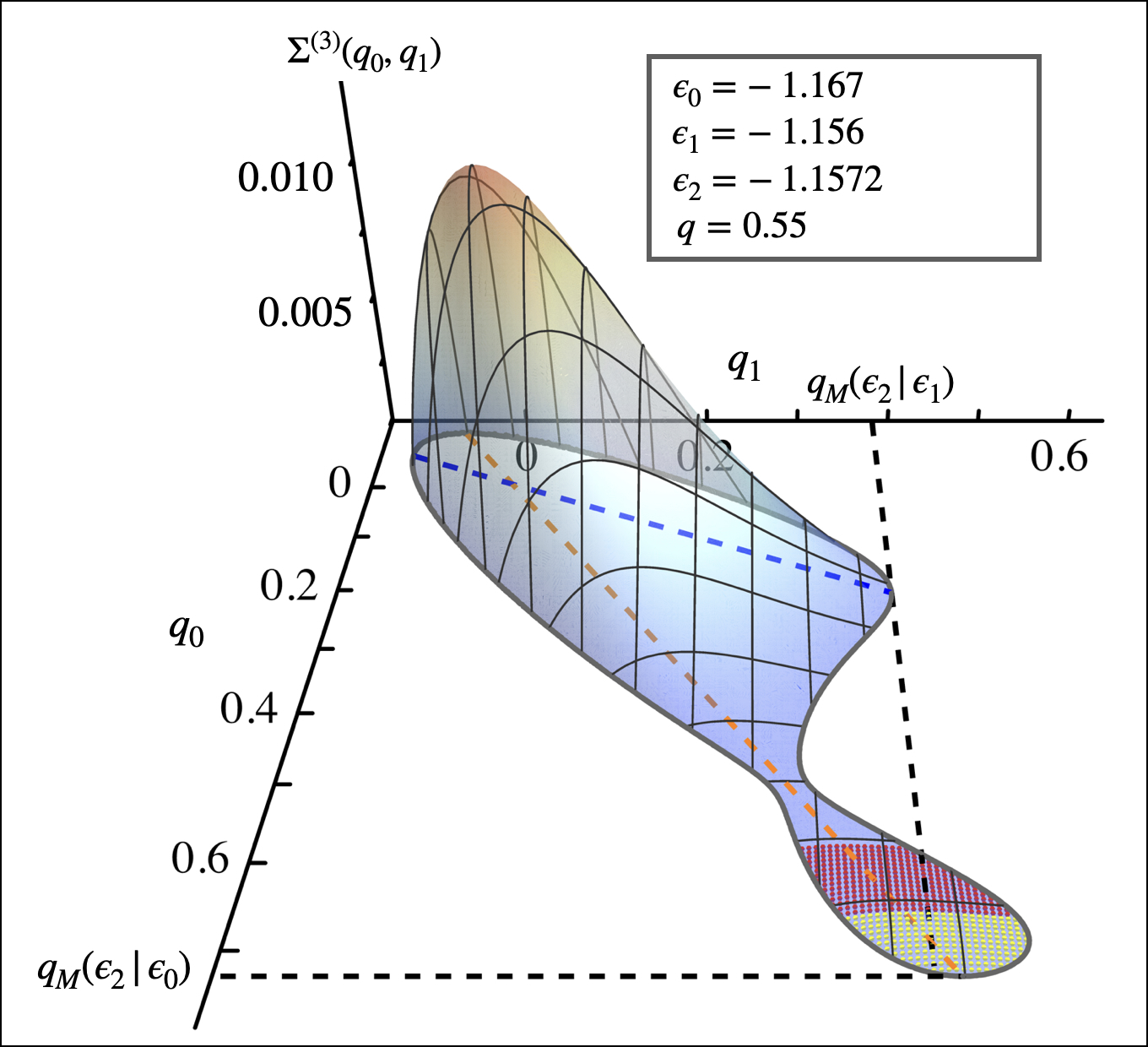}
\caption{}
 \end{subfigure}
 \caption{ Three-point complexity $\Sigma^{(3)}(\epsilon_2,q_0,q_1|\epsilon_1,\epsilon_0,q)$ as a function of overlaps $q_0$ and $q_1$, for specific choices of parameters referring to Fig.~\ref{fig:2D_plot_clustering}, with symbols $\bigstar, \blacktriangle,\blacksquare$. In all plots, the \textit{dashed orange line} in the $(q_0, q_1)$-plane follows $q_1=q q_0$, and the \textit{dashed blue line} follows $q_0=q q_1$. In (b), the orange points indicate the maxima of the complexity at fixed $q_0$, occurring at $q_1=q q_0$. The blue points indicate the maximum at fixed $q_1$, following $q_0=q q_1$ for small $q_1$, but jumping to higher $q_0$ values when $q_1 \approx 0.4$, thus showing the \textit{local accumulation} effect. Red and yellow zones indicate the regions where the stationary points counted, that is ${\bf s}_2$, are typically correlated minima and index-1 saddles, respectively. In all pictures, ${\bf s}_1$ is taken to be a minimum, except (c), where it is a correlated one (see Fig~\ref{fig:2D_plot_clustering}). Except for (a), where the maximum value $q_M(\epsilon_2|\epsilon_1)$ is not exceeded by the three-point complexity, all the other pictures present \textit{clustering} (meaning that such value is exceeded). In (d), we show an example of clustering where the energies are not equal.}
 \label{Fig:Plots3D}
\end{figure*}

\subsection{The doubly-annealed complexity, and its reduction to the two-point complexity}\label{sec:DoublyAnnealed}
The calculation of the doubly-annealed complexity is given in full details in Appendix~\ref{app:A_com_2A}. The resulting expression reads:
\begin{align}
\begin{split}\label{eq:ann_formulaComp2A}
&\Sigma^{(3)}_{2A}=\frac{Q_{2A}({\bf q})}{2}- f_{2A}({\bm \epsilon}, {\bf q})+I\left(\epsilon_2\sqrt{\frac{p}{p-1}}\right)
\end{split}
\end{align}
where now 
\begin{equation}
\begin{split}
    &Q_{2A}({\bf q})=1+\log\left(\frac{2 (p-1)(1-q^{2p-2})}{1-q^2}\right)+\\
    &\log\left|\frac{1-q^2-q_0^2-q_1^2+2q\,q_0\,q_1}{1-q^{2p-2}-q_0^{2p-2}-q_1^{2p-2}+2(q\, q_0 \,q_1)^{p-1}}\right|
    \end{split}
\end{equation}
and 
\begin{equation}\label{eq:FuncFDA}
\begin{split}
&f_{2A}({\bm \epsilon}, {\bf q})=  \epsilon_2^2Y^{(p)}_{2}({\bf q})+\epsilon_1\epsilon_2 \,Y_{12}^{(p)}({\bf q})+\epsilon_0\epsilon_2 Y_{02}^{(p)}({\bf q})+\\
&\epsilon_0\epsilon_1 [Y_{01}^{(p)}({\bf q})-U(q)]+\epsilon_1^2[Y_{1}^{(p)}({\bf q})-U_1(q)]+\\
&\epsilon_0^2[Y_{0}^{(p)}({\bf q})-U_0(q)-1].
\end{split}
\end{equation}
The functions $I$ and $U_0(q), U(q)$ and $U_1(q)$ are the same functions appearing in the two-point complexity, see Eq.~\eqref{eq:UC}. The remaining $Y^{(p)}({\bf q})$ are 
 functions of the overlaps ${\bf q}=(q, q_0, q_1)$. Their implicit definition is given in Appendix~\ref{app:pdf_analysis}. Since their explicit expression for general $p$ is rather cumbersome, we report it from Eqs.~\eqref{eq:yp31} to \eqref{app:eq:last_Y} only for the case $p=3$, which is the value of $p$ we consider for all plots in this paper.\\
We now discuss some special limits of this function. It can be checked explicitly that for $q_1$ fixed, when $q_0= q q_1$, the three-point complexity \eqref{eq:ann_formulaComp2A} reduces to a two point complexity, meaning that:
\begin{equation}\label{eq:Id1}
    \begin{split}
 &\Sigma^{(3)}_{2A}(\epsilon_2, q_0=q q_1, q_1 | \epsilon_1,\epsilon_0, q)\equiv \Sigma^{(2)}(\epsilon_2, q_1|\epsilon_1).
    \end{split}
\end{equation}
Indeed, when $q_0=q q_1$ one finds that all the coefficients multiplying $\epsilon_0$ in \eqref{eq:FuncFDA} vanish exactly, while $Y_1^{(p)}({\bf q})-U_1(q) \to U_0(q_1)$, $Y_2^{(p)}({\bf q})\to U_1(q_1)$ and $Y_{12}^{(p)}({\bf q})\to U(q_1)$. These relations can be checked explicitly for $p=3$, using the expressions~\eqref{eq:yp31} to \eqref{app:eq:last_Y}. As a result, the three-point complexity becomes independent of the configuration ${\bf s}_0$.
In an analogous manner, for given $q_0$ and $q_1= q q_0$, it holds
\begin{equation}\label{eq:Id2}
    \begin{split}
    \Sigma^{(3)}_{2A}(\epsilon_2, q_0, q_1=q q_0 | \epsilon_1,\epsilon_0, q)\equiv \Sigma^{(2)}(\epsilon_2, q_0|\epsilon_0).
    \end{split}
\end{equation}
When $q_1=q q_0$ the coefficients multiplying $\epsilon_1$ in \eqref{eq:FuncFDA} vanish, while $Y_0^{(p)}({\bf q})-U_0(q)-1 \to U_0(q_0)$, $Y_2^{(p)}({\bf q})\to U_1(q_0)$ and $Y_{02}^{(p)}({\bf q})\to U(q_0)$. Again, for $p=3$ these identities can be checked explicitly using Eqs.~\eqref{eq:yp31} to \eqref{app:eq:last_Y}. In this case, the three-point complexity becomes independent of the configuration ${\bf s}_1$. These two lines cross at $q_0=q_1=0$, where the complexity is maximal, independent of ${\bf q}=(q, q_0, q_1)$ and equal to  $\Sigma(\epsilon_2)$.\\
These two special lines in the $(q_0, q_1)$-plane have a simple entropic interpretation: in fact, if one asks what is the value of the overlap $q_0$ that maximizes the volume of the configuration space associated to ${\bf s}_2$, given the constraints that ${\bf s}_2 \cdot{\bf s}_1= N q_1$
 and ${\bf s}_1 \cdot{\bf s}_0= N q$, one finds that $q_0= q q_1$. To see this explicitly, we can parametrize ${\bf s}_2$ as:
 \begin{equation}\label{eq:s2}
     {\bf s}_2= q_1 {\bf s}_1 + \sqrt{1-q_1^2}\; {\bf u}_1, \quad \quad  {\bf u}_1 \cdot {\bf u}_1=N,
 \end{equation}
 where ${\bf u}_1 \in \mathcal{S}_N(\sqrt{N})$ is a vector  orthogonal to ${\bf s}_1$. The parametrization \eqref{eq:s2} ensures that ${\bf s}_2 \cdot {\bf s}_2=N$ and ${\bf s}_1 \cdot {\bf s}_2=N q_1$. Using this expression, one sees that:
 \begin{equation}
    {\bf s}_0 \cdot {\bf s}_2 \equiv N q_0=  N q q_1 +  \sqrt{1-q_1^2} \, {\bf s}_0 \cdot {\bf u}_1.
 \end{equation}
 The value $q_0= q q_1$ corresponds to ${\bf s}_0 \cdot {\bf u}_1=0$:
 in high dimension $N$, for any fixed vector ${\bf s}_0$, the number of vectors ${\bf u}_1$ satisfying ${\bf s}_0 \cdot {\bf u}_1=0$ is exponentially larger (in $N$) than that associated to any other value ${\bf s}_0 \cdot {\bf u}_1 \neq 0$. Therefore, the majority of the configurations ${\bf s}_2$ fulfilling the constraint on $q_1$  are found at overlap $q_0=q q_1$ with ${\bf s}_0$. Eq.~\eqref{eq:Id1} is stating that in that region of configuration space, the number of stationary points ${\bf s}_2$ coincides with the number that one would obtain forgetting the constraints on the stationary point ${\bf s}_0$.  Analogously, $q_1= q q_0$ maximizes the configuration space associated to ${\bf s}_2$, given the constraints that ${\bf s}_2 \cdot{\bf s}_0= N q_0$
 and ${\bf s}_1 \cdot{\bf s}_0= N q$. Similarly as above, writing now ${\bf s}_2$ as
 \begin{equation}\label{eq:s2bis}
     {\bf s}_2= q_0 {\bf s}_0 + \sqrt{1-q_0^2}\; {\bf u}_0, \quad \quad  {\bf u}_0 \cdot {\bf u}_0=N
 \end{equation}
 for some ${\bf u}_0$ orthogonal to ${\bf s}_0$,
 one sees that the exponential majority of the configurations ${\bf s}_2$ fulfilling the constraint on $q_0$ are found at overlap $q_1=q q_0$ with ${\bf s}_1$.\\

\subsection{The quenched complexity}\label{eq:SpecialLines}
We now discuss the results of the calculation of the quenched complexity, referring the reader interested in the technicalities of the calculation to Ref.~\cite{PaperLungo}. We fix $\epsilon_0=-1.167$ for the plots in this section. Some relevant values of parameters descending from the two-point complexity for this particular value of $\epsilon_0$ are recalled in Table~\ref{fig:table}. \\

{\bf When $\Sigma^{(3)}$ reduces to  $\Sigma^{(2)}$.} Representative plots of the quenched three-point complexity $\Sigma^{(3)}(\epsilon_2,q_0,q_1|\epsilon_1,\epsilon_0,q)$ (denoted with $\Sigma^{(3)}(q_0,q_1)$ in the plots for brevity) as a function of the overlaps $q_0, q_1$ are given in Fig.~\ref{Fig:Plots3D} for different values of parameters. One sees that the quenched complexity is always maximal for $q_0=0=q_1$, where it reduces to the one-point complexity $\Sigma(\epsilon_2)$. 
The orange and blue dashed lines in the $(q_0, q_1)$-planes correspond to $q_1= q q_0$ (for fixed $q_0$), and at $q_0= q q_1$ (for fixed  $q_1$), respectively. These are the lines along which the doubly-annealed complexity~\eqref{eq:ann_formulaComp2A} reduces to the two-point complexity, see Eqs.~\eqref{eq:Id1} and \eqref{eq:Id2}. We find that an analogous statement remains true in the quenched calculation: along these lines, the three-point \emph{quenched} complexity coincides with the three-point \emph{doubly-annealed} complexity, 
\begin{equation}
    \begin{split}
 &\Sigma^{(3)}(\epsilon_2, q_0, q_1=q q_0 )= \Sigma^{(3)}_{2A}(\epsilon_2, q_0, q_1=q q_0 ),\\       &\Sigma^{(3)}(\epsilon_2, q_0=q q_1, q_1 )= \Sigma^{(3)}_{2A}(\epsilon_2, q_0=q q_1, q_1),
    \end{split}
\end{equation}
implying:
\begin{equation}
    \begin{split}
 &\Sigma^{(3)}(\epsilon_2, q_0, q_1=q q_0 | \epsilon_1,\epsilon_0, q)= \Sigma^{(2)}(\epsilon_2, q_0|\epsilon_0),\\       &\Sigma^{(3)}(\epsilon_2, q_0=q q_1, q_1 | \epsilon_1,\epsilon_0, q)= \Sigma^{(2)}(\epsilon_2, q_1|\epsilon_1).
    \end{split}
\end{equation}
As explained in Sec.~\ref{sec:DoublyAnnealed}, these lines can be interpreted in terms of maximization of the volume of configuration space accessible to the third configuration ${\bf s}_2$. \\

{\bf The maxima of the complexity.} We find that for any choice of the parameters ${\bm \epsilon}=(\epsilon_0, \epsilon_1, \epsilon_2)$ and $q$, the quenched three-point complexity at fixed $q_0$ is in fact always maximal for $q_1=q q_0$: the highest number of stationary points ${\bf s}_2$ fulfilling the constraint on $q_0$ is found in the region of configuration space corresponding to the maximal entropy.  Fig.~\ref{Fig:Plots3D}(a) gives an example for which the analogous statement holds true at fixed $q_1$, the  complexity being maximal at $q_0=q q_1$. In this case, the maxima of the complexity are thus attained at values of the parameters $q_0$ and $q_1$ where the corresponding quenched complexity reduces to the annealed one, and it also coincides with a two-point complexity. Fig.~\ref{Fig:Plots3D}(b) represents instead a different scenario, in which one observes a jump in the value of $q_0$ maximizing the complexity at fixed $q_1$ (see blue points on the 3D plot). For small $q_1$ the maximum is attained at $q_0 = q q_1$; when $q_1$ exceeds a critical value (around $q_1\approx 0.4$ in Fig.~\ref{Fig:Plots3D}(b)) the point where the complexity is maximal jumps to a much larger $q_0$. This means that when ${\bf s}_2$ approaches ${\bf s}_1$, typically one finds a higher number of stationary points in the region where ${\bf s}_2$ is also close to ${\bf s}_0$. This corresponds to the \emph{local accumulation} defined in Sec.~\ref{subsec:defs_clustering}.\\

{\bf The boundaries of the domain.} The dashed black lines in Fig.~\ref{Fig:Plots3D} indicate the overlaps $q_M(\epsilon_2 | \epsilon_0)$ and $q_M(\epsilon_2 | \epsilon_1)$; these are two quantities related to the two-point complexity, see \eqref{eq:qMAx}. The first one, $q_M(\epsilon_2 | \epsilon_0)$, is the overlap at which one finds the stationary points at energy density $\epsilon_2$ that are  \emph{closest} (at highest overlap) to a stationary point at energy density $\epsilon_0$. One sees from the plots that this is always also the maximal overlap $q_0$ for which the three-point complexity is non-zero. Again, the analogous statement in $q_1$ does not necessarily hold true: except for Fig.~\ref{Fig:Plots3D}(a), in all the other cases the domain where $\Sigma^{(3)}>0$ exceeds $q_M(\epsilon_2 | \epsilon_1)$. 
This corresponds to the  \textit{clustering} defined in Sec.~\ref{subsec:defs_clustering}.\\

Away from the special lines $q_0=q q_1$ and $q_1=q q_0$, the three-point complexity can not be written in terms of the two-point one, meaning that genuine three-point correlations exist between the stationary points. These correlations give rise to several transitions in the structure of the landscape as we change the parameters ${\bm \epsilon}=(\epsilon_0, \epsilon_1, \epsilon_2)$ and ${\bf q}=(q,q_0, q_1)$ describing the properties of the stationary points. We discuss these transitions in Sec.~\ref{sec:LandscapeEvolution}.

\subsection{Stability of the stationary points ${\bf s}_2$}\label{sec:stability}
In the pure $p$-spin model, stationary points with energy density smaller than $\epsilon_{\rm th}$ are either local minima, or saddles with $\sim O(N^0)$ negative modes of the Hessian (i.e., with an intensive index). To determine whether, for a fixed choice of the parameters $\epsilon_2, q_0$ and $q_1$, the stationary points counted by the three-point complexity are predominantly local minima or saddles of low index, one has to study the typical spectrum of the Hessian matrices $\nabla_\perp^2 \mathcal{E}({\bf s}_2)$. This means that one has to determine the statistical properties (e.g., the eigenvalues distribution) of the matrices $\nabla_\perp^2 \mathcal{E}({\bf s}_2)$ evaluated at configurations ${\bf s}_2$ that are extracted with the measure:
\begin{equation}\label{eq:measureH}
    \mu({\bf s}_2):= \mathbb{E}\quadre{ \frac{\omega_{\epsilon_2, q_0, q_1}({\bf s}_2| {\bf s}_1, {\bf s}_0)}{\mathcal{N}_{{\bf s}_0, {\bf s}_1}(\epsilon_2, q_0, q_1)}}_{0,1}
\end{equation}
with the notation of Sec.~\ref{subsec:3_pt_def}.\\
It is simple to see that at large $N$, the Hessian   $\nabla_\perp^2\mathcal{E}({\bf s})$ at an arbitrary configuration (including stationary points) of energy density $\epsilon$ is statistically equivalent to  a random matrix extracted from a Gaussian Orthogonal Ensemble (GOE), properly rescaled and shifted by a diagonal matrix:
\begin{equation}\label{eq:OEUnconditioned}
  \nabla_\perp^2 \mathcal{E}({\bf s}) \stackrel{\text{ in law}}{\sim} \sqrt{\frac{p(p-1)}{2}} G -  \, p \epsilon \mathbb{I}
\end{equation}
where $G$ is a $(N-1)$-dimensional GOE matrix with $\mathbb{E}[G_{ij}]=0$, $\mathbb{E}[ G_{ij} G_{kl} ]= \frac{1}{N}[\delta_{ik} \delta_{jl}+ \delta_{il} \delta_{jk}]$. The eigenvalues of $G$ are distributed according to a semicircular law with support in $[-2,2]$: from \eqref{eq:OEUnconditioned} it immediately follows that stationary points with energy density $\epsilon$ are typically local minima when $\epsilon< - \sqrt{2 (p-1)/p}= \epsilon_{\rm th}$, since the Hessian is positive definite in this case (the semicircle has support lying in the positive semiaxis). \\
The effect of conditioning a stationary point to a fixed overlap with one other stationary point of given energy has been studied in \cite{ros2019complexity, ros2020distribution, paccoros}. This is the case of interest when computing the two-point complexity: in our formalism, it corresponds to characterizing the statistics of the Hessian at ${\bf s}_1$, conditioned to the properties of ${\bf s}_0$. One finds that $\nabla_\perp^2\mathcal{E}({\bf s}_1)$ is again a shifted GOE matrix, which is now perturbed by additive and multiplicative rank-1 perturbations, that arise from conditioning to the properties of ${\bf s}_0$ and that depend explicitly on $\epsilon_0, \epsilon_1$ and $q$. These perturbations affect the eigenvalue distribution of the Hessian, in that they (can) produce an outlier (isolated eigenvalue) that detaches from the rest of the spectrum, described by the shifted semicircular law. The outlier is produced for critical values of the parameters $\epsilon_0, \epsilon_1$ and $q$, in a transition akin to the celebrated BBP (Baik-Ben Arous-P\'{e}ch\'{e}) transition \cite{edwards1976eigenvalue, BBP}. When the parameters are such that this eigenvalue has a negative sign, then stationary points ${\bf s}_1$ with those parameters are rank-1 saddles. This corresponds to the yellow area in Fig.~\ref{fig:2D_plot_clustering}. We remark that this isolated eigenvalue does not arise from the rank-1 term appearing in the Hessian associated to the Thouless-Anderson-Palmer (TAP) free energy of the model \cite{thouless1977solution}, already discussed in \cite{leuzzi2003complexity}. Rather, it arises from the geometrical constraint on the overlap $q$ with another stationary point ${\bf s}_0$. We elucidate this point in Appendix~\ref{app:hessian_TAP}.\\
The properties of the Hessians $\nabla_\perp^2\mathcal{E}({\bf s}_1)$  have been characterized thoroughly (studying for instance large deviations of the outlier~\cite{ros2020distribution} or eigenvectors properties~\cite{paccoros}) also thanks to the fact that within the framework of the two-point complexity, quenched and annealed calculations coincide and the measure on ${\bf s}_1$ (conditioned to ${\bf s}_0$) can be treated without making use of the replica trick. 
For the three-point complexity, this is no longer true and in principle one should resort to the full replica formalism to express the measure~\eqref{eq:measureH}. Since the corresponding calculation is very involved \footnote{A calculation of the isolated eigenvalues within the quenched formalism is given in \cite{ros2019complex} for a slightly different problem.}, here we resort to the doubly-annealed approximation and study the Hessian on configurations extracted with the modified measure 
\begin{equation}
    \mu_{\rm 2A}({\bf s}_2):=\frac{\mathbb{E}\quadre{ \omega_{\epsilon_2, q_0, q_1}({\bf s}_2| {\bf s}_1, {\bf s}_0)}_{\rm 2A}}{\mathbb{E}\quadre{\mathcal{N}_{{\bf s}_0, {\bf s}_1}(\epsilon_2, q_0, q_1)}_{\rm 2A}}
\end{equation}
again with the notation of Sec.~\ref{subsec:3_pt_def} and Sec.~\ref{subsec:quenched_annealed}. The resulting calculation is significantly simpler, and discussed in Appendix~\ref{app:hessian_analysis}. Even though approximate, this calculation already gives important indications on the stability of the points counted by the three-point complexity.\\
We find that under the conditioning on the overlaps and on the properties of ${\bf s}_0, {\bf s}_1$ the Hessian at a  stationary point ${\bf s}_2$ is still distributed as a shifted GOE, this time perturbed by rank-2 additive and multiplicative perturbations along correlated directions. A precise description of its statistics within the annealed framework is given in Appendix~\ref{app:DistributionConditional}. Again, the finite-rank perturbations can give rise to isolated eigenvalues that do not belong to the support of the semicircular law describing the eigenvalue density of matrices of the form \eqref{eq:OEUnconditioned}. In the Appendix, we show that the outliers of the conditioned matrix $ \nabla^2_\perp \mathcal{E}({\bf s}_2)$, when they exist, are given by 
\begin{equation}\label{eq:IsoMain}
    \lambda_{\rm iso}({\bm \epsilon}, {\bf q})= 2^{-\frac{1}{2}}\, z^*({\bm \epsilon}, {\bf q}) - p \epsilon_2
\end{equation} 
where $z^*({\bm \epsilon}, {\bf q})$ are real solutions of the equation:
\begin{equation}
[z-\mu_0'-\Delta'_0 \mathfrak{g}(z)][z-\mu_1'-\Delta'_1 \mathfrak{g}(z)]-(\mu'_{01})^2 =0,  
\end{equation}
that do not belong to the support of the semicircle, i.e.,  $z \notin [- 2 \sigma, 2 \sigma]  $ with $\sigma=\sqrt{p(p-1)}$. The function $\mathfrak{g}(z)$ is the Stieltjes transform of GOE matrices with variance $\sigma^2$, evaluated on the real axis:
\begin{equation}
    \mathfrak{g}(z)=\frac{1}{2 \sigma^2} \tonde{z- \text{sign}(z) \sqrt{z^2 - 4 \sigma^2}}, \quad \quad z \in \mathbb{R}.
\end{equation} 
The functions $\Delta'_{\alpha}$ and $\mu'_\alpha$ depend on the parameters ${\bf q}$ and ${\bm \epsilon}$, and their explicit expression is given in Appendix~\ref{app:DistributionConditional}. 
When one (or more than one) isolated eigenvalue exist and they are all positive, then the stationary points ${\bf s}_2$ are correlated minima: the curvature of the landscape around the stationary points is everywhere positive, but there is at least one direction in configuration space where the curvature is softer; this direction has an $O(1)$ overlap with the directions connecting ${\bf s}_2$ to ${\bf s}_1, {\bf s}_0$ in configuration space. This happens in the region of parameters corresponding to the red zones in Fig.~\ref{Fig:Plots3D}. When at least one isolated eigenvalue exists and it is negative, then stationary points ${\bf s}_2$ with those parameters are typically saddles: there is one direction in configuration space along which the curvature of the landscape around ${\bf s}_2$ is negative; once more, this direction has an $O(1)$ overlap with the directions connecting ${\bf s}_2$ to ${\bf s}_1, {\bf s}_0$ in configuration space. This happens in the region of parameters corresponding to the yellow zones in Fig.~\ref{Fig:Plots3D}.\\
As seen in Fig.~\ref{Fig:Plots3D}, in the region in which clustering occurs, most of the stationary points are either saddles or correlated minima. There are however values of parameters for which clustering occurs, but the corresponding stationary points ${\bf s}_2$ have an Hessian spectrum that is exactly the same as the spectrum at typical stationary points at energy $\epsilon_2$, showing no correlations with ${\bf s}_1, {\bf s}_0$. This is particularly evident in Fig.~\ref{Fig:Plots3D}(b), which clearly shows a region at $q_1>q_M(\epsilon_2|\epsilon_1)$ which is neither red nor yellow. We do not know whether this feature remains robust when extending the calculation of the isolated eigenvalue(s) to the quenched formalism. \\

\begin{figure*}[t!]
  \begin{subfigure}[b]{1\columnwidth}
     \includegraphics[width=0.88
\textwidth, trim=5 5 5 5,clip]{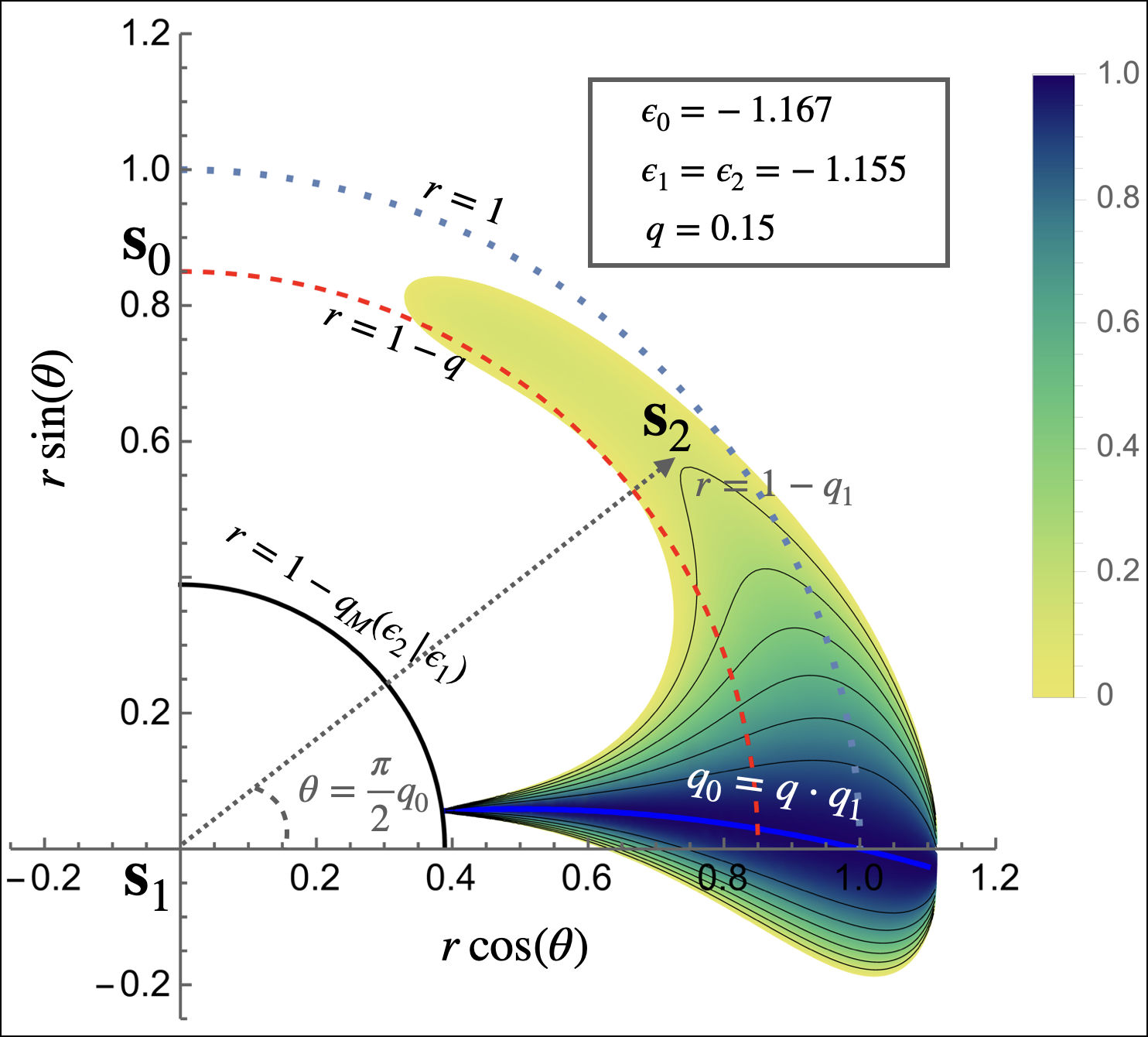}
\caption{}
 \end{subfigure}
 \hfill
  \begin{subfigure}[b]{1\columnwidth}
     \includegraphics[width=0.77
\textwidth, trim=4 4 4 4,clip]{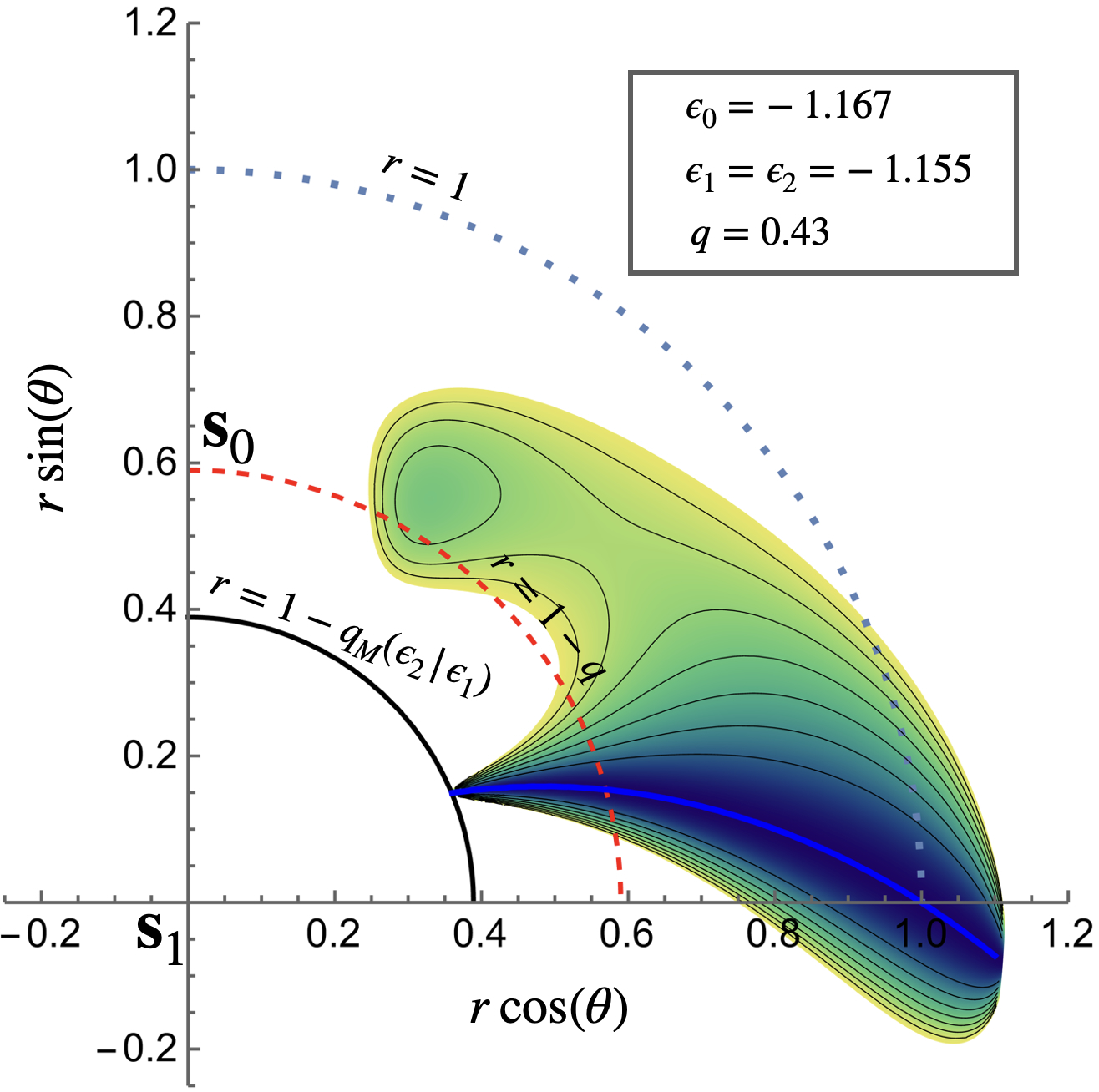}
\caption{}
 \end{subfigure}
 \medskip
 \begin{subfigure}[b]{1\columnwidth}
     \includegraphics[width=0.77
\textwidth, trim=4 4 4 4,clip]{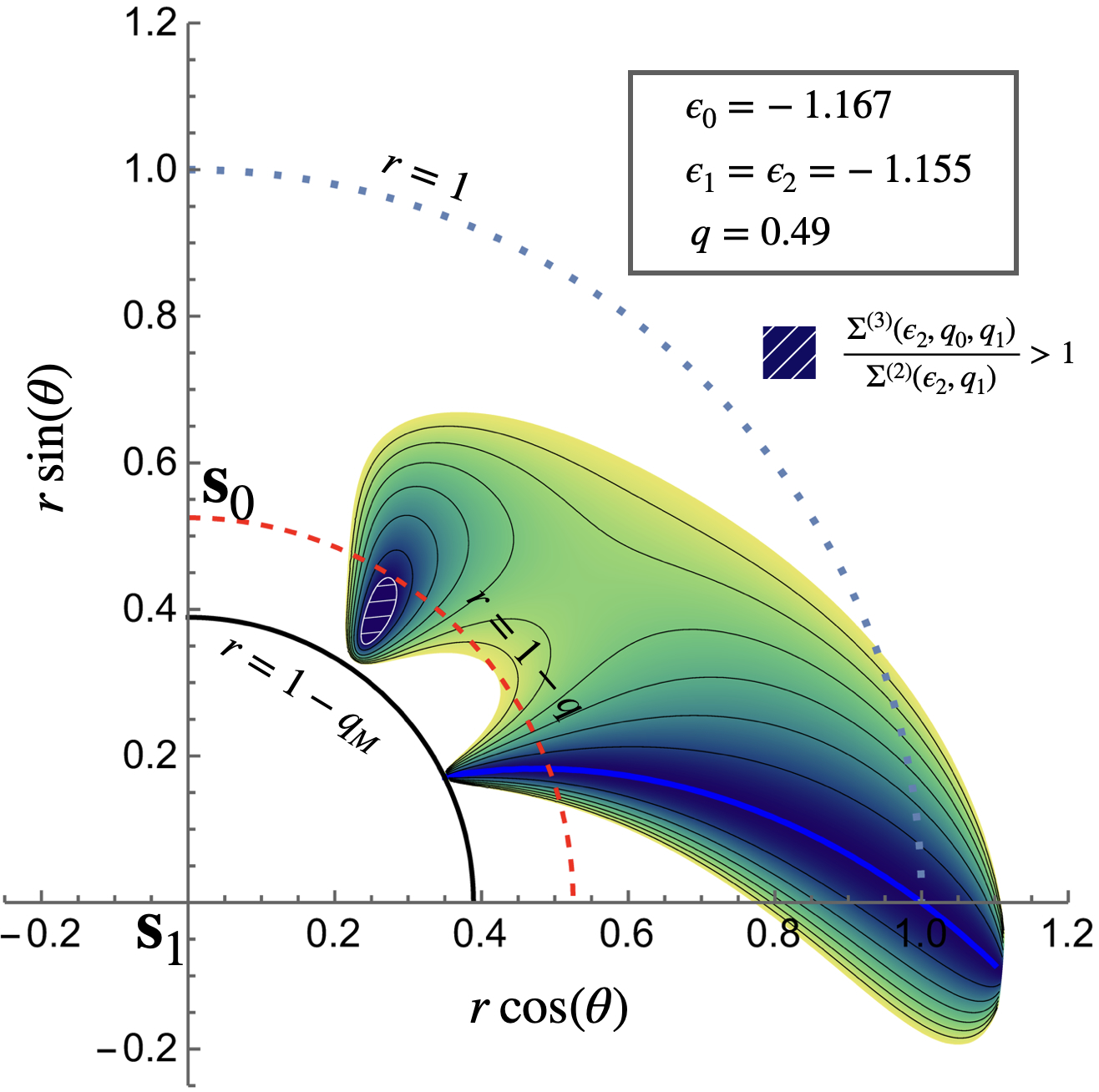}
\caption{}
 \end{subfigure}
 \hfill
  \begin{subfigure}[b]{1\columnwidth}
    \includegraphics[width=0.77
\textwidth, trim=4 4 4 4,clip]{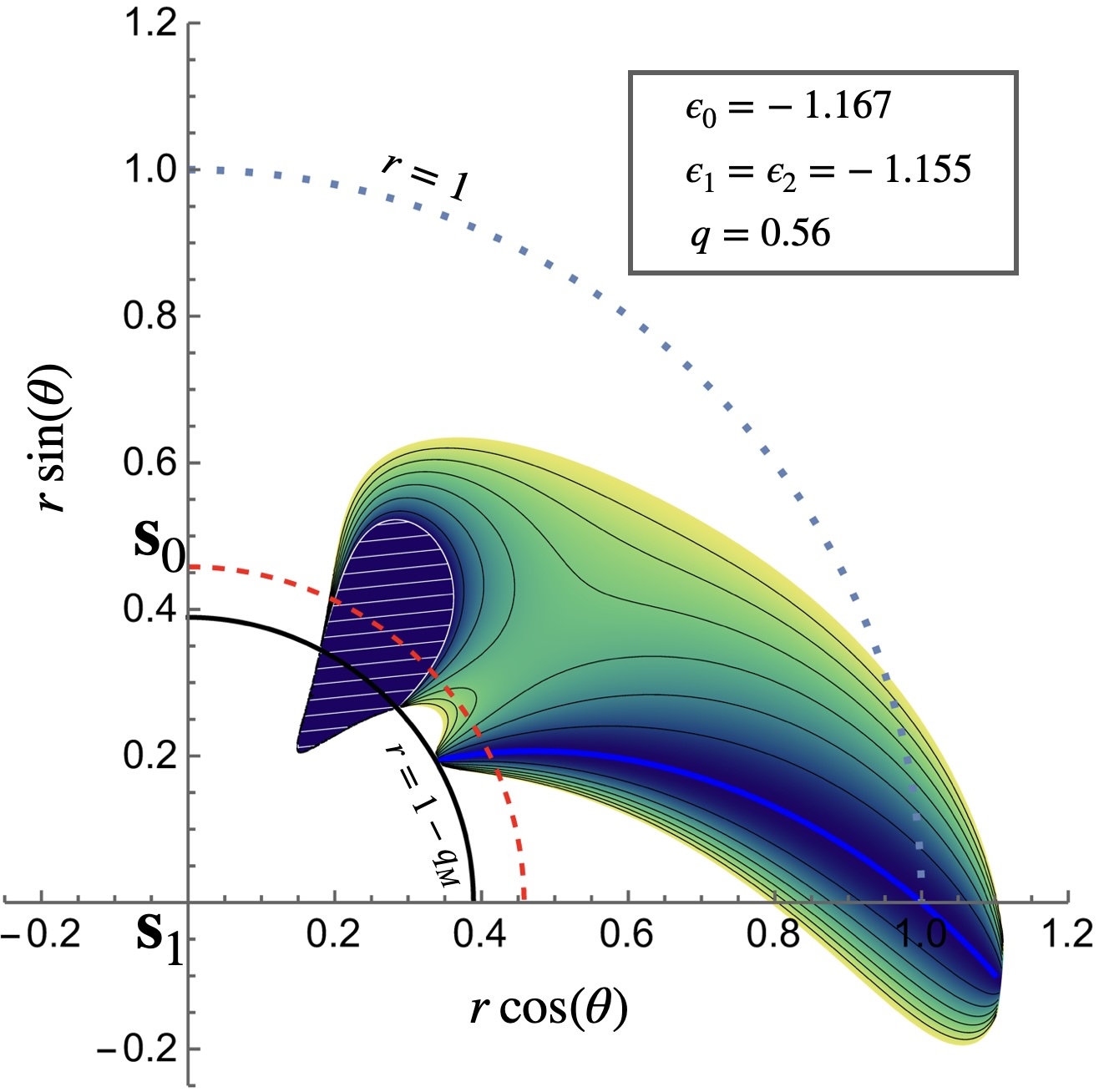}
\caption{}
 \end{subfigure}
\caption{ The figures show a density plot of the ratio $\frac{\Sigma^{(3)}(\epsilon_2,q_0,q_1)}{\Sigma^{(2)}(\epsilon_2,q_1)}$. Each point $(\theta, r)$ corresponds to ${\bf s}_2$ with overlaps $q_1=1-r$ and $q_0=2\theta/\pi$. The origin represents ${\bf s}_1$, and the point $(\theta=\pi/2, r=1-q)$ represents ${\bf s}_0$. The blue line follows $q_0=q\,q_1$, where the ratio reaches the value 1. The black arc marks $q_1=q_M(\epsilon_2|\epsilon_1)$, which is exceeded in the case of clustering (figure (d)). Regions with white hatching indicate density values greater than 1, implying \textit{accumulation}. From left to right, and top to bottom, $q$ increases while the other parameters are fixed. In (b), a second maximum emerges along constant $q_1$ curves; in (c), the corresponding area shows accumulation; and in (d), it surpasses $q_1=q_M(\epsilon_2|\epsilon_1)$, indicating \textit{clustering}.}
 \label{Fig:2Ddensityplots}
\end{figure*}

\subsection{Dependence on $q$: stationary points distribution and landscape's transitions}\label{sec:LandscapeEvolution}

To describe the local arrangement of stationary points encoded in the three-point complexity and unveil the presence of landscape's transitions, we consider first the case of fixed energy densities $\epsilon_0=-1.167$, $\epsilon_1= \epsilon_2=-1.155$, and study the landscape's evolution varying the parameter $q$, which measures the overlap between ${\bf s}_0$ and ${\bf s}_1$, in Fig.~\ref{Fig:2Ddensityplots}.  The figure represents a projection of configuration space, where the stationary point ${\bf s}_1$ is placed at the center of the reference frame, while ${\bf s}_0$ is aligned along the $y$ axis at a distance $r=1-q$ (red dashed quarter of a circle). The third stationary point has radial coordinate $r=1-q_1$ (measuring its distance to ${\bf s}_1$) and angular coordinate $\theta= \frac{\pi}{2} q_0$ (measuring its vicinity to ${\bf s}_0$). The colored area of the plot identifies the region of configuration space where we find an exponentially large population of stationary points ${\bf s}_2$ at energy density $\epsilon_2$, meaning that $\Sigma^{(3)}(\epsilon_2, q_0, q_1 | \epsilon_1,\epsilon_0, q)>0$. The intensity of the color plot corresponds to the value of the ratio $\Sigma^{(3)}(\epsilon_2, q_0, q_1 | \epsilon_1,\epsilon_0, q)/\Sigma^{(2)}(\epsilon_2, q_1 | \epsilon_1)$. The different plots in Fig.~\ref{Fig:2Ddensityplots} represent the evolution of the landscape as $q$ increases, meaning that ${\bf s}_0$ and ${\bf s}_1$ are chosen to be progressively closer to each others in configuration space. \\
With increasing $q$, we see the following different regimes:

\begin{itemize}
    \item[(a)] {\bf Depletion regime. } This corresponds to Fig.~\ref{Fig:2Ddensityplots}(a), i.e., to small values of $q$. In this regime the complexity of ${\bf s}_2$ is always maximal along the curve $q_0= q q_1$, which identifies the region of configuration space maximizing the entropy of configurations at fixed $q_1$; along this line (blue in the figures), $\Sigma^{(3)}(\epsilon_2, q_0, q_1 | \epsilon_1,\epsilon_0, q)=\Sigma^{(2)}(\epsilon_2, q_1 | \epsilon_1)$. The complexity of stationary points decreases when moving away from this line, and it is always smaller than $\Sigma^{(2)}(\epsilon_2, q_1 | \epsilon_1)$. In particular, it is progressively smaller as one looks at regions closer and closer to ${\bf s}_0$ (i.e., increasing $\theta$). In this regime, the population of stationary points with energy $\epsilon_2$ in the vicinity of ${\bf s}_0$ is \emph{depleted}. Moreover, no clustering occurs, as indicated by the fact that the colored area never exceeds $r=1-q_M(\epsilon_2|\epsilon_1)$ (black continuous quarter of a circle). An example of the complexity surface in this regime (for a different choice of parameters) is given also in Fig.~\ref{Fig:Plots3D}(a).

    \item[(b)] {\bf Non-monotonic regime. } This corresponds to Fig.~\ref{Fig:2Ddensityplots}(b). In this case, the highest concentration of stationary points ${\bf s}_2$ is again at $q_0= q q_1$. However, at fixed distance to ${\bf s}_1$ (i.e., for fixed $r$), the distribution of stationary points is non monotonic in $\theta$: in the vicinity of ${\bf s}_0$, the number of points ${\bf s}_2$ increases again, and the complexity $\Sigma^{(3)}(\epsilon_2, q_0, q_1 | \epsilon_1,\epsilon_0, q)$ has a local maximum. This is a sign of  correlations in the landscape, i.e., of the fact that the presence of ${\bf s}_0$ affects the distribution of ${\bf s}_2$.  

    \item[(c)] {\bf Local accumulation transition/regime. } This corresponds to Fig.~\ref{Fig:2Ddensityplots}(c). In this case, the highest number of stationary configurations ${\bf s}_2$ is found in the vicinity of ${\bf s}_0$, where the three-point complexity is \emph{larger} than the two-point one and Eq.~\eqref{eq:Bound2} is satisfied. The line $q_0=q q_1$ (continuous blue line) is now a line of maxima of the quenched complexity, that are only local and no longer global. \\
    In this regime, the correlations in the energy landscape become strong enough, to generate \emph{local accumulation} of stationary points, whose complexity exceeds $\Sigma^{(2)}(\epsilon_2, q_1| \epsilon_1)$, see the blue zone with white hatching; the majority of the stationary points ${\bf s}_2$ is no longer found in the region where the volume of configuration space accessible to them is maximal, but it is found closer to ${\bf s}_0$, meaning that energy correlations prevail on entropy.  
     The non-monotonic behavior in $q_0$ is particularly evident for large enough $q_1$ (small $r$), where by increasing $\theta= \frac{\pi}{2} q_0$ one goes through regions of configurations space at intermediate $q_0$ where there is no stationary point of energy $\epsilon_2$. 
    For fixed $q_1$, the landscape undergoes a transition at a critical value of $q$, where the maximum of $\Sigma^{(3)}(\epsilon_2, q_0, q_1 | \epsilon_1,\epsilon_0, q)$ as a function of $q_0$ jumps from $q_0=q q_1$ to a higher value, that depends on the energy densities ${\bm \epsilon}$.
     We refer to this transition as the “local accumulation transition".

    \item[(d)] {\bf Clustering transition/regime. } This corresponds to Fig.~\ref{Fig:2Ddensityplots}(d), i.e. to large values of $q$. This regime is characterized by clustering: the maximal value of $q_1$ (smallest value of $r$) for which the three-point complexity is non-negative is \emph{larger} than the value predicted by the two-point complexity in absence of ${\bf s}_0$, meaning that 
     \begin{equation}\label{eq:q1max}
   q_1^{\rm max}({\bm \epsilon}, q) :=\text{argmax}_{q_1} \left[\max_{q_0} \Sigma^{(3)}(\epsilon_2, q_0, q_1)\right]
    \end{equation}
   satisfies
    \begin{equation}
   q_1^{\rm max}({\bm \epsilon}, q) > q_M(\epsilon_2|\epsilon_1).
    \end{equation}
 This corresponds to the fact that the blue zone with white hatching extends within the black quarter of a circle in the picture: the three-point complexity in that zone is positive, while the two-point complexity $\Sigma^{(2)}(\epsilon_2,q_1|\epsilon_1)$ predicts that there are typically no stationary points at those energies and at those overlaps $q_1$ from a ${\bf s}_1$ extracted without conditioning on ${\bf s}_0$. 
 Examples of the complexity surfaces in this clustering regime are given in Fig.~\ref{Fig:Plots3D}(b-d): the three-point complexity is positive for a range of values of $q_1 > q_M(\epsilon_2|\epsilon_1)$. We call the associated transition a “clustering transition".
\end{itemize}

In summary, Fig.~\ref{Fig:2Ddensityplots} describes how the distribution of stationary points in configuration space evolves as one tunes the overlap $q$, for a fixed choice of $\epsilon_1= \epsilon_2 > \epsilon_0$. 
 Recall that in the inset of  Fig.~\ref{fig:2D_plot_clustering} we show the region where clustering exists (black hatched region), for fixed $\epsilon_0=-1.167$, and as a function of $\epsilon_1=\epsilon_2$ and $q$. For these values of parameters, clustering always occurs whenever ${\bf s}_1$ is a correlated minimum or an unstable saddle, but it also occurs when ${\bf s}_1$ is a local minimum with an Hessian that shows no correlations to ${\bf s}_0$. The parameters of Fig.~\ref{Fig:Plots3D}(b) are precisely chosen in such a way that  ${\bf s}_1$ is a stable minimum (see $\star$ in Fig.~\ref{fig:2D_plot_clustering}) and clustering is present. In the following section, we describe how much this scenario and the landscape's transitions depend on the choices of the energies $\epsilon_1, \epsilon_2.$

\subsection{Dependence on $\epsilon_\alpha$: critical energies}\label{sec:EnergyDependence}
For $\epsilon_1= \epsilon_2 > \epsilon_0$, as shown in Fig.~\ref{Fig:2Ddensityplots} 
a rather rich phenomenology occurs, with two distinct transitions (local accumulation and clustering) happening at different critical values of $q$. We now consider, for $\epsilon_0$ fixed, arbitrary values of $\epsilon_1, \epsilon_2$ in the range $[\epsilon_{\rm gs}, \epsilon_{\rm th}]$, and ask for which choices of the energies local accumulation and clustering occur for at least some values of the parameters ${\bf q}$. We focus in particular on clustering, which is a special case of local accumulations, see \eqref{eq:Bound2}.

\begin{figure}[t]
\centering
\includegraphics[width=0.46
\textwidth, trim=5 5 5 5,clip]{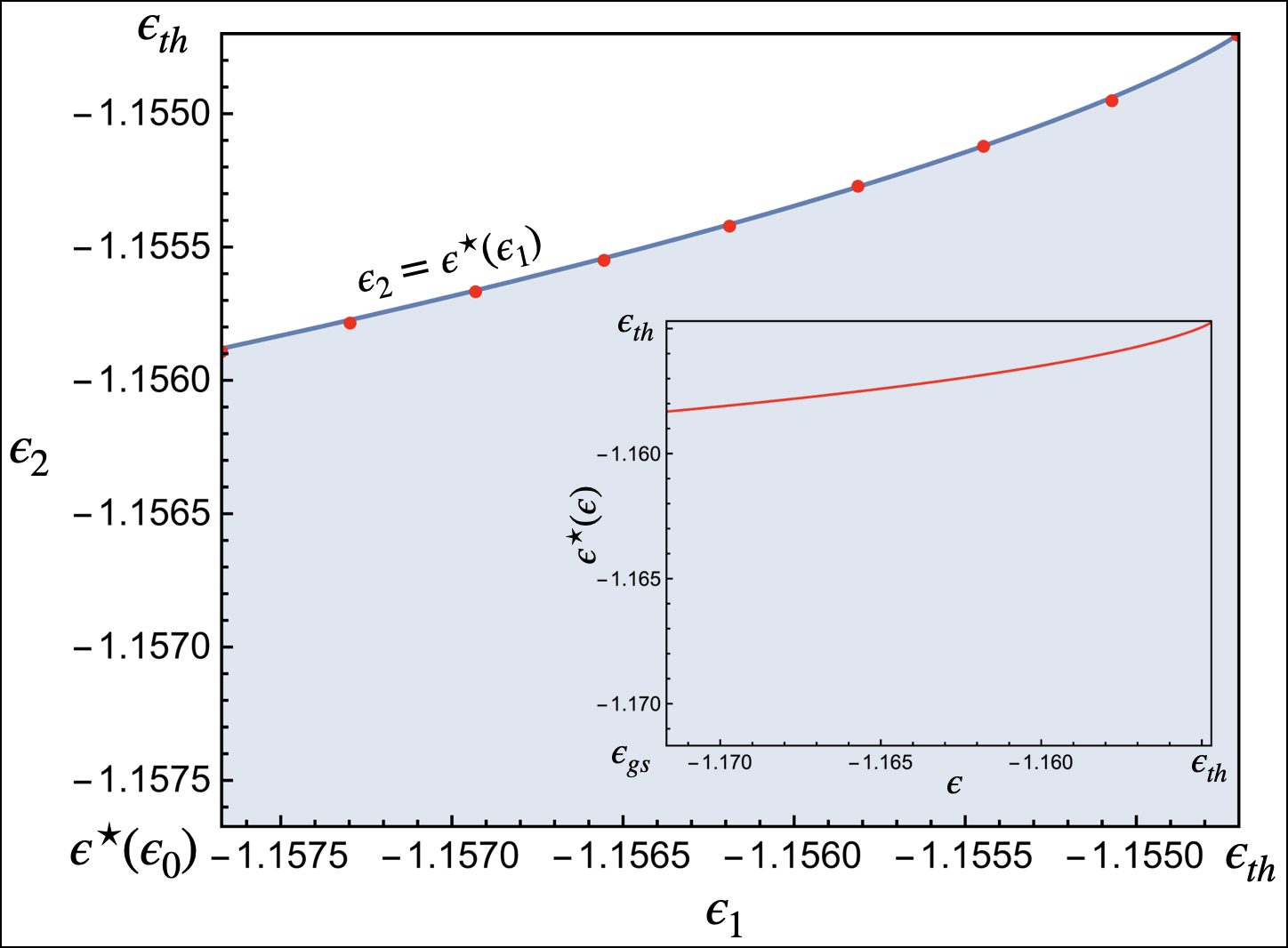}
\caption{The colored area identifies the range of energies $\epsilon_1, \epsilon_2 \geq \epsilon^*(\epsilon_0)$ for which clustering occurs, for $\epsilon_0=-1.167$. Red points are values extracted from the  analysis of the three-point complexity; the continuous line corresponds to $\epsilon_2=\epsilon^\star(\epsilon_1).$  Clustering is present whenever $\epsilon_1>\epsilon^\star(\epsilon_0)$ and $\epsilon_2\in[\epsilon^\star(\epsilon_0),\epsilon^\star(\epsilon_1)]$. {\it Inset.} Behavior of the critical energy $\epsilon^*(\epsilon)$ computed in \cite{ros2019complexity}.}
\label{fig:2D_plot_clustering_e1e2}
\end{figure}

Since correlations in the landscape become relevant for large values of the overlaps, to simplify the discussion in this section we set $q=q_M(\epsilon_1|\epsilon_0)$ for each choice  of $\epsilon_1$. We then ask for which values of $\epsilon_2$  clustering occurs in the energy landscape, meaning that there is at least one choice of $q_1, q_0$ for which 
$\Sigma^{(3)}(\epsilon_2, q_0, q_1 | \epsilon_1,\epsilon_0, q_M(\epsilon_1|\epsilon_0)) >0$ but $\Sigma^{(2)}(\epsilon_2, q_1 |\epsilon_1)=-\infty$. We find that a crucial role is played by the energy density $\epsilon^*(\epsilon)$. We recall that this critical energy is the one above which index-1 saddles or correlated minima appear in the landscape in the vicinity of a local minimum ${\bf s}$ of energy density $\epsilon$, in the two-point complexity, see Fig.~\ref{fig:2D_plot_clustering}. This critical energy acquires also another role in the context of our analysis:
we find indeed that clustering occurs only whenever two conditions on the energy densities are met: (i) $\epsilon_1, \epsilon_2 \geq \epsilon^*(\epsilon_0)$, and (ii) $\epsilon_2 \leq \epsilon^*(\epsilon_1)$. Therefore, the critical energies identified by the two-point complexity play also a crucial role in determining which stationary points are strongly correlated in the landscape. In Fig.~\ref{fig:2D_plot_clustering_e1e2} we show, for $\epsilon_1, \epsilon_2> \epsilon^*(\epsilon_0)$, the maximal values of $\epsilon_2$  for which clustering occurs (red points), derived by inspecting the three-point complexity: these value match perfectly with $\epsilon^*(\epsilon_1)$.

Notice that clustering appears discontinuously as a function of $\epsilon_1$: for $\epsilon_1 < \epsilon^*(\epsilon_0)$ there is no clustering (no matter what is the value of $\epsilon_2$), whereas as soon as $\epsilon_1\geq  \epsilon^*(\epsilon_0)$ there is a whole range of energies $\epsilon_2$ such that the closest stationary points at those energies are in the clustering region. When clustering occurs for $q=q_M(\epsilon_1|\epsilon_0)$, it is also in general present in the landscape for smaller values of $q$:  Fig.~\ref{Fig:Plots3D}(d) gives an example of clustering occurring when $ \epsilon_1> \epsilon_2> \epsilon^*(\epsilon_0)$, for a value of $q<q_M(\epsilon_1|\epsilon_0).$\\

Let us conclude this section by commenting on the connections between clustering and isolated modes in the spectrum of the Hessian of the stationary points. The necessary conditions on the energy densities that we have identified for clustering to occur involve the special energy $\epsilon^*(\epsilon)$,
which is related to the appearance of isolated eigenvalues in the Hessian of stationary points counted by the two-point complexity: for such isolated eigenvalue to be present, the energy density of the counted stationary points must be larger than this threshold. It is interesting that this energy density, that can be determined solely from the two-point complexity, also provides some information on the distribution of triplets of stationary points in the landscape. However, as remarked above, the occurrence of clustering \emph{is not} in one-to-one correspondence with the presence of isolated eigenvalues in the Hessian spectra of either ${\bf s}_1$ or ${\bf s}_2$: Fig.~\ref{Fig:Plots3D}(b) shows on one hand that stationary points cluster in the vicinity of ${\bf s}_1$ even when the latter is a local minimum with an Hessian that shows no isolated eigenvalues and no signatures of instability; on the other hand, the stationary points ${\bf s}_2$ that cluster are not necessarily unstable saddles or correlated minima, but can be stable minima with an Hessian that shows no signatures of correlations to ${\bf s}_0, {\bf s}_1$ (at least, within the annealed study of the Hessian statistics discussed in  Sec.~\ref{sec:stability}).

\section{Hints on activated dynamics}\label{sec:Dynamics}
The three-point complexity encodes the information on the conditional distribution of stationary points ${\bf s}_2$ that lie in the region of configuration space surrounding another pair of stationary points ${\bf s}_0, {\bf s}_1$. The analysis of Sec.~\ref{sec:LandscapeEvolution} allows us to understand to what extent the landscape in the vicinity of a deep minimum ${\bf s}_0$ ($\epsilon_0 < \epsilon_{\rm th}$) is affected by the presence of the minimum itself. This contains some potentially interesting information on activated dynamics in this model, as we discuss in this section.\\

{\bf Typical vs atypical regions of the landscape. } As we have stressed in Sec.~\ref{subsec:defs_clustering}, the knowledge of the two- and three-point complexity allows us to compare the structure of the landscape in the vicinity of (i) \emph{typical} stationary points ${\bf s}_1$ with energy density $\epsilon_1$, i.e., stationary points extracted with the uniform measure over all stationary points with that energy density and no additional constraint; (ii) \emph{conditioned} stationary points ${\bf s}_1$ with  energy density $\epsilon_1$, at overlap $q$ with another \textit{typical} stationary point ${\bf s}_0$ with energy density $\epsilon_0$. The absence of local accumulation and clustering implies that the energy landscape in the vicinity of ${\bf s}_1$, probed with ${\bf s}_2$, is not strongly affected by the conditioning to ${\bf s}_0$; rather, it is similar to the landscape in the vicinity of a typical stationary point ${\bf s}_1$ with the same energy density $\epsilon_1$. 
In particular, given a  sequence of three stationary points ${\bf s}_0$, ${\bf s}_1$ and ${\bf s}_2$ with fixed overlaps $q, q_1$ between the consecutive pairs, when there is no local accumulation then the third stationary point ${\bf s}_2$ is typically at overlap $q_0=q \cdot q_1$ with the first one (meaning that this is the overlap where the complexity is maximal, i.e., the exponential majority in $N$ of stationary points is found there); the corresponding complexity shows no dependence on ${\bf s}_0$, as it coincides precisely with the two-point complexity that one would get neglecting the conditioning to ${\bf s}_0$, see Sec.~\ref{sec:DoublyAnnealed}. Notice that this does not mean that the landscape is totally unaffected by the presence of ${\bf s}_0$: in fact, for all values of $q_0 \neq q q_1$ the three-point complexity is sensitive to the conditioning to ${\bf s}_0$, and it is smaller than the two-point complexity. However, optimizing over $q_0$ this dependence disappears. On the other hand, local accumulation and clustering occur when the landscape in the vicinity of ${\bf s}_1$ is strongly affected by ${\bf s}_0$, and it is characterized by a higher concentration of stationary points with respect to the concentration one finds around a typical point of the same energy density $\epsilon_1$. The local accumulation and clustering transitions can then be interpreted as decorrelation-correlation transitions in the energy landscape. \\

\begin{figure}[ht]
    \centering
\includegraphics[width=0.46
\textwidth, trim=4 4 4 4,clip]{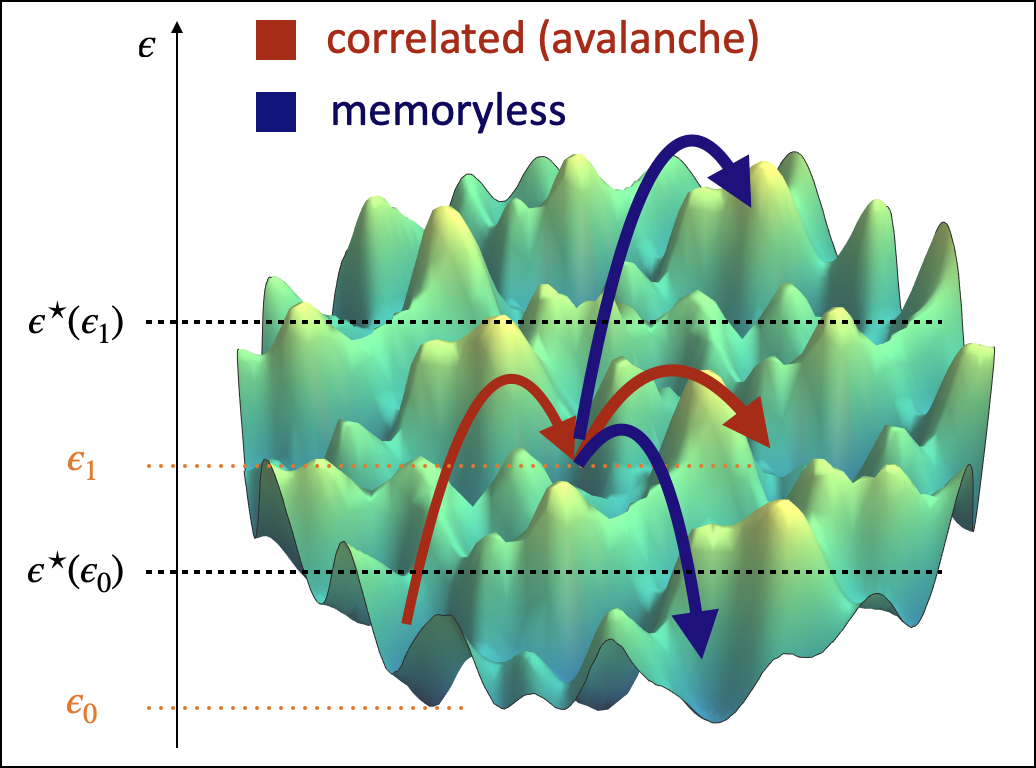}
    \caption{Pictorial representation of activated jumps in the landscape. The leftmost red arrow represents a jump from $\epsilon_0$ to $\epsilon_1$, while arrows starting from $\epsilon_1$ represent the second jump, to $\epsilon_2$. The avalanche scenario is realized for $\epsilon_1, \epsilon_2 \geq \epsilon^*(\epsilon_0)$ and $\epsilon_2< \epsilon^*(\epsilon_1)$; in all other cases, jumps are memoryless.}
   \label{fig:land_jumps}
\end{figure}

{\bf Memoryless jumps and avalanche precursors.} 
We now discuss some implication of these findings for the system's dynamics. 
Low-temperature activated dynamics in glassy landscapes is characterized by a separation of timescales: the system spends long periods fluctuating near a local minimum, until a rare noise-induced fluctuation pushes it over an energy barrier, allowing it to settle into a new local minimum. When modelling this dynamics, it is natural to neglect short-time fluctuations and  introduce effective models in which the dynamics is described as a stochastic jump process on a network of states (representing the local minima). Models of this type have been studied quite extensively, from the exactly solvable trap model~\cite{bouchaud1992weak, dyre1987master,monthus1996models, bouchaud1995aging}, to the step model~\cite{barrat1995phase}, up to their generalizations~\cite{margiotta2018spectral, margiotta2019glassy, bertin2003cross, tapias2020entropic}. Comparison with simulations of the dynamics in finite size systems with $p$-spin energy landscape with binary variables $s_i=\pm 1$ have also been performed recently~\cite{stariolo2019activated, stariolo2020barriers}. \\ 
These models assume a Markovian effective dynamics, in which the system jumps with transition rates that do not depend on the configurations (minima) previously visited by the system. Here, we revisit this in light of our results on the three-point complexity.
 We consider a short activated path $\mathcal{P}$ made of two consecutive jumps between stationary points/local minima of the landscapes, $\mathcal{P}={\bf s}_0 \to {\bf s}_1  \to {\bf s}_2$, of energy densities $\epsilon_\alpha$ with $\alpha=0, 1, 2$. This path is associated to a \textit{transition rate} $r(\mathcal{P})$, which is the product of the transition rates $r^{(1,2)}$ associated to each of the jumps. For Langevin dynamics at small temperature, the transition rates $r^{(1,2)}$ are expected to follow an Arrhenius scaling, i.e., to be exponentially suppressed in the energy barrier crossed by the system in the jump. Determining the statistics of these barriers is an open problem~\cite{pacco2024curvature, ros2023high, aspelmeier2022free, ros2019complexity, kent2024arrangement}. Given the isotropic statistics of the landscape (Eq.~\eqref{eq:Enp}), it is natural to assume that the resulting typical transition rates depend on the configurations only through their overlaps; we thus write the transition rate associated to our short path as
\begin{equation}\label{eq:TransRates}
\begin{split}
    &r(\mathcal{P})= r^{(1)}(\epsilon_1, q| \epsilon_0) r^{(2)}(\epsilon_2, q_1| \epsilon_1,\epsilon_0,q, q_0).
    \end{split}
\end{equation}
We say that the second jump is \emph{memoryless} if its rate does not depend on the properties of ${\bf s}_0$, 
\begin{equation}\label{eq:Markov}
r^{(2)}(\epsilon_2, q_1| \epsilon_1,\epsilon_0,q, q_{0})= r^{(1)}(\epsilon_2, q_1| \epsilon_1).
\end{equation}
In contrast, we say that the jump is \emph{avalanche-like} when 
\begin{equation}\label{eq:aval}
 r^{(2)}(\epsilon_2, q_1| \epsilon_1,\epsilon_0,q, q_{0})> r^{(1)}(\epsilon_2, q_1| \epsilon_1),
\end{equation}
meaning that the systems escaping from the deep minimum ${\bf s}_0$ visits configurations ${\bf s}_1$ from which jumps to ${\bf s}_2$ are “easier" (associated to lower energy barriers and thus larger transition rates) than what one would find starting the dynamics from a typical configuration with the same energy density as ${\bf s}_1$. This facilitation can happen whenever the minima at energy $\epsilon_1$ reached from ${\bf s}_0$ are not typical in the sense discussed above, and the landscape around them is strongly correlated to ${\bf s}_0$. \\
Our landscape analysis suggests that both types of jumps can happen in the $p$-spin energy landscape depending on the energy densities, but that memoryless jumps are preponderant. To relate more directly to the complexity calculation, we focus on paths in which at each step the system jumps to the \emph{closest} available stationary points/minima at the given energy density $\epsilon_1, \epsilon_2$. It is natural to expect that these processes are associated to the lowest energy barriers, since less rearrangements have to be performed to map the starting configuration into final one \footnote{For the $p$-spin energy landscape, this trend is confirmed by the calculation of proxies of the energy barriers, such as the maximum reached by the energy profile along predefined paths interpolating between local minima of the landscape~\cite{pacco2024curvature}.}. In this setting, for the dynamics to be memoryless, it must hold that the maximal possible value of the overlap $q_1^{\rm max}({\bm \epsilon},q)$ (see the definition in Eq.~\eqref{eq:q1max})  associated to the second jump is independent of $\epsilon_0$: this is precisely what happens when there is no clustering in the energy landscape, and $q_1^{\rm max}({\bm \epsilon},q)= q_M(\epsilon_2| \epsilon_1)$. Our results on the three-point complexity indicate that
 most of the jumps to the closest \emph{stationary points} (minima or saddles) are memoryless. In fact, it suffices that either ${\bf s}_1$ and ${\bf s}_2$ 
have energies that lie below $\epsilon^*(\epsilon_0)$, or that $\epsilon_1>\epsilon^*(\epsilon_0)$ and $\epsilon_2> \epsilon^*(\epsilon_1)$ to have no clustering in the landscape: the closest stationary points are found at $q=q_{M}(\epsilon_1| \epsilon_0)$ and $q_1^{\rm max}=q_{M}(\epsilon_2| \epsilon_1)$, and  correspondingly $q_0=q_{M}(\epsilon_1| \epsilon_0) q_{M}(\epsilon_2| \epsilon_1)$.  On the other hand, a necessary (but not sufficient) condition for correlations (i.e. avalanche-like behavior) to show up in the dynamics is that the two jumps bring and keep the system to stationary points at high-enough energies, larger than $\epsilon^*(\epsilon_0)$. This is summarized pictorially in Fig.~\ref{fig:land_jumps}.  Notice that when modeling activated dynamics, one is particularly interested in jumps between stable \emph{local minima} of the landscape, rather than stationary points in general. When $\epsilon_1 > \epsilon^*(\epsilon_0)$ (or  $\epsilon_2 > \epsilon^*(\epsilon_1)$), the closest local minima are not at overlap $q_M$ but rather at overlap $q_{\rm ms} (\epsilon_1| \epsilon_0)$  (or $q_{\rm ms} (\epsilon_2| \epsilon_1)$), see Fig.~\ref{fig:2D_plot_clustering}. In this case, the dynamics is avalanche-like whenever local minima are found at overlap $q_1 >q_{\rm ms} (\epsilon_2| \epsilon_1)$. We find that the same statement remains true also when restricting to jumps between local minima: when $\epsilon_1, \epsilon_2 > \epsilon^*(\epsilon_0)$ and $\epsilon_2<\epsilon^*(\epsilon_1)$, one can find local minima, both correlated and uncorrelated, at values of $q_1> q_M(\epsilon_2|\epsilon_1)$. On the other hand, as soon as $\epsilon_2> \epsilon^*(\epsilon_1)$, the closest local minima are found exactly at $q_{\rm ms}(\epsilon_2|\epsilon_1)$: it turns out that this is exactly the value of the overlap that corresponds to the vanishing of the isolated eigenvalue of the Hessian at the stationary point ${\bf s}_2$, computed in Appendix~\ref{app:spectralP}.\\

\section{Conclusions}
\label{sec:conclusion}
In this work we have characterized local correlations between stationary points of the energy landscape of the pure, spherical $p$-spin model. In particular, we have computed the three-point complexity, giving the typical number of stationary points ${\bf s}_2$ of arbitrary energy density that are found  at fixed distance (in configuration space) from a pair of stationary points ${\bf s}_1, {\bf s}_0$ at given energy densities and mutual distance. By tuning the parameters, this calculation allowed us to quantify to what extent the landscape around a stationary point ${\bf s}_1$ placed in the vicinity of a deep minimum ${\bf s}_0$ is different from the landscape surrounding typical stationary points of the same energy density as  ${\bf s}_1$; in other words, it allows us to quantify to what extent the landscape in the vicinity of a deep minimum is correlated to the minimum itself. 
We have shown how correlations get stronger as one gets closer to the deep minimum ${\bf s}_0$, generating phenomena such as local accumulation of stationary points, and their clustering. However, to see such effects one has to reach regions of the  landscapes that are at extensively higher energy with respect to ${\bf s}_0$. We have  argued that this implies that, in an effective description of the dynamics in which the system jumps between local minima of the landscape, most of the jumps are memoryless, i.e., independent of the minima previously visited by the system. Correlations in the transition rates should be accounted for only when considering sequences of jumps reaching minima at high-enough energies. 

Our analysis is motivated by the problem of activated dynamics in glassy landscapes, and in particular by the scenario of thermal avalanches occurring in low-dimensional systems. In systems such as elastic random manifolds, the statistics of energy barriers is obtained implementing numerical protocols in which the system transitions between configurations of the same energy, that are connected by the smallest possible number of rearrangements.   When applied to the $p$-spin random energy landscape, these protocols correspond to iso-energetic dynamical paths connecting the closest local minima at a given energy density. These paths are particularly interesting, since, 
 inspired by the idea of the thermally assisted flux~\cite{anderson1964hard,ioffe1987dynamics}, 
one expects that escape processes from a local minimum require to reach stable configurations of energy equal or smaller  than the departing one.
Our results imply that these iso-energetic paths between minima at energies below $\epsilon_{\rm th}$ are  memoryless: the transition rate associated to each jump is independent of the previous history of the path, and at each step the system reaches minima at the same typical distance from the starting configuration than in the previous steps. In this sense, in this mean-field model we do not find the analog of thermal avalanches, that would correspond to the scenario in which a first jump is followed by subsequent jumps to closest minima at the same energy, associated to larger transitions rate. On the other hand, our calculation shows some precursors of these avalanche-like correlations if one considers jumps from deep minima to minima at much higher energies. How much these results are specific of the pure $p$-spin model, and whether stronger correlations are present in mean-field energy landscapes associated to mixed $p$-spin models, in currently under investigation.  \\

\begin{acknowledgments}
We thank Jaron Kent-Dobias for interesting discussions on this topic. 
VR acknowledges funding by the Investissements d’Avenir LabEx PALM with reference ANR-10-LABX-0039-PALM, and by the French government under the France 2030 program (PhOM - Graduate School of Physics) with reference ANR-11-IDEX-0003. AR acknowledges funding by the ANR with reference ANR-23-CE30-0031-04.
\end{acknowledgments}

\bibliography{refs.bib}

\newpage
\onecolumngrid

\appendix

\section{The three-point complexity: doubly-annealed calculation}\label{app:A_com_2A}

\noindent In this Appendix we outline the main steps to derive the three-point complexity in the doubly-annealed setting, as introduced in Sec.~\ref{subsec:3_pt_def} and~\ref{subsec:quenched_annealed}. Formulas are derived for a general $p$,  with explicit expressions provided for $p=3$.

\subsection{Gradients and Hessians on the sphere, reference frames}
\label{app:tang_planes}

\noindent We define $\nabla_\perp\mathcal{E}({\bf s})$ and $\nabla_\perp^2\mathcal{E}({\bf s})$, the Riemannian gradient and Hessian of the function $\mathcal{E}$ restricted to the  hypersphere of dimension $N-1$ and radius $\sqrt{N}$. To align with previous works \cite{ros2019complexity, ros2019complex} and simplify the derivations, we introduce the  rescaled field:
\begin{align}\label{eq:rescFF}
    {\bm\sigma}:=\frac{\bf s}{\sqrt{N}},\quad\quad\quad h({\bm\sigma})=\sqrt{\frac{2}{N}}\mathcal{E}(\sqrt{N}{\bm\sigma}).
\end{align}

\noindent We denote with $\tau[{\bm \sigma}]$ the tangent plane to the hypersphere at the point ${\bm \sigma}$: this is the $(N-1)-$dimensional plane that is orthogonal to ${\bm \sigma}$. We indicate by $\mathcal{C}:=\{{\bf x}_1,\ldots,{\bf x}_N\}$ a canonical (orthonormal) basis of $\mathbb{R}^N$, and define a family of orthonormal local bases $\mathcal{B}[{\bm \sigma}]$ of $\mathbb{R}^N$, that contains an orthonormal set of vectors ${\bf e}_\alpha({\bm \sigma})$ spanning the tangent plane $\tau[{\bm \sigma}]$:

\begin{align}
\label{app:eq:local_basis}
    \mathcal{B}[{\bm \sigma}]&:=\bigg\{\underbrace{{\bf e}_1({\bm \sigma}),\ldots,{\bf e}_{N-1}({\bm\sigma})}_{\tau[{\bm \sigma}]},{\bf e}_N({\bm \sigma}):={\bm \sigma}\bigg\}
\end{align}

\noindent It is particularly useful to choose the set of basis vectors of the tangent planes of the three points ${\bm \sigma}_0,{\bm \sigma}_1,{\bm \sigma}_2$ to have a common intersection. In particular, we  define $S:=\text{span}({\bm \sigma}_0,{\bm \sigma}_1,{\bm \sigma}_2)$, to which is associated an orthogonal space, denoted $S^\perp$: $\mathbb{R}^N=S^\perp\oplus S$. The space $S^\perp$ is included in the tangent planes associated to all the three configurations. 
We choose an orthonormal basis of $S^\perp$, and declare that the first $N-3$ basis vectors in each $ \mathcal{B}[{\bm \sigma}_a]$ are equal:
\begin{align}
\begin{split}
&S^\perp=\text{span}({\bf e}_1,\ldots,{\bf e}_{N-3})\\
&{\bf e}_i({\bm \sigma}_a):={\bf e}_i,\quad a=0,1,2,\quad 1\leq i\leq N-3.
\end{split}
\end{align}
In this way, the sets of vectors $ \mathcal{B}[{\bm \sigma}_a]$ will only differ for the choice of their last three orthonormal vectors, which belong to $S$. A specific choice of these vectors is given in Sec.~\ref{app:sec:basis}.\\
\noindent The (unconstrained) gradient has components with respect to the basis $\mathcal{C}$ given by
\begin{equation}
\label{app:eq:def_grad}
  [\nabla h({\bm\sigma})]_i\equiv\nabla h({\bm \sigma})\cdot{\bf x}_i =\frac{\partial h({\bm \sigma})}{\partial \sigma_i}.
\end{equation}
In this specific model, the homogeneity of the field $\mathcal{E}$ implies that 
\begin{align}
\label{app:eq:first_grad_id}
    \nabla h({\bm \sigma})\cdot{\bm\sigma}=ph({\bm \sigma}).
\end{align}
This identity implies that we can write
\begin{align}
\label{app:eq:second_grad_id}
\nabla h({\bm\sigma})={\bf g}({\bm \sigma})+ph({\bm\sigma}){\bm \sigma}
\end{align}
where ${\bf g}({\bm \sigma})$ has components in the $\mathcal{B}[{\bm \sigma}]$ basis given by
\begin{align*}
&[{\bf g}({\bm \sigma})]_{\alpha<N}=\nabla h({\bm\sigma})\cdot {\bf e}_\alpha({\bm\sigma})\\
&[{\bf g}({\bm \sigma})]_N=0.
\end{align*}
The Riemannian gradient $\nabla_\perp h({\bm \sigma})$ is the $(N-1)-$dimensional vector obtained projecting ${\bf g}({\bm \sigma})$  on $\tau[{\bm \sigma}]$, that is, neglecting the null component. In the following, we sometimes denote with ${\bf g}({\bm \sigma})$ this $(N-1)- $dimensional projection as well, with a slight abuse of notation. \\
\noindent Similarly, we define by $\nabla^2 h({\bm \sigma})$ the $N \times N$-dimensional Hessian matrix with components in $\mathcal{C}$:
\begin{equation}
\label{app:eq:def_hes}
  [\nabla^2 h({\bm \sigma})]_{ij} \equiv  {\bf x}_i \cdot  \nabla^2 h({\bm \sigma}) \cdot {\bf x}_j= \frac{\partial^2 h({\bm \sigma})}{\partial \sigma_i \partial \sigma_j}.
\end{equation}
One can show that 
\begin{align}
\label{app:eq:first_hes_id}
    \nabla^2 h({\bm \sigma})\cdot{\bm \sigma}=(p-1)\nabla h({\bm \sigma}).
\end{align}

\noindent In order to obtain $\nabla^2_\perp h({\bm\sigma})$, it is convenient to enforce the constraint ${\bm\sigma}^2=1$ with a Lagrange multiplier: given $h_{\lambda}({\bm\sigma}):=h({\bm\sigma})-\frac{\lambda}{2}({\bm\sigma}^2-N)$, then $\nabla h_\lambda({\bm\sigma})=\nabla h({\bm\sigma})-\lambda{\bm\sigma}\overset{!}{=}0\Rightarrow \lambda={\bm\sigma}\cdot\nabla h({\bm\sigma})$. Therefore in the $\mathcal{B}[{\bm\sigma}]$ basis we have
\begin{align}
\label{app:eq:def_hes_riem}
[\nabla^2_\perp h({\bm \sigma})]_{\alpha\beta}\equiv \mathbf{e}_{\alpha}({\bm\sigma})^\top \nabla^2 h_\lambda({\bm\sigma})  \mathbf{e}_{\beta}({\bm \sigma})=\mathbf{e}_{\alpha}({\bm \sigma})^\top\nabla^2 h({\bm\sigma}) \mathbf{e}_{\beta}({\bm\sigma})-p h({\bm\sigma})\delta_{\alpha \beta},\quad\quad \alpha,\beta \leq N-1.
\end{align}
where we used the identity \eqref{app:eq:first_grad_id}. 
Eq.~\eqref{app:eq:def_hes_riem} shows that the Riemannian Hessian is derived from the unconstrained Hessian by shifting with a diagonal matrix proportional to $p h({\bm\sigma})$, and by projecting onto the local tangent plane. Thus, working with either the unconstrained or Riemannian Hessian is essentially equivalent, as long as the shift is taken into account. In the following, we use the simplified notation $h({\bm \sigma}_a)= h^a$, and similarly for the gradients and Hessians.

\subsection{The doubly-annealed complexity: a Kac-Rice formula}
\label{app:ann_computation}
\noindent The definition of the doubly-annealed three-point complexity is given in Sec. \ref{subsec:quenched_annealed}, where the difference with the annealed and quenched three-point complexity is explained. In essence, the standard annealed computation involves averaging the number of stationary points $\mathcal{N}_{{\bm \sigma_0} \, {\bm \sigma_1}}$ (and not its logarithm) over the population of stationary points ${\bm \sigma_0}, {\bm \sigma_1}$ and over the realizations of the landscape. Such a calculation requires to exploit the replica trick to treat appropriately the average over the configurations ${\bm \sigma}^0$ and ${\bm \sigma}^1$. In the doubly-annealed framework, thanks to the factorization assumed in \eqref{eq:averagesAnn}, replicas are no longer required since the denominators are averaged separately with respect to the numerators. The resulting calculation is much more straightforward, and can be easily extended to a higher number of stationary points. Moreover, for the spherical $p$-spin model we find that both prescriptions for the annealed calculation lead to the same result, as stated in Eq.~\eqref{eq:EqaAnn}. The resulting three-point annealed complexity is however different with respect to the quenched counterpart, and it serves as an upper bound to the latter.

\noindent With the notation introduced in Sec.~\ref{subsec:quenched_annealed}, the annealed three-point complexity reads: 

\begin{align}\label{eq:DAC1}
\Sigma_{2A}^{(3)}(\epsilon_2,q_0,q_1|\epsilon_1, \epsilon_0,q)= \lim_{N\to\infty}
\frac{1}{N}\log\frac{\mathbb{E}\left[
\int_{\mathcal{S}_N(1)^{\otimes 2}} d{\bm\sigma}_0  d{\bm \sigma}_1 \,  \omega_{\epsilon_0}({\bm \sigma}_0) \omega_{\epsilon_1,q}({\bm \sigma}_1|{\bm\sigma}_0) \; \mathcal{N}_{{\bm \sigma}_0,{\bm \sigma}_1}(\epsilon_2,q_0,q_1)\right]}{\mathbb{E}[\mathcal{N}(\epsilon_0)\,  \mathcal{N}_{{\bm \sigma}_0}(\epsilon_1,q)] },
\end{align}
where $\mathcal{N}_{{\bm \sigma}_0,{\bm \sigma}_1}(\epsilon_2,q_0,q_1)$is obtained from \eqref{eq:enne} by rescaling ${\bf s} \to {\bm \sigma}$. The measures $\omega_{\epsilon_0}({\bm \sigma_0})$, $\omega_{\epsilon_1,q}({\bm \sigma}_1|{\bm\sigma}_0) $ as well as $\omega_{\epsilon_1,q}({\bm \sigma}_1|{\bm\sigma}_0) $ depend on the random quantities $h({\bm \sigma}_a)$,   $\nabla_\perp h({\bm \sigma}_a)$ and $\nabla^2_\perp h({\bm \sigma}_a)$. Exchanging the average with the integration over the configuration space, one sees that \eqref{eq:DAC1} can be re-written (omitting dependence on $(\epsilon_1,\epsilon_0,q)$ for brevity) as: 
\begin{align}\label{eq:KAcR2A}
    \Sigma_{2A}^{(3)}(\epsilon_2,q_0,q_1|\epsilon_1, \epsilon_0,q)= \lim_{N\to\infty}
    \frac{1}{N}\log\left[\int_{\mathcal{S}_N(1)} d{\bm \sigma}_2\,\delta({\bm\sigma}_2\cdot {\bm\sigma}_0-q_0)\,\delta({\bm \sigma}_2\cdot {\bm\sigma}_1-q_1)\,\,P_{\bm\epsilon}({\bm\sigma}_2|{\bm\sigma}_0,{\bm\sigma}_1)\,H_{\bm\epsilon}({\bm \sigma}_2|{\bm\sigma}_0,{\bm \sigma}_1)\right]
\end{align}
where 
\begin{align}\label{eq:Proba}
    P_{{\bm \epsilon}}({\bm\sigma}_2|{\bm\sigma}_0,{\bm \sigma}_1):=\mathbb{E}\left[\delta(h({\bm\sigma}_2)-\sqrt{2N}\epsilon_2)\,\delta \tonde{{\bf g}({\bm\sigma}_2)}\bigg|\grafe{
    \begin{subarray}{l}
    h^a = \sqrt{2N}\epsilon_a\\ 
    {\bf g}^a={\bf 0}, \; \forall  a=0,1 \end{subarray}}
    \right]
\end{align}
measures the probability that ${\bm \sigma}_2$ is a stationary point with energy density $\epsilon_2$, conditioned to the fact that ${\bm \sigma}_
0$ and ${\bm \sigma}_
1$ are stationary points of the given energy densities, while 
\begin{align}
\label{app:eq:determinant_H}
    H_{{\bm \epsilon}}({\bm \sigma}_2|{\bm \sigma}_0,{\bm \sigma}_1)=\mathbb{E}\left[|\det\nabla^2_\perp h({\bm \sigma}_2)|\bigg|
    \grafe{
    \begin{subarray}{l}
    h^a = \sqrt{2N}\epsilon_a\\
   {\bf g}^a={\bf 0},\, \; \forall  a=0,1,2 \end{subarray}}
    \right]
\end{align}
is the expectation of the Hessian at $ {\bm \sigma}_2$, conditioned to the fact that each ${\bm \sigma}_a$ is a stationary point of given energy density. The subscript ${\bm \epsilon}$ indicate that these quantities depend on the energy densities ${\bm \epsilon}=(\epsilon_0, \epsilon_1, \epsilon_2)$. Eq. \eqref{eq:KAcR2A} is a Kac-Rice formula for the doubly-annealed three-point complexity. In the following subsection, we argue that the terms $P_{{\bm \epsilon}}, H_{{\bm \epsilon}}$ are isotropic: they depend on the configurations ${\bm \sigma}_a$ only through their overlaps, allowing us to simplify the formula for the complexity.

\subsection{Correlations, isotropy, and simplification of the Kac-Rice formula}
\label{app:grad_correls}

\noindent The quantities $h({\bm \sigma}_a)$,   $\nabla_\perp h({\bm \sigma}_a)$ and $\nabla^2_\perp h({\bm \sigma}_a)$ are a scalar, a vector and a matrix with Gaussian statistics (of each entry). The first and second moments of these quantities can be computed explicitly \cite{ros2019complex, subag2017complexity}, and we report them here for completeness. As before, we use the simplified notation $h({\bm \sigma}_\alpha)= h^\alpha$, and similarly for the gradients and Hessians. For arbitrary vectors ${\bf u}_i$ and configurations ${\bm \sigma}_a$ it holds that:
 \begin{equation}\label{app:eq:AvGrad}
 \begin{split}
 & \mathbb{E}\left[ \tonde{{\bm \nabla} h^a \cdot {\bf u}_1} h^b  \right]= p ({\bm \sigma}_a \cdot {\bm \sigma}_b)^{p-1} \tonde{{\bf u}_1 \cdot {\bm \sigma}_b},\\
  & \mathbb{E}\left[\tonde{{\bf u}_1 \hspace{-0.1cm}\cdot \hspace{-0.1cm} {\bm \nabla}^2 h^a \hspace{-0.1cm} \cdot \hspace{-0.05cm} {\bf u}_2}  h^b \right]=p(p\hspace{-0.05cm}-\hspace{-0.05cm}1)({\bm \sigma}_a \hspace{-0.05cm}\cdot\hspace{-0.05cm} {\bm \sigma}_b)^{p-2}({\bf u}_1 \hspace{-0.05cm}\cdot\hspace{-0.05cm} {\bm \sigma}_b) ({\bf u}_2 \hspace{-0.05cm}\cdot\hspace{-0.05cm} {\bm \sigma}_b).
  \end{split}
 \end{equation}
Between the gradient components we have instead:
 \begin{equation}\label{app:eq:CovGradComp}
 \begin{split}
  & \mathbb{E}\left[ \tonde{{\bm \nabla} h^a \cdot {\bf u}_1} \tonde{{\bm \nabla} h^b \cdot {\bf u}_2} \right] =p ({\bm \sigma}_a \cdot {\bm \sigma}_b)^{p-1} \tonde{{\bf u}_1 \cdot {\bf u}_2}+p(p-1)({\bm \sigma}_a \cdot {\bm \sigma}_b)^{p-2} \tonde{{\bf u}_2 \cdot {\bm \sigma}_a} \tonde{{\bf u}_1 \cdot {\bm \sigma}_b}.
  \end{split}
 \end{equation}
 For what concerns the Hessians, one gets:
\begin{equation}\label{app:eq:HessTang}
\begin{split}
& \mathbb{E}\left[\tonde{ {\bf u}_1 \cdot  {\bm \nabla}^2 h^a \cdot {\bf u}_2} \tonde{ {\bf u}_3 \cdot {\bm \nabla}^2 h^b \cdot {\bf u}_4} \right]=\frac{p! ({\bm \sigma}_a \cdot {\bm \sigma}_b)^{p-4}}{(p-4)!}({\bf u}_1 \cdot {\bm \sigma}_b) ({\bf u}_2\cdot {\bm \sigma}_b) ({\bf u}_3\cdot {\bm \sigma}_a)( {\bf u}_4 \cdot {\bm \sigma}_a)+\\
 &\frac{p!}{(p-3)!}({\bm \sigma}_a \cdot {\bm \sigma}_b)^{p-3} \,{({\bf u}_1 \cdot {\bf u}_4 )( {\bf u}_2 \cdot {\bm \sigma}_b)( {\bf u}_3 \cdot {\bm \sigma}_a)}+\frac{p!}{(p-3)!}({\bm \sigma}_a \cdot {\bm \sigma}_b)^{p-3} \,{({\bf u}_2 \cdot {\bf u}_4)( {\bf u}_1 \cdot {\bm \sigma}_b)( {\bf u}_3\cdot {\bm \sigma}_a)}+\\
  &\frac{p!}{(p-3)!}({\bm \sigma}_a \cdot {\bm \sigma}_b)^{p-3}\, {({\bf u}_1 \cdot {\bf u}_3)( {\bf u}_2 \cdot {\bm \sigma}_b)( {\bf u}_4 \cdot {\bm \sigma}_a)}+\frac{p!}{(p-3)!}({\bm \sigma}_a \cdot {\bm \sigma}_b)^{p-3}\, {({\bf u}_2 \cdot {\bf u}_3)( {\bf u}_1 \cdot {\bm \sigma}_b)( {\bf u}_4\cdot {\bm \sigma}_a)}+\\
 &\frac{p! ({\bm \sigma}_a \cdot {\bm \sigma}_b)^{p-2} }{(p-2)!}\quadre{( {\bf u}_1\cdot  {\bf u}_3)( {\bf u}_2\cdot  {\bf u}_4)+( {\bf u}_1\cdot  {\bf u}_4)( {\bf u}_2\cdot  {\bf u}_3)}.
 \end{split}
\end{equation}
Finally, the correlations between Hessians and gradients read: 
\begin{equation}\label{app:eq:CorelationsHessianGrad}
 \begin{split}
& \mathbb{E}\left[   \tonde{ {\bf u}_1 \cdot  {\bm \nabla}^2 h^a \cdot {\bf u}_2} \tonde{{\bm \nabla} h^b \cdot  {\bf u}_3 } \right]
 =p(p-1)(p-2) ({\bm \sigma}_a \cdot {\bm \sigma}_b)^{p-3} ({\bf u}_1 \cdot {\bm \sigma}_b) ({\bf u}_2 \cdot {\bm \sigma}_b)  ({\bf u}_3\cdot {\bm \sigma}_a)+ \\
 & p(p-1) ({\bm \sigma}_a \cdot {\bm \sigma}_b)^{p-2} ({\bf u}_1 \cdot {\bf u}_3) ({\bf u}_2 \cdot {\bm \sigma}_b)+  p(p-1) ({\bm \sigma}_a \cdot {\bm \sigma}_b)^{p-2}({\bf u}_2 \cdot {\bf u}_3) ({\bf u}_1 \cdot {\bm \sigma}_b) .
 \end{split}
\end{equation}
We now assume that the vectors ${\bf u}_i$ multiplying each $\nabla h^a$ and $\nabla^2 h^a$ belong to the local bases $\mathcal{B}[{\bm \sigma}_a]$ defined in \eqref{app:eq:local_basis}. Then, one easily realizes that (i) the correlations between the components of $\nabla h^a$, $\nabla^2 h^a$ along directions that belong to the subspace $S^\perp$ simplify drastically and are isotropic (independent of the configuration where gradient and Hessian are evaluated), since these directions are orthogonal to the ${\bm \sigma}_a$; (ii) the correlations between the components along the remaining directions in $S$ depend on the configurations ${\bm \sigma}_a$, but are  functions only of their overlaps: this follows from the fact that the vectors in $S$ are a linear combination of the ${\bm \sigma}_a$. It follows from this that both the probability \eqref{eq:Proba} and the expectation \eqref{app:eq:determinant_H} depend on the configurations ${\bm \sigma}_a$ only through the parameters $q, q_0, q_1$. We can therefore set $P_{{\bm \epsilon}}({\bm\sigma}_2|{\bm\sigma}_0,{\bm \sigma}_1) \to  P(\epsilon_2, q_0, q_1| \epsilon_1, \epsilon_0, q)$ and $H_{{\bm \epsilon}}({\bm\sigma}_2|{\bm\sigma}_0,{\bm \sigma}_1) \to H(\epsilon_2, q_0, q_1| \epsilon_1, \epsilon_0, q)$. As a consequence of this isotropy,  the expression of $\Sigma^{(3)}_{2A}(\epsilon_2, q_0, q_1|\epsilon_1, \epsilon_0,q)$ simplifies to:
\begin{align}\label{eq:KR_iso}
    \Sigma_{2A}^{(3)}(\epsilon_2,q_0,q_1|\epsilon_1, \epsilon_0,q)= \lim_{N\to\infty}\frac{1}{N}\log\left[V({\bf q})\,\cdot  P(\epsilon_2, q_0, q_1| \epsilon_1, \epsilon_0, q)\,\cdot  H(\epsilon_2, q_0, q_1| \epsilon_1, \epsilon_0, q)\right]
\end{align}
where $V({\bf q})$ is the volume of the configuration  space associated to the configuration ${\bm \sigma}_2$, that is conditioned to be at overlaps $q_0$ and $q_1$ with two other configurations ${\bm \sigma}_0$ and ${\bm \sigma}_1$ that are at overlap $q$ with each others. Here, ${\bf q}=(q, q_0, q_1)$.\\
\noindent Below, we derive an explicit expression for each of the three terms in \eqref{eq:KR_iso}, to leading exponential order in $N$. Combining all the terms discussed in the subsections below, we get
 equation Eq.~\eqref{eq:ann_formulaComp2A}.

\subsection{The phase space factor}
\label{app:phase_factor}
\noindent Here, we compute the volume term $V({\bf q})$. We begin with a remark. In general, for $f:\mathbb{R}^N\to\mathbb{R}$, the coarea formula implies that
\begin{align}
    \int_{\mathbb{R}^N}d{\bf x}\,\delta(f({\bf x}))\,||\nabla f({\bf x})||\,\phi({\bf x})=\int_{S=\{{\bf x}|f({\bf x})=0\}}
    \phi({\bf x})dS({\bf x})
\end{align}
with $dS$ the surface measure. For $\phi({\bf x})=1$ and $f({\bf x})=||{\bf x}||-1$, one obtains the surface of the $N$-dimensional unit sphere by integrating
\begin{align}
    \int_{\mathbb{R}^N}d{\bf x}\,\delta(||{\bf x}||-1).
\end{align}
If instead one wants to work with $f({\bf x})={\bf x}^2-1$, it is necessary to introduce a factor 2, and compute
\begin{align}
    2\int_{\mathbb{R}^N}d{\bf x}\,\delta(||{\bf x}||^2-1).
\end{align}
Since here we are interested in the exponential behaviour to leading order in $N$, we can neglect proportionality factors. Therefore, the phase space factor, which should in principle be integrated on the sphere $\mathcal{S}_N(1)$, can be written as:
\begin{align}
V({\bf q})\propto \int_{\mathbb{R}^N} d{\bm \sigma}_2\,\delta({\bm \sigma}_2\cdot{\bm\sigma}_2-1)\,\delta({\bm \sigma}_2\cdot {\bm \sigma}_0-q_0)\,\delta({\bm \sigma}_2\cdot{\bm \sigma}_1-q_1).
\end{align}
To proceed, we insert the Fourier representation of the Dirac delta,  $\delta(f(x))=\frac{1}{2\pi}\int\,dt\,e^{i t f(x)}$ and perform some rescalings; keeping only the terms that contribute exponentially in $N$, we get:
\begin{align}
\begin{split}
V({\bf q})&\propto \int d\lambda_0d\lambda_1d\lambda\,
    e^{N[\lambda_0q_0+\lambda_1q_1+\lambda]}
    \int d{\bm\sigma}_2\,e^{-\lambda_0{\bm \sigma}_0\cdot{\bm \sigma}_2-\lambda_1{\bm \sigma}_1\cdot{\bm\sigma}_2-\lambda{\bm \sigma}_2\cdot{\bm\sigma}_2}\\
&\propto \int d\lambda_0d\lambda_1d\lambda\,
    e^{N[\lambda_0q_0+\lambda_1q_1+\lambda]}
    \prod_{i=1}^N\left[\int dx\,e^{-x(\lambda_0{\bm \sigma}_{0,i}+\lambda_1{\bm \sigma}_{1,i})-\lambda x^2}\right]\\
&\propto \pi^{\frac{N}{2}}N^{-\frac{N}{2}}\int d\lambda_0d\lambda_1 d\lambda e^{\frac{N}{2}\left[ -\log(\lambda)+2\lambda_0q_0+2\lambda_1q_1+2\lambda+\frac{1}{2\lambda}(\lambda_0^2+\lambda_1^2+2\lambda_0\lambda_1q) \right]}.
\end{split}
\end{align}
Evaluating the integrals with a saddle point approximation, we find
\begin{align}
    V({\bf q})=e^{\frac{N}{2}\left[1+\log\left(\frac{2\pi}{N}\right)+\log\left(\frac{1-q^2-q_0^2-q_1^2+2q\,q_0\,q_1}{1-q^2}\right)
    \right]+o(N)}.
\end{align}
We observe that this factor already imposes a geometrical constraint on the position of our configurations on the sphere:
$1-q^2-q_0^2-q_1^2+2q\,q_0\,q_1>0$. We also remark that the term scaling faster than exponential, as $N\log(N)$ in the exponent, will cancel out when combining this with the other contributions to the complexity.

\subsection{The conditional probability of the energies and gradients}
\label{app:pdf_analysis}
\noindent We now turn to the conditional probability density 
 $P(\epsilon_2, q_0, q_1| \epsilon_1, \epsilon_0, q)$. The conditional probability can be written as
\begin{align}\label{eq:DistRatio}
   P(\epsilon_2, q_0, q_1| \epsilon_1, \epsilon_0, q)=\frac{P_3({\bm \epsilon}, {\bf q})}{P_2(\epsilon_0, \epsilon_1, q)}
\end{align}
where $P_3({\bm \epsilon}, {\bf q})$ is the joint probability density of ${\bf g}^a=0$ and $h^a= \sqrt{2N} \epsilon_a$ for $a=0,1,2$, while $P_2(\epsilon_0, \epsilon_1, q)$ is the analogous quantity for $a=0,1$. The term $P_2(\epsilon_0, \epsilon_1, q)$ can be read from the calculation of the two-point complexity in \cite{ros2019complexity}. We therefore focus on the calculation of the numerator, following the method introduced in \cite{ros2019complex}. While within the quenched formalism the calculation of the joint probability is quite involved due to the presence of replicas, in the doubly-annealed approximation the method is more straightforward: in general, the calculation of the joint distribution of gradients and energies computed at $n$ distinct configurations requires to evaluate and invert a $n^2\times n^2$ matrix ($n=3$ here). We use the fact that for the $p$-spin model it holds $\nabla h({\bm \sigma})= {\bf g}({\bm\sigma}) + p\,h({\bm \sigma}){\bf e}_N({\bm \sigma})$, meaning that we can include the conditioning on the value of the energies as part of the conditioning on the values of the (unconstrained) gradients. Hence, we have to compute the joint probability density function of the three gradients  evaluated at
\begin{align*}
\nabla h({\bm \sigma}_a)={\bf 0}+p\sqrt{2N}\epsilon_a\,{\bf e}_N({\bm \sigma}_a).
\end{align*}
Using that the gradients components are centered Gaussian random variables, we get:
\begin{align}
    \begin{split}
   P_3({\bm \epsilon}, {\bf q})=     \mathbb{P}(\{\nabla h({\bm \sigma}_a)=p\sqrt{2N}\epsilon_a\,{\bf e}_N({\bm\sigma}_a)\}_{a=0}^2)=\frac{e^{-\frac{1}{2}\sum_{a,b=0}^2\nabla h^a \cdot\left[\hat{C}^{-1}\right]^{ab}\cdot\nabla h^b}}{(2\pi)^{\frac{3N}{2}}|\det\hat{C}|^{\frac{1}{2}}}=\frac{e^{-N\,F({\bm\epsilon},{\bf q})}}
        {(2\pi)^{\frac{3N}{2}}|\det\hat{C}|^{\frac{1}{2}}}
    \end{split}
\end{align}
where 
\begin{align}
\begin{split}\label{eq:effe}
F({\bm\epsilon},{\bf q})&=\epsilon_0^2\,Y_{0}^{(p)}({\bf q})+\epsilon_0\epsilon_1\,Y_{01}^{(p)}({\bf q})+\epsilon_0\epsilon_2\,Y_{02}^{(p)}({\bf q})
+\epsilon_1\epsilon_2\,Y_{12}^{(p)}({\bf q})+\epsilon_1^2Y_{1}^{(p)}({\bf q})+\epsilon_2^2Y_{2}^{(p)}({\bf q})
\end{split}
\end{align}
 with $Y_{ab}^{(p)}({\bf q}):=p^2\left[{\bm \sigma}_a\cdot[\hat{C}^{-1}]^{ab}\cdot{\bm \sigma}_b +
 {\bm \sigma}_b\cdot[\hat{C}^{-1}]^{ba}\cdot{\bm \sigma}_a\right]$ for $a<b$ (and only one term for $a=b$) and where the correlation matrix $\hat{C}$ takes the following form (in any orthonormal basis of $\mathbb{R}^N$), by Eq.~\eqref{app:eq:CovGradComp}:
\begin{align}
    C_{ij}^{ab}=\mathbb{E}\left[  \nabla h_i^a\nabla h^b_j\right]=p\, Q^{p-1}_{ab}\delta_{ij}+p(p-1)Q^{p-2}_{ab}\sigma_j^a\sigma_i^b, \quad \quad Q_{ab}={\bm\sigma}_a\cdot{\bm\sigma}_b.
\end{align}
 The matrix $\hat C$ is very high-dimensional ($3N \times 3N$); since $F({\bm \epsilon},{\bf q})$ requires to compute only contractions of its inverse with the vectors ${\bm \sigma}_a$, a convenient way to proceed is to identify subsets of vectors (including the ${\bm \sigma}_a$) that are closed under the action of  $\hat{C}$: this allows to invert the correlation matrix only in the corresponding subspace, reducing the complexity of the inversion. This method was introduced in \cite{ros2019complex}.  We therefore proceed with the following steps: (i) find  a set of vectors that are closed under the action of $\hat{C}$, (ii) orthogonalize them, (iii)  invert the matrix $\hat{C}$ (or better its action on this set of vectors), and (iv) extract the quantities ${\bm\sigma}_a\cdot[\hat{C}]^{ab}\cdot{\bm\sigma}_b$ for any $a,b=0,1,2$.\\
\noindent It can be easily checked that the following set of $(3N)$-dimensional vectors is closed under the action of $\hat{C}$:
\begin{align}
\begin{split}
    &{\bm \xi}^1=({\bm \sigma}_0,{\bf 0},{\bf 0})\quad\quad{\bm \xi}^2=({\bm \sigma}_1,{\bf 0},{\bf 0})\quad \quad {\bm \xi}^3=({\bm \sigma}_2,{\bf 0},{\bf 0})\\
    &{\bm \xi}^4=({\bf 0},{\bm \sigma}_0,{\bf 0})\quad\quad{\bm \xi}^5=({\bf 0},{\bm \sigma}_1,{\bf 0})\quad \quad {\bm \xi}^6=({\bf 0},{\bm \sigma}_2,{\bf 0})\\
    &{\bm \xi}^7=({\bf 0},{\bf 0},{\bm \sigma}_0)\quad\quad{\bm \xi}^8=({\bf 0},{\bf 0},{\bm \sigma}_1)\quad \quad {\bm \xi}^9=({\bf 0},{\bf 0},{\bm \sigma}_2).
\end{split}
\end{align}
In our notation, the vector ${\bm \xi}^1=({\bm \sigma}_0,{\bf 0},{\bf 0})$ is obtained concatenating the $N$-dimensional vector ${\bm \sigma}_0$ with two other $N$-dimensional vectors with zero entries, and similar for the others. The vectors in this family are linearly independent, but not orthogonal to each others. We therefore introduce a set of orthogonal $(3N)$-dimensional vectors ${\bm\chi}$:
\begin{align}
\begin{aligned}
 {\bm\chi}^{1} & = {\bm\xi}^1 \quad\quad & {\bm\chi}^4 & = {\bm\xi}^4 - q{\bm\xi}^5 \quad\quad & {\bm\chi}^7 & = {\bm\xi}^7 - q_0{\bm\xi}^9 \\
 {\bm\chi}^2 & = {\bm\xi}^2 - q{\bm\xi}^1 \quad\quad & {\bm\chi}^5 & = {\bm\xi}^5 \quad\quad & {\bm\chi}^8 & = {\bm\xi}^8 + c_1{\bm\xi}^7 + d_1{\bm\xi}^9 \\
 {\bm\chi}^3 & = {\bm\xi}^3 + c{\bm\xi}^1 + d{\bm\xi}^2 \quad\quad & {\bm\chi}^6 & = {\bm\xi}^6 + c{\bm\xi}^4 + d{\bm\xi}^5 \quad\quad & {\bm\chi}^9 & = {\bm\xi}^9,
\end{aligned}
\end{align}
with 
\begin{align}
\begin{aligned}
c&=\frac{q q_1-q_0 }{1 - q^2}\quad\quad d=\frac{qq_0 - q_1}{1 - q^2}\quad\quad c_1=\frac{ q_0 q_1-q}{1 - q_0^2}\quad\quad d_1=\frac{qq_0 -  q_1}{1 - q_0^2}.
\end{aligned}
\end{align}
\noindent The key trick to compute the action of $\hat C^{-1}$ on these vectors is to write the correlation matrix as $\hat{C}=p(\hat{D}+\hat{O})$,  where 
\begin{align}
    \begin{split}
        &D_{ij}^{ab}=\delta_{ab}[\delta_{ij}+(p-1)\sigma_{a,i}\sigma_{a,j}]\\
        &O_{ij}^{ab}=(1-\delta_{ab})[\delta_{ij}Q_{ab}^{p-1}+(p-1)Q_{ab}^{p-2}\sigma_{b,i}\sigma_{a,j}].
    \end{split}
\end{align}
This implies that $\hat{C}^{-1}=p^{-1}\hat{D}^{-1}[1+\hat{O}\hat{D}^{-1}]^{-1}$;
moreover, the Sherman-Morrison formula allows us to write
\begin{align}
    [\hat{D}^{-1}]^{ab}_{ij}=\delta_{ab}[\delta_{ij}-(p-1)p^{-1}\sigma_{a,i}\sigma_{a,j}].
\end{align}
\noindent Our goal is to determine what is the action of the matrices $\hat{D}^{-1}$ and $[1+\hat{O}\hat{D}^{-1}]$ upon the vectors ${\bm \chi}^i$, and to then invert the matrix $[1+\hat{O}\hat{D}^{-1}]$ restricted to this subspace. 
One might notice that, at variance with \cite{ros2019complex}, we are working here with vectors ${\bm \chi}^i$ that are orthogonal but not necessarily of unit norm, except for those vectors that only contain one configuration ${\bm \sigma}_a$ and, that are ${\bm\chi}^1,{\bm\chi}^5,{\bm\chi}^9$. This is due to the fact that the quadratic form \eqref{eq:effe} is a function of the matrix elements of $\hat C^{-1}$ only along these three directions: the normalization of the remaining ${\bm \chi}^i$ will not enter in the final result, and it can therefore be safely neglected, simplifying the  formalism.\\

\noindent To compute the action of $\hat{C}$ on the ${\bm\chi}$ vectors, it is convenient to first do it on the ${\bm\xi}$ vectors and then to make a change of basis. For convenience, we introduce the following $(3N)$-dimensional vector:
\begin{align}
    {\bf v}(n,\beta):=(0,\ldots,\underbrace{{\bm\sigma}_n}_{\beta},0,\ldots)\quad\quad n,\beta\in\{0,1,2\}
\end{align}
where we basically put the $n$-th configuration in the location $\beta$ (with $n,\beta$ starting at 0). In this way, each ${\bm\xi}$ vector is written as a ${\bf v}$ vector (notice that this can be generalized to an arbitrary number of configurations). Then, one gets
\begin{align}
    \hat{D}^{-1}{\bf v}(n,\beta)={\bf v}(n,\beta)-\frac{p-1}{p}Q_{n\beta}\,{\bf v}(\beta,\beta)
\end{align}
and similarly
\begin{align}
\hat{O}{\bf v}(n,\beta)=\sum_{\alpha\neq\beta}\left[Q_{\alpha\beta}^{p-1}{\bf v}(n,\alpha)+(p-1)Q_{\alpha\beta}^{p-2}Q_{\alpha n}\,{\bf v}(\beta,\alpha)\right].
\end{align}
To be explicit, let us express such actions in full extent. We begin with the action of $\hat{D}^{-1}$ on the ${\bm\xi}$ basis:
\begin{align}
[\hat{D}^{-1}]_{{\bm\xi}}=\begin{pmatrix}
p^{-1} & (p^{-1}-1) q & (p^{-1}-1) q_{0} & 0 & 0 & 0 & 0 & 0 & 0 \\
0 & 1 & 0 & 0 & 0 & 0 & 0 & 0 & 0 \\
0 & 0 & 1 & 0 & 0 & 0 & 0 & 0 & 0 \\
0 & 0 & 0 & 1 & 0 & 0 & 0 & 0 & 0 \\
0 & 0 & 0 & (p^{-1}-1) q & p^{-1} & (p^{-1}-1) q_{1} & 0 & 0 & 0 \\
0 & 0 & 0 & 0 & 0 & 1 & 0 & 0 & 0 \\
0 & 0 & 0 & 0 & 0 & 0 & 1 & 0 & 0 \\
0 & 0 & 0 & 0 & 0 & 0 & 0 & 1 & 0 \\
0 & 0 & 0 & 0 & 0 & 0 & (p^{-1}-1) q_{0} & (p^{-1}-1) q_{1} & p^{-1}
\end{pmatrix}
\end{align}

\noindent Similarly we can get the $\hat{O}$ action:

\[
[\hat{O}]_{{\bm\xi}}=
\left(
\begin{array}{ccccc}
0 & 0 & 0 & q^{p-1} & 0 \\
0 & 0 & 0 & (p-1) q^{p-2} & p q^{p-1}\\
0 & 0 & 0 & 0 & 0 \\
p q^{p-1} & (p-1) q^{p-2} & (p-1) q^{p-2} q_1 & 0 & 0\\
0 & q^{p-1} & 0 & 0 & 0 \\
0 & 0 & q^{p-1} & 0 & 0\\
p q_0^{p-1} & (p-1) q_0^{p-2} q_1 & (p-1) q_0^{p-2} q_1^{p-1} & 0\\
0 & q_0^{p-1} & 0 & (p-1) q_0 q_1^{p-2} & p q_1^{p-1}\\
0 & 0 & q_0^{p-1} & 0 & 0\\
\end{array}
\right.
\]
\[
\left.
\begin{array}{cccc}
0 & q_0^{p-1} & 0 & 0\\
(p-1) q^{p-2} q_0 & 0 & q_0^{p-1} & 0\\
q^{p-1} & (p-1) q_0^{p-2} & (p-1) q q_0^{p-2} & p q_0^{p-1}\\
0 & q_1^{p-1} & 0 & 0 \\
0 & 0 & q_1^{p-1} & 0 \\
0 & (p-1) q q_1^{p-2} & (p-1) q_1^{p-2} & p q_1^{p-1} \\
0 & 0 & 0 & 0 \\
(p-1) q_1^{p-2} & 0 & 0 & 0\\
q_1^{p-1} & 0 & 0 & 0
\end{array}
\right).
\]

\noindent Now, one proceeds by multiplying the above matrices to obtain $[\hat{O}\hat{D}^{-1}]_{{\bm\xi}}$, and then performing a change of basis from ${\bm\xi}$ to ${\bm\chi}$, with the matrix of basis change $U$:
\begin{align}
U_{\xi\to\chi}=\begin{pmatrix}
1 & -q & \frac{  q q_{1}-q_{0}}{1 - q^2} & 0 & 0 & 0 & 0 & 0 & 0 \\
0 & 1 & \frac{q q_{0} - q_{1}}{1 - q^2} & 0 & 0 & 0 & 0 & 0 & 0 \\
0 & 0 & 1 & 0 & 0 & 0 & 0 & 0 & 0 \\
0 & 0 & 0 & 1 & 0 & \frac{  q q_{1}-q_{0}}{1 - q^2} & 0 & 0 & 0 \\
0 & 0 & 0 & -q & 1 & \frac{q q_{0} - q_{1}}{1 - q^2} & 0 & 0 & 0 \\
0 & 0 & 0 & 0 & 0 & 1 & 0 & 0 & 0 \\
0 & 0 & 0 & 0 & 0 & 0 & 1 & \frac{ q_{0} q_{1}-q}{1 - q_{0}^2} & 0 \\
0 & 0 & 0 & 0 & 0 & 0 & 0 & 1 & 0 \\
0 & 0 & 0 & 0 & 0 & 0 & -q_{0} & \frac{q q_{0} - q_{1}}{1 - q_{0}^2} & 1 \\
\end{pmatrix}.
\end{align}
In such way, we have that 
\begin{align}
    [\hat{O}\hat{D}^{-1}]_{{\bm\chi}}=U_{\xi\to\chi}^{-1}[\hat{O}\hat{D}^{-1}]_{{\bm\xi}}U_{\xi\to\chi}
\end{align}
The only thing left to do is to invert now the following matrix:
\begin{align}
    \hat{M}=\mathds{1}_{9\times 9} + [\hat{O}\hat{D}^{-1}]_{{\bm\chi}}.
\end{align}
and then use that $[\hat{C}]_{{\bm\chi}}=p^{-1}\hat{D}^{-1}\hat{M}^{-1}$. Ultimately, we are interested in ${\bm\chi}^a\cdot [\hat{C}]_{\bm\chi}\cdot {\bm\chi}^b$ for $a,b\in\{1,5,9\}$. Since the action of $\hat{D}$ on such vectors is trivial (it just gives back a factor $p^{-1}$), we immediately see that 
\begin{align}
\begin{split}
&Y_{0}^{(p)}({\bf q})=[\hat{M}^{-1}]_{11}\\
&Y_{1}^{(p)}({\bf q})=[\hat{M}^{-1}]_{55}\\
&Y_{2}^{(p)}({\bf q})=[\hat{M}^{-1}]_{99}\\
&Y_{01}^{(p)}({\bf q})=[\hat{M}^{-1}]_{15}+[\hat{M}^{-1}]_{51}\\
&Y_{02}^{(p)}({\bf q})=[\hat{M}^{-1}]_{19}+[\hat{M}^{-1}]_{91}\\
&Y_{12}^{(p)}({\bf q})=[\hat{M}^{-1}]_{59}+[\hat{M}^{-1}]_{95}.\\
\end{split}
\end{align}
The expression of these matrix elements for generic values of $p$ is too long to be reported here. We report here the explicit expressions for the case $p=3$, which is the specif case considered in all the plots in this work. One finds:
\begin{equation}\label{eq:yp31}
\begin{split}
Y_0^{(3)}({\bf q})=&-((6 q^6 + 6 q_0^6 + 2 q_0^2 (-1 + q_1^2)^3 - (-1 + q_1^2)^3 (1 + q_1^2) - 
     12 q^5 q_0 q_1 (2 + q_1^2) + 3 q_0^4 (1 - 4 q_1^2 + 3 q_1^4)\\
     &- 
     4 q q_0 q_1 (-(-1 + q_1^2)^3 + 3 q_0^4 (2 + q_1^2) + 
        3 q_0^2 (-1 + q_1^4)) + 
     12 q^3 q_0 q_1 (1 - q_1^4\\
     &+ q_0^2 (1 - 12 q_1^2 + q_1^4)) + 
     3 q^4 (1 - 4 q_1^2 + 3 q_1^4 + q_0^2 (-4 + 28 q_1^2 + 6 q_1^4))+ 
     2 q^2 ((-1 + q_1^2)^3\\
     &+ q_0^4 (-6 + 42 q_1^2 + 9 q_1^4) + 
        q_0^2 (3 - 30 q_1^2 + 33 q_1^4 - 6 q_1^6)))/((-1 + q^2 + q_0^2 - 
       2 q q_0 q_1 + q_1^2)^2\\
       &(-1 + q^4 + q_0^4 + 8 q q_0 q_1 - 
       4 q_0^2 q_1^2 + q_1^4 + 2 q^2 (-2 q_1^2 + q_0^2 (-2 + q_1^2)))))
 \end{split}
 \end{equation}

 \begin{equation}
 \begin{split}
Y_1^{(3)}({\bf q})=&
(-6 q^6 + (-1 + q_0^2)^3 (1 + q_0^2) + 12 q^5 q_0 (2 + q_0^2) q_1 - 
   2 (-1 + q_0^2)^3 q_1^2 - 3 (1 - 4 q_0^2 + 3 q_0^4) q_1^4 - 6 q_1^6\\& -
   12 q^3 q_0 q_1 (1 - q_0^4 + (1 - 12 q_0^2 + q_0^4) q_1^2) + 
   4 q q_0 q_1 (-(-1 + q_0^2)^3 + 3 (-1 + q_0^4) q_1^2 + 
      3 (2 + q_0^2) q_1^4) \\
      &+ 
   2 q^2 (-(-1 + q_0^2)^3 + 
      3 (-1 + q_0) (1 + q_0) (1 - 9 q_0^2 + 2 q_0^4) q_1^2 - 
      3 (-2 + 14 q_0^2 + 3 q_0^4) q_1^4) \\&- 3 q^4 (1 - 4 q_1^2 + q_0^4 (3 + 6 q_1^2) + 
      4 q_0^2 (-1 + 7 q_1^2)))/((-1 + q^2 + q_0^2 - 2 q q_0 q_1\\& + 
     q_1^2)^2 (-1 + q^4 + q_0^4 + 8 q q_0 q_1 - 4 q_0^2 q_1^2 + q_1^4 + 
     2 q^2 (-2 q_1^2 + q_0^2 (-2 + q_1^2))))
      \end{split}
 \end{equation}

 \begin{equation}
 \begin{split}
Y_2^{(3)}({\bf q})=&(-1 + 2 q^2 - 2 q^6 + q^8 + 2 q_0^2 - 6 q^2 q_0^2 + 6 q^4 q_0^2 - 
   2 q^6 q_0^2 - 3 q_0^4 + 12 q^2 q_0^4 - 9 q^4 q_0^4 - 6 q_0^6\\& - 
   4 q q_0 ((-1 + q^2)^3 - 3 (-1 + q^4) q_0^2 - 3 (2 + q^2) q_0^4) q_1 + 
   2 (-(-1 + q^2)^3\\
   &+ 3 (-1 + q) (1 + q) (1 - 9 q^2 + 2 q^4) q_0^2 - 
      3 (-2 + 14 q^2 + 3 q^4) q_0^4) q_1^2\\& - 
   12 q q_0 (1 - q^4 + (1 - 12 q^2 + q^4) q_0^2) q_1^3 - 
   3 (1 - 4 q_0^2 + q^4 (3 + 6 q_0^2)\\& + 4 q^2 (-1 + 7 q_0^2)) q_1^4 + 
   12 q (2 + q^2) q_0 q_1^5 - 
   6 q_1^6)/((-1 + q^2 + q_0^2\\& - 2 q q_0 q_1 + q_1^2)^2 (-1 + q^4 + q_0^4 + 
     8 q q_0 q_1 - 4 q_0^2 q_1^2 + q_1^4 + 
     2 q^2 (-2 q_1^2 + q_0^2 (-2 + q_1^2))))
     \end{split}
     \end{equation}

     \begin{equation}\label{eq:yp32}
     \begin{split}
Y_{01}^{(3)}({\bf q})=&(2 (q - q_0 q_1)^2 (3 q^5 - 6 q^4 q_0 q_1 + 2 q_0 q_1 (1 + q_0^2 + q_1^2) - 
     2 q^2 q_0 q_1 (-2 + 3 q_0^2 + 3 q_1^2)\\& - 
     q (-1 + 3 q_0^4 + 2 q_1^2 + 3 q_1^4 + q_0^2 (2 - 4 q_1^2)) + 
     2 q^3 (2 - 3 q_1^2 + q_0^2 (-3 + 9 q_1^2))))/((-1 + q^2 + q_0^2\\& - 
     2 q q_0 q_1 + q_1^2)^2 (-1 + q^4 + q_0^4 + 8 q q_0 q_1 - 4 q_0^2 q_1^2 + 
     q_1^4 + 2 q^2 (-2 q_1^2 + q_0^2 (-2 + q_1^2))))    \end{split}
 \end{equation}

 \begin{equation}
 \begin{split}
Y_{12}^{(3)}({\bf q})=&-((2 (-q q_0 + 
       q_1)^2 (-2 q q_0 (1 + q^2 + q_0^2) + (-1 + 3 q^4 + 2 q_0^2 + 
          3 q_0^4 + q^2 (2 - 4 q_0^2)) q_1\\& + 
       2 q q_0 (-2 + 3 q^2 + 3 q_0^2) q_1^2 + 
       2 (-2 + 3 q_0^2 + q^2 (3 - 9 q_0^2)) q_1^3 + 6 q q_0 q_1^4 - 
       3 q_1^5))/((-1 + q^2 \\&+ q_0^2 - 2 q q_0 q_1 + q_1^2)^2 (-1 + q^4 + 
       q_0^4 + 8 q q_0 q_1 - 4 q_0^2 q_1^2 + q_1^4 + 
       2 q^2 (-2 q_1^2 + q_0^2 (-2 + q_1^2)))))
           \end{split}
 \end{equation}

 \begin{equation}
 \label{app:eq:last_Y}
 \begin{split}
Y_{02}^{(3)}({\bf q})=&(2 (q_0 - q q_1)^2 (3 q_0^5 - 6 q q_0^4 q_1 + 2 q q_1 (1 + q^2 + q_1^2) - 
     2 q q_0^2 q_1 (-2 + 3 q^2 + 3 q_1^2) - 
     q_0 (-1 + 3 q^4 + 2 q_1^2 + 3 q_1^4 \\&+ q^2 (2 - 4 q_1^2)) + 
     2 q_0^3 (2 - 3 q_1^2 + q^2 (-3 + 9 q_1^2))))/((-1 + q^2\\& + q_0^2 - 
     2 q q_0 q_1 + q_1^2)^2 (-1 + q^4 + q_0^4 + 8 q q_0 q_1 - 4 q_0^2 q_1^2 + 
     q_1^4 + 2 q^2 (-2 q_1^2 + q_0^2 (-2 + q_1^2)))).
\end{split}
\end{equation}
To complete the derivation of the joint distribution, it remains to compute the determinant of the correlation matrix. To do so we place ourselves in an orthonormal basis that respects the direct product $\mathbb{R}^N\simeq S\oplus S^\perp$.
Then, we can write $\hat{C}^{ab}=\text{diag}(\hat{A}^{ab},\hat{B}^{ab})$, with $\hat{A}$ containing the components of the gradient along directions belonging to $S^\perp$, and $\hat{B}$ containing the gradient components in $S$ (the mixed components are zero from the expression of $\hat{C}$). We can see that $\hat{A}^{ab}_{ij}=p\,\delta_{ij}\,Q_{ab}^{p-1}$. To leading order in $N$, we can neglect $\hat{B}$, which is only a $3\times 3$ matrix and it does not contribute exponentially in $N$ to the probability. For any $i$ we find:
\begin{align}
    \hat{A}^{ab}_{ii}=p
    \begin{pmatrix}
        1 & q^{p-1} & q_0^{p-1}\\
        q^{p-1} & 1 & q_1^{p-1}\\
        q_0^{p-1} & q_1^{p-1} & 1
    \end{pmatrix}
\end{align}
and a simple computation gives:
\begin{align}
    |\det\hat{C}|=p^{3N}\left[1-q^{2(p-1)}-q_0^{2(p-1)}-q_1^{2(p-1)}+2(q\, q_0 \,q_1)^{p-1}\right]^{N} e^{o(N)}.
\end{align}
As we see, this expression also implies a geometric constraint on the overlaps, i.e., that the above must be positive.
\noindent Combining these terms, we find
\begin{align}
    \begin{split}
   P_3({\bm \epsilon}, {\bf q})= e^{-N\, \grafe{F({\bm\epsilon},{\bf q})- \frac{3 }{2} \log (2 \pi p)- \frac{1}{2} \log [1-q^{2(p-1)}-q_0^{2(p-1)}-q_1^{2(p-1)}+2(q\, q_0 \,q_1)^{p-1}]}+ o(N)}. 
    \end{split}
\end{align}
The  denominator $P_2(\epsilon_0, \epsilon_1, q)$ in \eqref{eq:DistRatio} instead reads \cite{ros2019complexity}:
\begin{align}
  P_2(\epsilon_0, \epsilon_1, q)=e^{-N \grafe{\epsilon_0^2(1+U_0(q)) + \epsilon_0\epsilon_1 U(q)+\epsilon_1^2 U_1(q) +\log(2\pi p)+\frac{1}{2} \log(1-q^{2p-2}) }+ o(N)},
\end{align}
with the functions defined in Eq. \eqref{eq:UC} in the main text. 

\subsection{The conditional expectation of the Hessian's determinant}
\noindent It remains to determine the conditional expectation of the Hessian:
\begin{equation}\label{eq:detee}
  H(\epsilon_2, q_0, q_1| \epsilon_1, \epsilon_0, q)=\mathbb{E}\left[|\det\nabla^2_\perp h({\bm \sigma}_2)|\bigg|
    \grafe{
    \begin{subarray}{l}
    h^a = \sqrt{2N}\epsilon_a\\
   {\bf g}^a={\bf 0},\, \; \forall  a=0,1,2 \end{subarray}}
    \right].
\end{equation}
The statistics of the Hessian matrix $\nabla^2_\perp h({\bm \sigma}_2)$ conditioned to $ h^a = \sqrt{2N}\epsilon_a,
   {\bf g}^a={\bf 0}$ is discussed in detail in Appendix \ref{app:hessian_analysis}. We anticipate that the conditional Hessian is statistically equivalent to a random matrix extracted from a GOE ensemble with variance $\sigma^2=p(p-1)$, modified by some finite-rank perturbations. To leading order in $N$, the expected value \eqref{eq:detee} depends only on the  continuous part of the eigenvalue distribution of the Hessian, that is unaffected by the presence of the finite-rank perturbations and coincides with the density of a GOE matrix. Using this and following the same steps elucidated in \cite{ros2019complexity}, we find that 
   \begin{align}\label{eq:HessianAyntot}
     H(\epsilon_2, q_0, q_1| \epsilon_1, \epsilon_0, q)=e^{\frac{N}{2}\left[\log{N}+\log(2\,p(p-1))+2\,I\left(\epsilon_2\sqrt{\frac{p}{p-1}}\right)\right]+ o(N)},
\end{align}
with $I$ defined in Eq.~\eqref{eq:IDef}.

\section{The Hessian at the stationary points: distribution and spectral properties}\label{app:hessian_analysis}
\noindent In this Appendix, we study the statistical properties of the Hessian matrices at the stationary points ${\bm \sigma}_2$. In Sec. \ref{app:DistributionConditional}, we derive the distribution of the Hessian, conditioned on the gradients and energies of ${\bm\sigma}_a$ with $a=0,1,2$. In Sec. \ref{app:spectralP}, we analyse the spectral properties of these Hessians, in the annealed setting. We recall that the overlaps between the configurations are defined to be ${\bm\sigma}_0\cdot{\bm\sigma}_1=q$, ${\bm\sigma}_0\cdot{\bm\sigma}_2=q_0$ and ${\bm\sigma}_1\cdot{\bm\sigma}_2=q_1$. Let us define $M=N-\dim(S)=N-3$;
we recall that the Riemannian Hessian at a point ${\bm\sigma}_a$ is a $(N-1)\times (N-1)$ matrix whose components in a basis ${\bf e}_i^a={\bf e}_i[{\bm \sigma}_a]$ of the tangent plane $\tau[{\bm\sigma}_a]$ read:
\begin{align*}
\begin{split}
    \nabla^2_\perp h^a_{ij}&=[{\bf e}^a_i]^\top\left(\nabla^2h^a-(p h^a)\mathds{1} \right){\bf e}^a_j \equiv \mathcal{M}^a_{ij}-(p h^a)\delta_{ij} \quad \quad\text{ for }\quad 0\leq i,j\leq N-1.
    \end{split}
\end{align*}
We denote with  $\tilde{\mathcal{H}}^2$ the matrix following the conditional law of $\nabla^2_\perp h({\bm\sigma}_2)$ conditioned to ${\bf g}^a=0$ and $h^a=\sqrt{2N}\epsilon_a$ for $a=0,1,2$. We also define $\tilde{\mathcal{M}}^2$ to be the $(N-1)\times (N-1)$ matrices that follow the law of $\mathcal{M}^2$  with such conditioning. Since the conditioning imposes $p h({\bm\sigma}_2) \to p\sqrt{2N}\epsilon_2$, then
\begin{align}\label{eq:ConDHess}
\tilde{\mathcal{H}}^2_{ij}=\tilde{\mathcal{M}}^2_{ij}-p\epsilon_2\sqrt{2N} \,\delta_{ij}.
\end{align}
In Sec. \ref{app:DistributionConditional}, we compute the mean and covariances of the entries of $\tilde{\mathcal{M}}^2$. 

\subsection{The conditional law of the Hessian}\label{app:DistributionConditional}
\noindent The statistics of the entries of the unconditioned matrix $\mathcal{M}^2_{ij}$ can be read from \eqref{app:eq:HessTang},  setting $a=b$ and choosing vectors ${\bf u}$ belonging to the tangent plane $\tau[{\bm\sigma}_2]$. Since we are working with Gaussian variables, we need to take such variables and condition them on the (Gaussian) laws of the energies and gradients. We discuss this in the following subsection.

\subsubsection{ Gaussian conditioning: the general structure}
\noindent We begin by recalling the formula for Gaussian conditioning.
Given a multivariate normal vector distributed as $Y\sim\mathcal{N}({\bm \mu}, \Sigma)$, and given the  partition:
\begin{align*}
    \begin{split}
        &Y=\begin{pmatrix}
            Y_1\\
            Y_2
        \end{pmatrix}\quad\quad{\bm\mu}=\begin{pmatrix}
            {\bm\mu}_1\\
            {\bm\mu}_2
        \end{pmatrix}\quad\quad\Sigma=\begin{pmatrix}
            \Sigma_{11} & \Sigma_{12}\\
            \Sigma_{21} & \Sigma_{22}
        \end{pmatrix},
    \end{split}
\end{align*}
then the conditional law of $(Y_1|Y_2={\bf a})$ is a multivariate normal $\mathcal{N}(\tilde{{\bm\mu}},\tilde{\Sigma})$ with parameters 
\begin{align}
\begin{split}
&\tilde{{\bm\mu}}={\bm\mu}_1+\Sigma_{12}\Sigma_{22}^{-1}({\bf a}-{\bm\mu}_2)\\
&\tilde{\Sigma}=\Sigma_{11}-\Sigma_{12}\Sigma_{22}^{-1}\Sigma_{21}.
\end{split}
\end{align}

\noindent To apply this formalism to the Hessian, it is convenient to group all the independent components of $\mathcal{M}^2$ into a $N(N+1)/2$ dimensional vector ${\bf M}=({\bf M}_0, {\bf M}_{1/2},{\bf M}_1)^\top$, where  (recall that $M=N-3$)
\begin{align}
\begin{split}
    &{\bf M}_0=(\{\mathcal{M}^2_{ij}\}_{1\leq i\leq j\leq M})^\top\\
    &{\bf M}_{1/2}=(\{\mathcal{M}^2_{ij}\}_{1\leq i\leq M, \, M+1\leq j\leq N-1})^\top\\
    &{\bf M}_1=(\{\mathcal{M}^2_{ij}\}_{M+1\leq i\leq j\leq N-1})^\top,\\
\end{split}
\end{align}
i.e. ${\bf M}_0$ contains elements with both indices in $S^\perp$,  ${\bf M}_1$ with both indices in $S$, and ${\bf M}_{1/2}$ mixed indices, all of which with respect to basis vectors spanning $\tau[{\bm\sigma^2}]$. We also define the $3N$-dimensional Gaussian vector upon which we condition, namely $\tilde{{\bf g}}=(\tilde{{\bf g}}_0,\tilde{{\bf g}}_1)$ where the $0$ component has indices in $S^\perp$ and the 1 component has indices in $S$, in such a way that $\tilde{{\bf g}}_\gamma=(\tilde{{\bf g}}_\gamma^0,\tilde{{\bf g}}_\gamma^{1},\tilde{{\bf g}}_\gamma^2)^\top$ for $\gamma\in\{0,1\}$ and
\begin{align}
\begin{split}
    &\tilde{{\bf g}}_0^a=(g_1^a,\ldots,g_M^a)^\top\quad\quad\quad\text{for }a\in\{0,1,2\}\\
    &\tilde{{\bf g}}_1^a=(g_{M+1}^a,\ldots,g_N^a)^\top\\
    &g_i^a=\nabla h^a\cdot{\bf e}_i({\bm\sigma}_a)\quad\quad\quad\,\,\, 1\leq i\leq N-1\\
    &g_N^a=\nabla h^a\cdot{\bm\sigma}_a=p\,h^a\quad\quad\text{to include the energy as part of the gradient}
\end{split}
\end{align}
Notice that ${\bf M}$ is $(N-1) \times (N-1)$-dimensional and it is expressed in the tangent plane basis $\tau[{\bm\sigma}_2]$, whereas $\tilde{{\bf g}}$ is $N$-dimensional and it is expressed in its own local basis $\mathcal{B}[{\bm\sigma}_a]$. Using Appendix~\ref{app:grad_correls} we see that the covariances between elements of $\bf M$ are equal to zero whenever the two chosen elements $\mathcal{M}^2_{ij},\mathcal{M}^2_{kl}$ don't belong to the same group (i.e. either 0,1/2 or 1). Hence, this means that the covariance matrix of the vector $\bf M$ is block diagonal, and takes value 0 outside these three blocks:
\begin{align}
    \hat{\Sigma}_{{\bf M}{\bf M}}=
    \begin{pmatrix}
        \hat{\Sigma}_{{\bf M}{\bf M}}^0 & 0 & 0\\
        0 & \hat{\Sigma}_{{\bf M}{\bf M}}^{1/2} & 0\\
        0 & 0 & \hat{\Sigma}_{{\bf M}{\bf M}}^1
    \end{pmatrix}.
\end{align}
Similarly we see that for the covariances of elements belonging to $\tilde{{\bf g}}$, only those belonging to the same group (i.e. 0 or 1) are non-zero:
\begin{align}
    \hat{\Sigma}_{\tilde{\bf g}\tilde{\bf g}}=
    \begin{pmatrix}
        \hat{\Sigma}_{\tilde{\bf g}\tilde{\bf g}}^0 & 0\\
        0 & \hat{\Sigma}_{\tilde{\bf g}\tilde{\bf g}}^1
    \end{pmatrix}.
\end{align}
Moreover we see that $\bf M$ and $\tilde{\bf g}$ are correlated only when we take elements of $\bf M$ belonging to the group $1/2$ and elements of $\tilde{\bf g}$ belonging to the group $0$, or when both elements are taken from group $1$. Hence:
\begin{align}
\label{app:eq:sigma_conditional}
    \hat{\Sigma}_{{\bf M}\tilde{{\bf g}}}=
    \begin{pmatrix}
        0 & 0\\
        \hat{\Sigma}_{{\bf M}\tilde{{\bf g}}}^{\frac{1}{2}0} & 0\\
        0 & \hat{\Sigma}_{{\bf M}\tilde{{\bf g}}}^{11}
    \end{pmatrix}.
\end{align}
If now we use the formula for the conditioning of Gaussian vectors, we obtain
\begin{align}\label{eq:CondCov}
\begin{split}
    \hat{\Sigma}_{{\bf M}|\tilde{\bf g}}&=
    \hat{\Sigma}_{{\bf M}{\bf M}}-\hat{\Sigma}_{{\bf M}\tilde{{\bf g}}}\hat{\Sigma}_{\tilde{\bf g}\tilde{\bf g}}^{-1}\Sigma_{{\bf M}\tilde{{\bf g}}}^\top\\
    &=
    \begin{pmatrix}
        \hat{\Sigma}_{{\bf M}{\bf M}}^0 & 0 & 0\\
        0 & \hat{\Sigma}_{{\bf M}{\bf M}}^{1/2}-\hat{\Sigma}_{{\bf M}\tilde{{\bf g}}}^{\frac{1}{2}0}\left(\hat{\Sigma}_{\tilde{\bf g}\tilde{\bf g}}^{0}\right)^{-1}\left(\hat{\Sigma}_{{\bf M}\tilde{{\bf g}}}^{\frac{1}{2}0}\right)^\top   & 0\\
        0 & 0 & \hat{\Sigma}_{{\bf M}{\bf M}}^1 -\hat{\Sigma}_{{\bf M}\tilde{{\bf g}}}^{11}\left(\hat{\Sigma}_{\tilde{\bf g}\tilde{\bf g}}^1\right)^{-1}\left(\hat{\Sigma}_{{\bf M}\tilde{{\bf g}}}^{11}\right)^\top
    \end{pmatrix}.
\end{split}
\end{align}
Hence, we see that the new conditioned variables conserve the block structure, where correlations only appear between elements of the same group (0 ,1 or 1/2).  \\
\noindent We now turn our attention to the averages. We have that 
\begin{align*}
    {\bm\mu}_{{\bf M}|\tilde{{\bf g}}}={\bm\mu}_{{\bf M}}+\hat{\Sigma}_{{\bf M}\tilde{{\bf g}}}\hat{\Sigma}_{\tilde{{\bf g}}\tilde{{\bf g}}}^{-1}(\tilde{\bf g}|_c-{\bm\mu}_{\tilde{\bf g}})
\end{align*}
with $\tilde{\bf g}|_c$ the conditional vector, which is zero everywhere expect at the energies of the configurations. Moreover, since all variables here are centered Gaussian, we have that their (unconditional) means are zero, and hence we can write
\begin{equation}\label{eq:CondAverages}
\begin{split}
{\bm\mu}_{{\bf M}|\tilde{{\bf g}}}&=\hat{\Sigma}_{{\bf M}\tilde{{\bf g}}}\hat{\Sigma}_{\tilde{{\bf g}}\tilde{{\bf g}}}^{-1}\tilde{\bf g}|_c=
\begin{pmatrix}
0 & 0\\
\hat{\Sigma}_{{\bf M}\tilde{{\bf g}}}^{\frac{1}{2}0} & 0\\
0 & \hat{\Sigma}_{{\bf M}\tilde{{\bf g}}}^{11}
\end{pmatrix}
\begin{pmatrix}
\left(\hat{\Sigma}_{\tilde{\bf g}\tilde{\bf g}}^0\right)^{-1} & 0\\
0 & \left(\hat{\Sigma}_{\tilde{\bf g}\tilde{\bf g}}^1\right)^{-1}
\end{pmatrix}
\begin{pmatrix}
{\bf 0}\\
\tilde{{\bf g}}_1|_c
\end{pmatrix}
=\begin{pmatrix}
    {\bf 0}\\
    {\bf 0}\\
    \hat{\Sigma}^{11}_{{\bf M}\tilde{\bf g}}\left(\hat{\Sigma}_{\tilde{\bf g}\tilde{\bf g}}^1\right)^{-1}\tilde{{\bf g}}_1|_c
\end{pmatrix}.
\end{split}
\end{equation}
As we can see, only the matrix elements with both components in $S$ has non-zero averages. \\
\noindent In the following, we compute an explicit expression of the covariance matrices appearing in these formulas. It will be convenient to make an explicit, configuration-dependent choice of the basis in the three dimensional subspace $S$, which we report below. 

\subsubsection{ Explicit parametrization of the basis vectors }\label{app:sec:basis}
\noindent In order to obtain closed formulas from the above expressions, we need to choose a specific basis for the $S$ components of $\mathcal{B}[{\bm\sigma}_a]$ for $a=0,1,2$. The following are acceptable choices, where the coefficients are found by imposing that the vectors in a specific local basis are orthonormal:
\begin{align}
\begin{split}
&{\bf e}_{N}^2={\bm\sigma}_2\\
&{\bf e}_{N-1}^2=\frac{1}{\sqrt{1-q_0^2}}\left(q_0{\bm\sigma}_2-{\bm\sigma}_0\right)\\
&{\bf e}_{N-2}^2=\frac{1}{\sqrt{1 - q_0^2} \sqrt{1 - q^2 - q_0^2 + 2 q q_0 q_1 - q_1^2}}\left[(q_1-q q_0){\bm\sigma}_2+(q - q_0 q_1) {\bm\sigma}_0 -(1-q_0^2){\bm\sigma}_1 \right]\\
\end{split}
\end{align}

\begin{align}
\begin{split}
&{\bf e}_{N}^1={\bm\sigma}_1\\
&{\bf e}_{N-1}^1=\frac{1}{\sqrt{1-q^2}}\left(q{\bm\sigma}_1-{\bm\sigma}_0\right)\\
&{\bf e}_{N-2}^1=\frac{1}{\sqrt{1 - q^2} \sqrt{1 - q^2 - q_0^2 + 2 q q_0 q_1 - q_1^2}}\left[(q_1-q q_0){\bm\sigma}_1+(q_0 - q q_1) {\bm\sigma}_0 -(1-q^2){\bm\sigma}_2 \right]\\
\end{split}
\end{align}

\begin{align}
\begin{split}
&{\bf e}_{N}^0={\bm\sigma}_0\\
&{\bf e}_{N-1}^0=\frac{1}{\sqrt{1-q_0^2}}\left(q_0{\bm\sigma}_0-{\bm\sigma}_2\right)\\
&{\bf e}_{N-2}^0=\frac{1}{\sqrt{1 - q_0^2} \sqrt{1 - q^2 - q_0^2 + 2 q q_0 q_1 - q_1^2}}\left[(q-q_1 q_0){\bm\sigma}_0+(q_1 - q_0 q) {\bm\sigma}_2 - (1-q_0^2){\bm\sigma}_1 \right]\\
\end{split}
\end{align}

\noindent We now determine the statistics of the conditional Hessian components, with respect to this specific choice of basis vectors.

\subsubsection{Conditional law: the covariances}
\noindent From \eqref{eq:CondCov} it appears that the covariances between elements in the first block of the Hessian, labeled by $0$, are left untouched by the conditioning. Hence, from Eq.~\eqref{app:eq:HessTang} one sees that the components $\mathcal{M}^2_{1\leq i\leq j\leq M}$ have the following covariance matrix
\begin{align}
    \left(\hat{\Sigma}_{{\bf M}{\bf M}}^0\right)_{ij,kl}=\mathbb{E}\left[  \mathcal{M}^2_{ij}\mathcal{M}^2_{kl}\right]=p(p-1)(\delta_{ik}\delta_{jl}+\delta_{il}\delta_{jk})
\end{align}
which is precisely the structure of the correlations of a GOE random matrix with $\sigma^2=p(p-1)$. Hence, the bulk of such Hessians is a $M\times M$ matrix with law GOE$(\sigma^2=p(p-1))$. This information alone is sufficient to determine the expression of the determinant to leading order in $N$, given in \eqref{eq:HessianAyntot}. On the other hand, in order to characterize the fine structure of the eigenvalues distribution and to discuss the stability of the stationary points which we are counting with the complexity, we have to find the conditional distribution of the components in the $1/2$ group (the components in the group labelled by $1$ are only a few, and their variance will not impact the spectral properties of the Hessian to the order in $N$ that we are interested in). 
We aim therefore at determining $\hat{\Sigma}_{{\bf M}|\tilde{\bf g}}^{1/2}$.
At first, from Eq.~\eqref{app:eq:HessTang} it follows:
\begin{align}
   \left(\hat{\Sigma}_{{\bf M}{\bf M}}^{1/2}\right)_{ij,kl}=\mathbb{E}\left[ \mathcal{M}^2_{ij}\mathcal{M}^2_{kl}\right]=\delta_{ik}\delta_{jl}\, p(p-1)\quad\quad 1\leq i,k\leq M,\quad M+1\leq j,l\leq N-1.
\end{align}
Then, we have to study $\hat{\Sigma}_{{\bf M}\tilde{{\bf g}}}^{\frac{1}{2}0}$, which using again Eq.~\eqref{app:eq:CorelationsHessianGrad}, gives (here $1\leq i,k\leq M$  and $M+1\leq j\leq N-1$ ):
\begin{align}
    \left(\hat{\Sigma}_{{\bf M}\tilde{{\bf g}}}^{\frac{1}{2}0}\right)^{a}_{ij,k}=\mathbb{E}\left[ \mathcal{M}_{ij}^2g_k^a\right]=\delta_{ik}p(p-1)Q_{2 a}^{p-2}({\bf e}_j^2\cdot {\bm\sigma}_a).
\end{align}
And finally we find the expression for $\left(\hat{\Sigma}_{\tilde{\bf g}\tilde{\bf g}}^0\right)^{-1}$ using Eq.~\eqref{app:eq:CovGradComp} (here $1\leq i,j\leq M$):
\begin{align}
    &\left(\hat{\Sigma}_{\tilde{\bf g}\tilde{\bf g}}^0\right)^{ab}_{i,j}
    =pQ_{ab}^{p-1}\delta_{ij}
    \Rightarrow \hat{\Sigma}_{\tilde{\bf g}\tilde{\bf g}}^0=
    p
\begin{pmatrix}
\mathds{1} & q^{p-1}\mathds{1} & q_0^{p-1}\mathds{1}\\
q^{p-1}\mathds{1} & \mathds{1} & q_1^{p-1}\mathds{1}\\
q_0^{p-1}\mathds{1} & q_1^{p-1}\mathds{1} & \mathds{1}
\end{pmatrix}
\end{align}
where we have $9$ blocks, each of which has dimension $M\times M$. In order to invert this, we make an ansatz and prove that it gives the right inverse by solving for the parameters, and showing that they are unique:
\begin{align}\label{eq:Pmat}
\left(\hat{\Sigma}_{\tilde{\bf g}\tilde{\bf g}}^0\right)^{-1}=\frac{1}{p}
\begin{pmatrix}
\tilde{a}\mathds{1} &  \tilde{b}\mathds{1} &  \tilde{c}\mathds{1} \\
\tilde{b}\mathds{1} &  \tilde{e}\mathds{1} &  \tilde{d}\mathds{1}\\
\tilde{c}\mathds{1} &  \tilde{d}\mathds{1} &  \tilde{f}\mathds{1}\\
\end{pmatrix}, \quad \quad \quad \bf P=
\begin{pmatrix}
\tilde{a} &  \tilde{b} &  \tilde{c} \\
\tilde{b} &  \tilde{e} &  \tilde{d}\\
\tilde{c} &  \tilde{d} &  \tilde{f}\\
\end{pmatrix}.
\end{align}
This inversion gives 6 equations, that solved give the unique solution:
\begin{align*}
&\tilde{a}({\bf q})=\frac{1 - q_1^{-2 + 2 p}}{1 - q^{-2 + 2 p} - q_0^{-2 + 2 p}- q_1^{-2 + 2 p} +2 (q q_0q_1)^{p-1} }\\
&\tilde{b}({\bf q})=\frac{-q^{p-1} + (q_0 q_1)^{p-1}}{1 - q^{-2 + 2 p} - q_0^{-2 + 2 p}- q_1^{-2 + 2 p} +2 (q q_0q_1)^{p-1} }\\
&\tilde{c}({\bf q})=\frac{-q_0^{p-1} + (q q_1)^{p-1}}{1 - q^{-2 + 2 p} - q_0^{-2 + 2 p}- q_1^{-2 + 2 p} +2 (q q_0q_1)^{p-1} }\\
\end{align*}
\begin{align*}
&\tilde{d}({\bf q})=\frac{-q_1^{p-1} + (q_0 q)^{p-1}}{1 - q^{-2 + 2 p} - q_0^{-2 + 2 p}- q_1^{-2 + 2 p} +2 (q q_0q_1)^{p-1} }\\
&\tilde{e}({\bf q})=\frac{1 - q_0^{-2 + 2 p}}{1 - q^{-2 + 2 p} - q_0^{-2 + 2 p}- q_1^{-2 + 2 p} +2 (q q_0q_1)^{p-1} }\\
&\tilde{f}({\bf q})=\frac{1 - q^{-2 + 2 p}}{1 - q^{-2 + 2 p} - q_0^{-2 + 2 p}- q_1^{-2 + 2 p} +2 (q q_0q_1)^{p-1} }.\\
\end{align*}
We defined ${\bf P}$ to be the $3\times 3$ matrix with these elements, see Eq.~\eqref{eq:Pmat}. Now, we need to carry out the final product to obtain the relevant conditional covariance:
\begin{align}
     \hat{\Sigma}^{1/2}_{{\bf M}|\tilde{\bf g}}= \hat{\Sigma}_{{\bf M}{\bf M}}^{1/2}-\hat{\Sigma}_{{\bf M}\tilde{{\bf g}}}^{\frac{1}{2}0}\left(\hat{\Sigma}_{\tilde{\bf g}\tilde{\bf g}}^{0}\right)^{-1}\left(\hat{\Sigma}_{{\bf M}\tilde{{\bf g}}}^{\frac{1}{2}0}\right)^\top 
\end{align}
which, componentwise, reads
\begin{align}
    \left(\hat{\Sigma}^{1/2}_{{\bf M}|\tilde{\bf g}} \right)_{ik,jl}^{22}=\left(\hat{\Sigma}^{1/2}_{{\bf M}{\bf M}}\right)_{ik,jl}^{2d}-\sum_{a,b=0}^{2}\sum_{\gamma,\delta=1}^{M}\left(\hat{\Sigma}^{\frac{1}{2}0}_{{\bf M}\tilde{\bf g}}\right)_{ik,\gamma}^{ca}\left[\left(\hat{\Sigma}_{\tilde{\bf g}\tilde{\bf g}}^{0}  \right)^{-1}\right]^{ab}_{\gamma,\delta}\left(\hat{\Sigma}^{\frac{1}{2}0}_{{\bf M}\tilde{\bf g}}\right)^{b2}_{jl,\delta}.
\end{align}
The second term is
\begin{align*}
&\sum_{a,b=0}^{2}\sum_{\gamma,\delta=1}^{M}\left(\hat{\Sigma}^{\frac{1}{2}0}_{{\bf M}\tilde{\bf g}}\right)_{ik,\gamma}^{2a}\left[\left(\hat{\Sigma}_{\tilde{\bf g}\tilde{\bf g}}^{0}  \right)^{-1}\right]^{ab}_{\gamma,\delta}\left(\hat{\Sigma}^{\frac{1}{2}0}_{{\bf M}\tilde{\bf g}}\right)^{b2}_{jl,\delta}\\
&=\sum_{a,b=0}^{2}\sum_{\gamma,\delta=1}^{M}
\delta_{i\gamma}p(p-1)Q_{2 a}^{p-2}({\bf e}_k^2\cdot {\bm\sigma}_a)
\delta_{j\delta}p(p-1)Q_{2 b}^{p-2}({\bf e}_l^2\cdot {\bm\sigma}_b)\frac{1}{p}{\bf P}^{ab}\delta_{\gamma\delta}\\
&=\delta_{ij}p(p-1)^2\sum_{a,b=0}^{2}
Q_{2 a}^{p-2}({\bf e}_k^2\cdot {\bm\sigma}_a)
Q_{2 b}^{p-2}({\bf e}_l^2\cdot {\bm\sigma}_b){\bf P}^{ab}\\
&=\delta_{ij}p(p-1)^2\Bigg\{
q_0^{2p-4}({\bf e}_k^2\cdot {\bm\sigma}_0)({\bf e}_l^2\cdot {\bm\sigma}_0)\tilde{a}({\bf q}) + (q_1q_0)^{p-2}\left[({\bf e}_k^2\cdot {\bm\sigma}_0)({\bf e}_l^2\cdot {\bm\sigma}_1)+({\bf e}_k^2\cdot {\bm\sigma}_1)({\bf e}_l^2\cdot {\bm\sigma}_0)\right]\tilde{b}({\bf q})\\
&+q_1^{2p-4}({\bf e}_k^2\cdot {\bm\sigma}_1)({\bf e}_l^2\cdot {\bm\sigma}_1)\tilde{e}({\bf q})
\Bigg\}.
\end{align*}
Hence,  
\begin{align}
\begin{split}
\left(\hat{\Sigma}^{1/2}_{{\bf M}|\tilde{\bf g}} \right)_{ik,jl}&=
\delta_{ij}\delta_{kl}p(p-1)-\delta_{ij}p(p-1)^2\Bigg\{
q_0^{2p-4}({\bf e}_k^2\cdot {\bm\sigma}_0)({\bf e}_l^2\cdot {\bm\sigma}_0)\tilde{a}({\bf q})\\
&+ (q_1q_0)^{p-2}\left[({\bf e}_k^2\cdot {\bm\sigma}_0)({\bf e}_l^2\cdot {\bm\sigma}_1)+({\bf e}_k^2\cdot {\bm\sigma}_1)({\bf e}_l^2\cdot {\bm\sigma}_0)\right]\tilde{b}({\bf q})
+q_1^{2p-4}({\bf e}_k^2\cdot {\bm\sigma}_1)({\bf e}_l^2\cdot {\bm\sigma}_1)\tilde{e}({\bf q})
\Bigg\}.
\end{split}
\end{align}

\noindent Now, we want to express this  quantities in the basis introduced in Sec. \ref{app:sec:basis}. 
The covariances are identical for any $i$. We define.
\begin{align}
\begin{split}
&\Delta_1({\bf q})\equiv \left(\hat{\Sigma}^{1/2}_{{\bf M}|\tilde{\bf g}} \right)_{i(N-2),i(N-2)}\\
&\Delta_0 ({\bf q})\equiv \left(\hat{\Sigma}^{1/2}_{{\bf M}|\tilde{\bf g}} \right)_{i(N-1),i(N-1)}\\
&\Delta_{01}({\bf q}) \equiv \left(\hat{\Sigma}^{1/2}_{{\bf M}|\tilde{\bf g}} \right)_{i(N-2),i(N-1)}.
\end{split}
\end{align}
Using the basis vectors described above, we get 
\begin{equation}\label{eq:Deltas}
\begin{split}
    &\Delta_1({\bf q})= p(p-1) \quadre{1-\frac{(p-1) q_0^2(1-q_0^{2 p-2}) q_1^{2 p-2} \left(1-q^2-q_0^2-q_1^2+2
   q q_0 q_1\right)}{(1-q_0^2)
   \left(2 q^{p-1} q_0^{p+1} q_1^{p+1}+   q_0^{2} q_1^{2}-q_0^2 q_1^2 q^{2 p-2}- q_1^2 q_0^{2 p}-
  q_0^2 q_1^{2
   p}\right)}}\\
& \Delta_0({\bf q})= p(p-1) \quadre{1-\frac{(p-1)A_0({\bf q})}{\left(1-q_0^2\right) \left(1-q^{2 p-2}-q_0^{2
   p-2}-q_1^{2 p-2}+2 (q q_0 q_1)^{p-1}\right)}}\\
  & A_0({\bf q})= q_0^{2 p-4} (1-q_1^{2 p-2})(1-q_0^2)^2+(1-q_0^{2
   p-2}) q_1^{2 p-4} (q-q_0 q_1)^2   -2 (1-q_0^2) (q_0 q_1)^{p-2} (q-q_0
   q_1)[q^{p-1}-(q_0q_1)^{p-1}]\\
   &\Delta_{01}({\bf q})= p(p-1)^2 \frac{ \sqrt{1-q^2-q_0^2-q_1^2+2 q q_0 q_1}  [ 
   q(1-q_0^2)  (q q_0 q_1)^p+q^2 q_1^{2 p-2} 
   \left(q_0^{2 p-1} (q q_0-q_1)+q_0^2 (q_0
   q_1-q)\right)]}{   \left(1-q_0^2\right)\left(2 q^{p+1} q_0^{p+1} q_1^{p+1}+  q^{2} q_0^{2} q_1^{2}-q_0^2 q_1^2 q^{2 p}-q^2 q_1^2 q_0^{2 p}-q^2
  q_0^2 q_1^{2
   p}\right)}.
   \end{split}
\end{equation}
Let us discuss some limits of these expressions.  When $q\to 0, q_1\to 0$, the configuration ${\bm\sigma}_1$ decouples to the others, and ${\bm\sigma}_2$ is conditioned only to its overlap $q_0$ with ${\bm\sigma}_0$. The bass vectors in $\tau[{\bm \sigma}_2]$ reduce to ${\bf e}^2_{N-1} \propto {\bm \sigma}_0$ and ${\bf e}^2_{N-2} \propto {\bm \sigma}_1$, see Sec. \ref{app:sec:basis}. Due to the absence of conditioning to ${\bm\sigma}_1$, the components corresponding to the direction ${\bf e}^2_{N-2}$ become statistically equivalent to all other components of the $(N-3) \times (N-3)$ GOE block of the Hessian; the expressions \eqref{eq:Deltas} reduce indeed to 
\begin{align}
    &\Delta_0({\bf q})\stackrel{q,q_1 \to 0}{\longrightarrow}(p-1) p \left(1 - \frac{(p-1) (1-q_0^2)q_0^{2p-4}}{1 - q_0^{2p - 2}}\right)\\
    &\Delta_{01}({\bf q})\stackrel{q,q_1 \to 0}{\longrightarrow}=0\\
    &  \Delta_{1}({\bf q})\stackrel{q,q_1 \to 0}{\longrightarrow}=p(p-1)
\end{align}
and one can check that $\Delta_0({\bf q})$ reproduces the function obtained in the calculation of the two-point complexity \cite{ros2019complexity}, as expected. The analogous holds true when $q, q_0 \to 0$, with $q_1$ replacing $q_0$.

\subsubsection{Conditional law: the averages}\label{app:averages}
\noindent We now come to the calculation of the averages \eqref{eq:CondAverages} induced by the conditioning. The only components that acquire a non-zero average are those corresponding to the entries in the group labelled by $1$, 
\begin{align}
\label{app:eq:mean_vector}
\begin{split}
    \underbrace{{\bm \mu}^1_{{\bf M}|\tilde{{\bf g}}}}_{3\times 1}=\underbrace{\hat{\Sigma}^{11}_{{\bf M}\tilde{{\bf g}}}}_{3\times 9}\underbrace{\left(\hat{\Sigma}_{\tilde{{\bf g}}\tilde{{\bf g}}}^1\right)^{-1}}_{9\times 9}\underbrace{\tilde{{\bf g}}_1|_c}_{9\times 1}\\
\end{split}
\end{align}
where 
\begin{align*}
    &\tilde{\bf g}_1|_c=(\tilde{{\bf g}}_1^0,\tilde{{\bf g}}_1^1,\tilde{{\bf g}}_1^{2})^\top\\
    &\tilde{{\bf g}}_1^a|_c=\underbrace{(0,0,p\sqrt{2N}\epsilon_a)^\top}_{3}
\end{align*}
and where all components are expressed in the local bases of the corresponding tangent plane $\tau[{\bm\sigma}_a]$ of each configuration. Let us rewrite Eq.~\eqref{app:eq:mean_vector} as:
\begin{align}
\label{app:eq:means}
    \frac{{\bm \mu}^1_{{\bf M}|\tilde{{\bf g}}}}{\sqrt{N}} &= \hat{\Sigma}^{11}_{{\bf M}\tilde{{\bf g}}}\left(\hat{\Sigma}_{\tilde{{\bf g}}\tilde{{\bf g}}}^1\right)^{-1}\sqrt{2}p\left(\epsilon_0{\bm \zeta}_0+\epsilon_1{\bm\zeta}_1+\epsilon_2{\bm\zeta}_2 \right)=:\begin{pmatrix}
    \mu_1\\
    \mu_{01}\\
    \mu_0
\end{pmatrix}
\end{align}
where
\begin{align}
\begin{split}
    &{\bm\zeta}_0=(0,0,1,0,0,0,0,0,0)^\top\\
    &{\bm\zeta}_1=(0,0,0,0,0,1,0,0,0)^\top\\
    &{\bm\zeta}_2=(0,0,0,0,0,0,0,0,1)^\top.\\
\end{split}
\end{align}

Now, we have that from Eq.~\eqref{app:eq:CorelationsHessianGrad}  ($N-2\leq i\leq j\leq N-1$ and $N-2\leq k\leq N$):
\begin{align}
\begin{split}
    \left(\hat{\Sigma}_{{\bf M}\tilde{\bf g}}^{11}\right)^{2a}_{ij,k}&=p(p-1)(p-2)Q_{2a}^{p-3}({\bf e}_i^2\cdot{\bm\sigma}_a)({\bf e}_j^2\cdot{\bm\sigma}_a)({\bf e}_k^a\cdot{\bm\sigma}_2)\\
    &+p(p-1)Q_{2 a}^{p-2}\left[({\bf e}_i^2\cdot{\bf e}_k^a)({\bf e}_j^2\cdot{\bm\sigma}_a)+({\bf e}_j^2\cdot{\bf e}_k^a)({\bf e}_i^2\cdot{\bm\sigma}_a) \right].
\end{split}
\end{align}
We have moreover from Eq.\eqref{app:eq:CovGradComp} ($N-2\leq i,j\leq N$ and $0\leq a,b\leq 2$):
\begin{align}
    \begin{split}
    \left(\hat{\Sigma}^1_{\tilde{\bf g}\tilde{\bf g}}\right)^{ab}_{i,j}=pQ_{ab}^{p-1}({\bf e}_i^a\cdot{\bf e}_j^b)+p(p-1)Q_{ab}^{p-2}({\bm\sigma}_a\cdot{\bf e}_j^b)
        ({\bm\sigma}_b\cdot{\bf e}_i^a).
    \end{split}
\end{align}
The fastest way to perform this computation is to write the explicit form of the matrices involved, using the specific choice of basis described above.  In such basis, we get:

\[
\hat{\Sigma}^1_{{\bf \tilde g}{\bf\tilde g}}=\left(
\begin{array}{ccccc}
p &  0 &  0 &  \frac{p q^{p - 1} (q q_0 - q_1)}{\beta} \\
  0 & p & 0 & p q^{p - 1}\frac{\alpha}{\beta} \\
  0 & 0 & p^2 &  0  \\
  \frac{p q^{p-1} (q q_0 - q_1)}{\beta} & p q^{p - 1}\frac{\alpha}{\beta} &  0 &
   p \\
   p q^{p - 2} (p-1 - p q^2)\frac{\alpha}{\beta} & \frac{
  p q^{p - 2} (1 - p + p q^2) (q q_0 - q_1)}{\beta} & -p^2 q^{p - 1}\sqrt{1 - q^2} & 0 \\
  \frac{-p^2 q^{p - 1} \alpha}{\sqrt{1 - q_0^2}} & \frac{p^2 q^{p - 1} (q q_0 - q_1)}{\sqrt{1 - q_0^2}} & p^2 q^p & 0 \\
  p q_0^{p - 1} & 0 & 0 & \frac{p q_1^{p - 2} \left( q^2 + q_0^2 - q q_0 q_1 + p \alpha^2-1 \right)}{\beta}  \\
0 & p q_0^{p - 2} \left(p-1 - p q_0^2\right) & -p^2 q_0^{p - 1} \sqrt{1 - q_0^2} & \frac{p q_1^{p - 2} \left((p-1) q - p q_0 q_1\right) \alpha}{\beta}  \\
0 & -p^2 q_0^{p - 1} \sqrt{1 - q_0^2} & p^2 q_0^p & \frac{-p^2 q_1^{p - 1} \alpha}{\sqrt{1 - q^2}} 
\end{array}
\right.
\]
\[
\left.\begin{array}{ccccc}
 p q^{p - 2} (p-1 - p q^2)\frac{\alpha}{\beta} & -\frac{p^2 q^{p-1} \alpha}{\sqrt{1-q_0^2}} & p q_0^{p-1} & 
  0 & 0\\
  \frac{
  p q^{p - 2} (1 - p + p q^2) (q q_0 - q_1)}{\beta} & \frac{
  p^2 q^{p-1} (q q_0 - q_1)}{\sqrt{1 - q_0^2}} &  0 &
  p q_0^{p-2} (p-1 - p q_0^2) & -p^2 q_0^{p-1} \sqrt{1 - q_0^2}\\
  -p^2 q^{p - 1}\sqrt{1 - q^2} & p^2 q^p & 
  0 & -p^2 q_0^{p-1}\sqrt{1 - q_0^2} & p^2 q_0^p\\
  0 & 0 & \frac{p q_1^{p-2} (  q^2 + q_0^2 - q q_0 q_1 + p\alpha^2-1)}{\beta} &
  p q_1^{p-2} ((p-1) q - p q_0 q_1)\frac{\alpha}{\beta} & \frac{-p^2 q_1^{p-1} \alpha}{\sqrt{1 - q^2}}\\
  p & 0 & -p q_1^{p-2} (q_0 - p q_0 + p q q_1)\frac{\alpha}{\beta} & -\frac{p q_1^{p-2} \gamma}{\beta} & \frac{p^2 q_1^{p-1} (-q_0 + q q_1)}{\sqrt{1 - q^2}}\\
0 & p^2 & -p^2 q_1^{p - 1} \alpha \sqrt{\frac{1}{1 - q_0^2}} & \frac{p^2 q_1^{p - 1} ( q_0 q_1-q)}{\sqrt{1 - q_0^2}} & p^2 q_1^p \\
-\frac{p q_1^{p - 2} (q_0 - p q_0 + p q q_1) \alpha}{\beta} & -p^2 q_1^{p - 1} \alpha \sqrt{\frac{1}{1 - q_0^2}} & p & 0 & 0 \\
-\frac{p q_1^{p - 2} \gamma}{\beta} & \frac{p^2 q_1^{p - 1} ( q_0 q_1-q)}{\sqrt{1 - q_0^2}} & 0 & p & 0 \\
\frac{p^2 q_1^{p - 1} ( q q_1-q_0)}{\sqrt{1 - q^2}} & p^2 q_1^p & 0 & 0 & p^2

\end{array}\right)
\]
where we introduced the functions
\begin{align*}
&\alpha({\bf q})=\sqrt{1 - q^2 - q_0^2 + 2 q q_0 q_1 - q_1^2}\\
&\beta({\bf q})=\sqrt{(1-q^2)(1-q_0^2)}\\
&\gamma({\bf q})=p q^2 q_1 + (-1 + p q_0^2) q_1 - q q_0 (p-1 + p q_1^2).
\end{align*}
Moreover,
\[
\hat{\Sigma}_{{\bf M}{\bf \tilde g}}^{11}=
\left(
\begin{array}{cccc}
0 & 0 & 0 & \delta \\
(1 - p) p q_0^{p - 2} \sqrt{1 - q_0^2} & 0 & 0 & \eta \\
0 & (p-1) p q_0^{p - 3} \sqrt{1 - q_0^2} \left[2 + p (-1 + q_0^2)\right] & (p - 1) p^2 q_0^{p - 2} (1 - q_0^2) & \theta
\end{array}\right.
\]
\[\left.
\begin{array}{ccccc}
\frac{(p-1) p q_1^{p - 3} \left[(2 - p) q_0 + p q q_1\right] \alpha^2}{\sqrt{1 - q^2}(1 - q_0^2)} & \frac{(p-1) p^2 q_1^{p - 2} \alpha^2}{1 - q_0^2} & 0 & 0 & 0 \\
\frac{(p-1) p q_1^{p - 3} \alpha \left[p q^2 q_1 + (-1 + (p-1) q_0^2) q_1 - q q_0 (-2 + p + p q_1^2)\right]}{(1 - q_0^2) \sqrt{1 - q^2}} & \frac{(p-1) p^2 q_1^{p - 2} (q - q_0 q_1) \alpha}{1 - q_0^2} & 0 & 0 & 0 \\
\frac{(p-1) p q_1^{p - 3} (q - q_0 q_1) \left[p q^2 q_1 + (-2 + p q_0^2) q_1 - q q_0 (-2 + p + p q_1^2)\right]}{\sqrt{1 - q^2}(1 - q_0^2)} & \frac{(p-1) p^2 q_1^{(p - 2)} (q - q_0 q_1)^2}{1 - q_0^2} & 0 & 0 & 0
\end{array}\right)
\]
with 
\begin{align*}
&\delta({\bf q})=\frac{(p-1) p q_1^{p - 3} \left[(-2 + p)(-1 + q^2 + q_0^2) - 2 (p-1) q q_0 q_1 + p q_1^2\right] \alpha({\bf q})}{\sqrt{1 - q^2}(1 - q_0^2)}\\
&\eta({\bf q})=\frac{(p-1) p q_1^{p - 3} \left[(p -2) q (-1 + q^2 + q_0^2) - q_0 (1 - p - 4 q^2 + 3 p q^2 + (p-1) q_0^2) q_1 - q (1 + q_0^2 - p (1 + 2 q_0^2)) q_1^2 - p q_0 q_1^3 \right]}{\sqrt{1 - q^2} (1 - q_0^2)}\\
&\theta({\bf q})=-\frac{(p-1) p q_1^{p - 3} (q - q_0 q_1) \left[(p-2) q - p q_0 q_1\right] \alpha({\bf q})}{\sqrt{1 - q^2}(1 - q_0^2)}.
\end{align*}
The functions $\mu_1, \mu_{01}$ and $\mu_0$ are obtained inverting the first of these matrices, and contracting the inverse with the second matrix. The resulting expressions are very lengthy for general values of $p$, too long to be reported here. 
In the particular case of $p=3$, such expressions can be simplified; one gets:
\begin{align}
    \mu_0({\bm \epsilon},{\bf q}) \Big|_{p=3}=\frac{6 \sqrt{2} q_0 \left( \epsilon_0 - q_0 \epsilon_2 \right)}{1 - q_0^2}, 
\end{align}
\begin{align}
    \mu_1({\bm \epsilon},{\bf q}) \Big|_{p=3}=\frac{A({\bm \epsilon},{\bf q}) }{B({\bm \epsilon},{\bf q}) }
\end{align}
with
\begin{align*}
    A({\bm \epsilon},{\bf q}) &=- 6 \sqrt{2} \{q^6 q_0 (  q_0 \epsilon_2-\epsilon_0) + 2 q^5 (q_0 \epsilon_1 - q_0^3 \epsilon_1 + q_1 (\epsilon_0 - q_0^3 \epsilon_2)) - 2 q^3 (-q_0 (1 - q_0^2)^2 \epsilon_1 + 5 q_0 (-1 + q_0^4) q_1^2 \epsilon_1\\& - 
q_1 (\epsilon_0 - 9 q_0^4 \epsilon_0 + 2 q_0^3 (3 + q_0^2) \epsilon_2)+ 
q_1^3 ((1 - 2 q_0^2 + 3 q_0^4) \epsilon_0 - 2 q_0 (1 - q_0^2 + q_0^4) \epsilon_2)) + 
q^4 (q_0^3 (6 + 5 q_1^2) \epsilon_0 - 
2 q_0 (\epsilon_0 + 2 q_1^2 \epsilon_0)\\& + 
q_0^4 (7 q_1 \epsilon_1 - 2 \epsilon_2) - 
2 q_0^2 (2 q_1 \epsilon_1 + \epsilon_2) - 
q_1 (3 \epsilon_1 + q_1 \epsilon_2)) + 
q_1 (q_0^7 q_1 \epsilon_0 + 2 q_0^5 q_1^3 \epsilon_0 \\&- 
q_0^3 q_1 (-1 + q_1^2)^2 \epsilon_0 - 
q_0^8 \epsilon_1 + 
2 q_0^2 (-1 + q_1^2)(\epsilon_1 - q_1 \epsilon_2) + 
2 q_0^6 (\epsilon_1 + q_1^2 \epsilon_1 - q_1 \epsilon_2) \\&+ 
(-1 + q_1^4)(-\epsilon_1 + q_1 \epsilon_2) + 
q_0^4 q_1 (-4 q_1 \epsilon_1 + q_1^3 \epsilon_1 + \epsilon_2 - 
2 q_1^2 \epsilon_2)) + 
2 q q_0 q_1 (q_0^3 q_1^2 (-4 + q_1^2) \epsilon_0 \\&+ 
q_0 (\epsilon_0 - q_1^2 \epsilon_0) - 
q_0^5 (\epsilon_0 + 3 q_1^2 \epsilon_0) + 
q_0^6 (q_1 \epsilon_1 + \epsilon_2) - 
2 q_1 (-1 + q_1^2)(-\epsilon_1 + q_1 \epsilon_2)\\& + 
q_0^4 q_1 (-4 \epsilon_1 - 3 q_1^2 \epsilon_1 + 
4 q_1 \epsilon_2) + 
q_0^2 (5 q_1 \epsilon_1 + q_1^3 \epsilon_1 - \epsilon_2 + 
2 q_1^2 \epsilon_2 + q_1^4 \epsilon_2)) \\&+ 
q^2 (-4 q_0^3 q_1^2 (-2 + q_1^2) \epsilon_0 + 
q_0^5 (1 + 18 q_1^2 + 2 q_1^4) \epsilon_0 + 
q_0 (-1 - 2 q_1^2 + 3 q_1^4) \epsilon_0 - 
2 q_1 (-1 + q_1^2)(-\epsilon_1 + q_1 \epsilon_2) \\&+ 
q_0^6 (2 q_1^3 \epsilon_1 - \epsilon_2 - 6 q_1^2 \epsilon_2) + 
q_0^2 (4 q_1 \epsilon_1 - 22 q_1^3 \epsilon_1 + \epsilon_2 - 
14 q_1^2 \epsilon_2 + 7 q_1^4 \epsilon_2) - 
2 q_0^4 q_1 (\epsilon_1 - 9 q_1^2 \epsilon_1 \\&+ 
3 q_1 (1 + q_1^2) \epsilon_2)\}
\end{align*}

\begin{align*}
B({\bm \epsilon},{\bf q}) =(1 - q_0^2) (1 - q^2 - q_0^2- q_1^2 + 2 q q_0 q_1 ) [ q^4 + q_0^4 + 
     8 q q_0 q_1 - 4 q_0^2 q_1^2 + q_1^4 - 
     2 q^2 (2 q_1^2 + q_0^2 (2 - q_1^2))-1].
\end{align*}
Finally,
\begin{align}
    \mu_{01}({\bm \epsilon},{\bf q})\Big|_{p=3} =\frac{C({\bm \epsilon},{\bf q}) }{D({\bm \epsilon},{\bf q}) }
\end{align}
where
\begin{align*}
C({\bm \epsilon},{\bf q}) &=- 6 \sqrt{2} (q q_0 - q_1) (q - q_0 q_1) \{-3 q_0^3 q_1 \epsilon_0 - 
  q_0 q_1 (1 - q_1^2) \epsilon_0 + q_0^4 \epsilon_1 + 
  q^3 (-\epsilon_0 + q_0 \epsilon_2) \\&- 
  q_0^2 ((2 + q_1^2) \epsilon_1 - 5 q_1 \epsilon_2) + 
  (1 + q_1^2)(\epsilon_1 - q_1 \epsilon_2) + 
  q^2 (\epsilon_1 - q_0 (q_1 \epsilon_0 + q_0 \epsilon_1) + 
    q_1 \epsilon_2)\\& + 
  q ((-1 + 5 q_0^2 + q_1^2) \epsilon_0 + 
    2 q_0 (-1 + q_0^2) q_1 \epsilon_1 - 
    q_0 (1 + 3 q_0^2 + q_1^2) \epsilon_2)\}
\end{align*}
and 
\begin{align*}
    D({\bm \epsilon},{\bf q}) =(-1 + q_0^2)\sqrt{
   1 - q^2 - q_0^2- q_1^2 + 2 q q_0 q_1 } [ q^4 + q_0^4 + 8 q q_0 q_1 - 4 q_0^2 q_1^2 + q_1^4 - 
     2 q^2 (2 q_1^2 + q_0^2 (2 - q_1^2))-1].
\end{align*}

\noindent We conclude this subsection by checking that these lengthy expressions reproduce the corresponding quantities obtained within the calculation of the two-point complexity, once the limits    $q\to 0, q_1\to 0$ are taken. For the case $p=3$ this is particularly straightforward to check, since $\mu_0 ({\bm \epsilon},{\bf q})$ already coincides with the corresponding quantity derived in \cite{ros2019complexity}, whereas $\mu_1({\bm \epsilon},{\bf q}) \to 0$ and $\mu_{01}({\bm \epsilon},{\bf q}) \to 0$ when  $q\to 0, q_1\to 0$. For generic $p$, one gets that 
\begin{align*}
\lim_{q,q_1 \to 0}\mu_{01}({\bm \epsilon},{\bf q}) =0, \quad \quad 
\lim_{q,q_1 \to 0}\mu_1({\bm \epsilon},{\bf q})=0, \quad \quad 
\lim_{q,q_1 \to 0}\mu_0({\bm \epsilon},{\bf q})=\frac{\sqrt{2}(p-1)p(1-q^2)(a_0(q_0)\epsilon_0 - a_1(q_0)\epsilon_2)}{a_2(q_0)}
\end{align*}
with 
\begin{align*}
&a_1(q_0)=q_0^{3p}+q_0^{p+2}(p-2-(p-1)q_0^2)\\
&a_0(q_0)=q_0^4+q_0^{2p}(1-p+(p-2)q_0^2)\\
&a_2(q_0)=q_0^{6-p}+q_0^{3p+2}-q_0^{p+2}((p-1)^2(q_0^4+1)-2(p-2)pq_0^2),
\end{align*}
which corresponds to what is found in \cite{ros2019complexity}. An analogous statement holds true for $q, q_0 \to 0$, with $q_1$ replacing $q_0$ in the formulas above.

\subsubsection{The conditioned Hessian statistics}\label{app:HessianSTat}
\noindent Combining the results of this section, we see that the fluctuating part of the conditioned Hessian 
$\tilde{\mathcal{H}}^2_{ij}=\tilde{\mathcal{M}}^2_{ij}-p\epsilon_2\sqrt{2N} \,\delta_{ij}$ 
is a matrix with the following structure: 
\begin{align}\label{eq:MTilde}
\frac{\tilde{{\mathcal{M}}}^2}{\sqrt{N}}=
\left(
\begin{array}{c c c c c| c c} 
      &  & & &  & v_1 & u_1\\
      &  &  & & & \vdots & \vdots\\
      & & {\bf T} & & & \vdots & \vdots\\
      &  &  & & & v_{N-2} & u_{N-2}\\
      &  &  & & & v_{N-3} & u_{N-3}\\
    \hline
     v_1 & v_2 & \ldots & \ldots & v_{N-3} & x_1 +\mu_1 & x_{01}+\mu_{01}\\
     u_1 & u_2 & \ldots & \ldots & u_{N-3} & x_{01}+ \mu_{01} & x_0+\mu_0\\
\end{array}
\right).
\end{align}
Here, ${\bf T}$ is a $(N-3) \times (N-3)$ symmetric block with independent Gaussian entries of mean zero and average
\begin{align*}
\mathbb{E}\left[  {\bf T}_{ij}{\bf T}_{kl}\right] =\frac{\sigma^2}{N}(\delta_{ik}\delta_{jl}+\delta_{il}\delta_{jk}), \quad \quad \sigma^2=p(p-1).
\end{align*}
This block has GOE statistics, and it is invariant with respect to the choice of the basis in the corresponding subspace $S^\perp$. The two remaining lines and columns have instead a special statistics, with averages and variances that depend explicitly on the choice of the basis in the subspace $S$. In the basis that we have chosen, the components $v_{i}, u_i$ satisfy
\begin{align}
\mathbb{E}\left[  v_i\right]=\mathbb{E}\left[ u_i\right] =0, \quad \quad 
\mathbb{E}\left[  v_iv_j\right]=\delta_{ij}\frac{\Delta_1({\bf q})}{N}, \quad \quad \mathbb{E}\left[  u_iu_j\right]=\delta_{ij}\frac{\Delta_0({\bf q})}{N}, \quad \quad \mathbb{E}\left[  v_iu_j\right] =\delta_{ij}\frac{\Delta_{01}({\bf q})}{N}
\end{align}
where the corresponding functions are given in \eqref{eq:Deltas}. The $x_1, x_0, x_{01}$ are random Gaussian numbers with zero average and variance that we have not determined explicitly, since it will not affect the results on the spectral properties of the Hessian that we are interested in. Finally,  $\mu_0({\bm \epsilon}, {\bf q}), \mu_1({\bm \epsilon}, {\bf q}), \mu_{01}({\bm \epsilon}, {\bf q})$ are the deterministic function determined in Appendix~\ref{app:averages}. Notice that this matrix is a GOE, perturbed with additive and multiplicative finite-rank perturbations, whose effect is to modify the averages and variances of the entries of the last two lines and columns of the matrix.

\subsection{Spectral properties of the conditioned Hessian}\label{app:spectralP}
\noindent In this Section, we discuss the spectral properties of random matrices having the statistics described in Sec.~\ref{app:HessianSTat}, in particular their eigenvalue distribution. Such distribution is contributed by a continuous density, and, possibly, by some outliers (or  isolated eigenvalues). The continuous eigenvalue density is essentially determined by the GOE block of the matrix, while the outliers are generated by the two special lines and columns, which can be interpreted as finite-rank perturbations to the GOE. Isolated eigenvalues correspond to $1/N$
 corrections to the eigenvalue distribution. Determining their value is crucial to study the stability of the stationary points ${\bm \sigma}_2$, which we discuss in Section \ref{sec:stability} in the main text. We begin this Section by recalling general results in  random matrix theory, and conclude with the calculation of the spectral properties of the Hessians. We stress that the characterization of the isolated eigenvalues is performed within the annealed formalism: in essence, this corresponds to the fact that the Hessian matrices at ${\bm \sigma}_2$ are not replicated. For a thorough discussion of the difference between quenched and annealed formalism in this context, we refer the reader to \cite{pacco2024curvature}.

\subsubsection{Distribution of eigenvalues for symmetric random matrices}\label{app.GeneralRM}
\noindent Before diving into the derivation of the spectrum of the Hessian matrices, let us remind a few things about the computation of the spectrum for symmetric random matrices. Consider a generic symmetric random matrix ${\bm\Lambda}$ of dimension $N\times N$ with $N\to\infty$, drawn from some probability distribution. Let its eigenvalues be $\lambda_1\leq\ldots\leq \lambda_N$, which are all real, the matrix being symmetric. The resolvent operator is defined as
\begin{align}
\begin{split}
{\bf G}:\mathbb{C}\setminus\{\lambda_1,\ldots,\lambda_N\}&\to\mathbb{C}^{N\times N}, \quad \quad z\to (z\mathds{1}-{\bm\Lambda})^{-1}
\end{split}
\end{align}
and the Stieltjes transform as 
\begin{align}
\begin{split}
\label{app:eq:def_stielt}
\mathfrak{g}_N:\mathbb{C}\setminus\{\lambda_1,\ldots,\lambda_N\}&\to\mathbb{C}\quad \quad 
z\to \frac{1}{N}\Tr{\bf G}(z)
\end{split}
\end{align}
Now, consider $\lambda\in\mathbb{R}$, and consider $\epsilon\in\mathbb{R}$. The Stieltjes transform is well defined on $\lambda-i\epsilon$, since this number is not real, and hence not an eigenvalue. Then we have that 
\begin{align*}
    &\lim_{\epsilon\to 0^+}\text{Im}\,\mathfrak{g}_N(\lambda-i\epsilon)=\lim_{\epsilon\to 0^+}\text{Im}\,\frac{1}{N}\sum_j\frac{1}{\lambda-i\epsilon-\lambda_j}=
    \frac{1}{N}\sum_j\lim_{\epsilon\to 0^+}\frac{\epsilon}{(\lambda-\lambda_j)^2+\epsilon^2}=\frac{\pi}{N}\sum_j\delta(\lambda-\lambda_j)\underset{N\to\infty}{=}\pi \rho(\lambda)
\end{align*}
where we used the delta function representation as 
\begin{align}
    \delta(x)=\lim_{\epsilon\to 0^+}\frac{\epsilon}{x^2+\epsilon^2}
\end{align}
and we defined the empirical spectral density:
\begin{align}
    \rho_N(\lambda)=\frac{1}{N}\sum_j\delta(\lambda-\lambda_j), \quad \quad \lim_{N \to \infty}  \rho_N(\lambda)=  \rho(\lambda).
\end{align}
The Stieltjes inversion formula thus reads
\begin{align}
\label{app:eq:spectrum_from_stielt}
    \rho(\lambda)=\frac{1}{\pi}\lim_{\epsilon\to 0^+}\text{Im}\,\mathfrak{g}(\lambda-i\epsilon), \quad \quad \lim_{N \to \infty} \mathfrak{g}_N(z)=\mathfrak{g}(z).
\end{align}
This holds true for any specific matrix. We now consider matrices that are random; for most random matrix ensembles, it turns out that this result  is independent on the realization of the particular random matrix (which is drawn from some probability distribution). In the limit of $N\to\infty$, indeed, all the aforementioned quantities are self-averaging, meaning that they converge towards their expected values. Hence, as $N\to\infty$, we can perform the average of the Stieltjes transform and the result is asymptotically the same for any particular realization of the random matrix from the ensemble. For $N\to\infty$, we will refer to the resolvent and the Stieltjes transform as ensemble dependent quantities, that is, as their averaged quantities. \\\\

\noindent Since we are ultimately interested in isolated eigenvalues, let us remark that these can be found by inspecting poles of $\mathfrak{g}_N$ to order $1/N$. Isolated eigenvalues $\lambda_{\rm iso}$ are poles of the empirical density $\rho_N(\lambda)$ that do not accumulate in the limit $N \to \infty$, but remains, indeed, isolated, without contributing to the continuous part of the eigenvalue density described by $\rho(\lambda)$. If the Stieltjes transform contains a term of the form $\mathfrak{g}_N(\lambda) \propto \frac{\alpha(\lambda)}{N}\frac{1}{\lambda-\lambda_{\rm iso}}$ then by the formulas above we get a contribution to the eigenvalue distribution of the form:
\begin{align}
    \rho_N(\lambda) \propto \frac{1}{N \pi}\delta(\lambda-\lambda_{\rm iso})\alpha(\lambda).
\end{align}

\subsubsection{The GOE ensemble: eigenvalues density}
\noindent For GOE matrices, the continuous part of the eigenvalue distribution $\rho(\lambda)$ is given by the well known Wigner's semicircle.   For completeness, we provide a quick proof of the GOE Wigner's law, following \cite{BurdaSpectrum2010}.
\noindent A matrix ${\bm \Lambda}$ belongs to the GOE ensemble if $\mathbb{E} \Lambda_{ij}=0$ and 
\begin{align}
    \mathbb{E}\left[\Lambda_{ij}\Lambda_{kl}\right] =\frac{\sigma^2}{N}(\delta_{ik}\delta_{jl}+\delta_{il}\delta_{jk}).
\end{align}
We write the resolvent as
\begin{align}
    \mathbf{G}(z)=\mathbb{E}\left[(Z-{\bf \Lambda})^{-1}\right]
\end{align}
with $Z=z\mathds{1}$, and observe that
\begin{align*}
    {\bf G}(z)=Z^{-1}+\mathbb{E}\left[Z^{-1}{\bm\Lambda}Z^{-1}{\bm\Lambda}Z^{-1}\right]+\mathbb{E}\left[Z^{-1}{\bm\Lambda}Z^{-1}{\bm\Lambda}Z^{-1}{\bm\Lambda}Z^{-1}{\bm\Lambda}Z^{-1}\right]+\ldots
\end{align*}
where odd products of Gaussian variables give zero contribution. The averages above can be computed using Wick's theorem (recalled below). This means that each average of the type $\mathbb{E}\left[\Lambda_{i_1,i_2}\ldots\Lambda_{i_{2k-1},i_{2k}}\right]$ is a sum of products of $k$ two-point functions, $\mathbb{E}[\Lambda_{ij}\Lambda_{kl}]$. It is useful to visualize the above in terms of diagrams. Each matrix carries two indices, and we have contractions of such matrices: hence we can use lines, where the endpoints of a line must have equal value (i.e. index), and points connecting two lines are summed over. The matrix $Z$ is proportional to the identity matrix, hence one line suffices:\\

$$
\begin{tikzpicture}
  \filldraw (0, 0) circle (3pt) node[below=8pt] {$i$};  
  \filldraw (3, 0) circle (3pt) node[below=8pt] {$j$};  
  \draw (0, 0) -- (3, 0);
  \node at (4.3, 0) {$= z^{-1}\delta_{ij}$};
\end{tikzpicture}
$$
and for the two point function we draw:
$$
\begin{tikzpicture}
  \filldraw (0, 0) circle (2pt) node[below=2pt] {$i$};  
  \filldraw (0.5, 0) circle (2pt) node[below=2pt] {$j$};  
  \filldraw (1.5, 0) circle (2pt) node[below=2pt] {$k$};  
  \filldraw (2, 0) circle (2pt) node[below=2pt] {$l$};  
  \draw (0, 0) to[out=60, in=120] (2, 0);
  \draw (0.5, 0) to[out=60, in=120] (1.5, 0);
  \node at (3, 0) {+};
  \filldraw (4, 0) circle (2pt) node[below=2pt] {$i$};  
  \filldraw (4.5, 0) circle (2pt) node[below=2pt] {$j$};  
  \filldraw (5.5, 0) circle (2pt) node[below=2pt] {$k$};  
  \filldraw (6, 0) circle (2pt) node[below=2pt] {$l$};  
  \draw (4, 0) to[out=60, in=120] (5.5, 0);
  \draw (4.5, 0) to[out=60, in=120] (6, 0);
  \node at (9, 0) {$=\mathbb{E}[\Lambda_{ij}\Lambda_{kl}]=\frac{\sigma^2}{N}(\delta_{il}\delta_{jk}+\delta_{ik}\delta_{jl})$};
\end{tikzpicture}
$$
Now, it is quick to see that for $N\to\infty$, each term of the expansion must be of order $1$ to give a meaningful contribution. It is also easy to see that if a diagram (made of a succession of lines and arcs) has crossings (i.e. arcs crossing), then we are not maximizing the number of traces obtained in the end, and hence we are counting subleading contributions. Therefore, only non crossing diagrams contribute to ${\bf G}$. Hence we can write, for some matrix $\Sigma$:\\

\begin{tikzpicture}
  \filldraw (0, 0) circle (2pt) node[below=2pt] {$i$};
  \filldraw (2, 0) circle (2pt) node[below=2pt] {$j$};
  \draw (0, 0) -- (0.5, 0);
  \draw (1.5, 0) -- (2, 0);
  \filldraw[fill=lightgray] (1, 0) circle (0.5cm);  
  \node at (1, 0) {\textbf{G}(z)};
  \node at (2.5, 0) {=};
  \filldraw (3, 0) circle (2pt) node[below=2pt] {$i$};
  \draw (3, 0) -- (4, 0);
  \filldraw (4, 0) circle (2pt) node[below=2pt] {$j$};
  \node at (4.5, 0) {+};
  \draw (5, 0) -- (7, 0);
  \filldraw[fill=lightgray] (6.5, 0) arc[start angle=0, end angle=180, radius=0.5cm] ; 
  \node at (6, 0.2) {${\bm\Sigma}$};
  \filldraw (5, 0) circle (2pt) node[below=2pt] {$i$};
  \filldraw (5.5, 0) circle (2pt) node[below=2pt] {};
  \filldraw (6.5, 0) circle (2pt) node[below=2pt] {};
  \filldraw (7, 0) circle (2pt) node[below=2pt] {$j$};
   \node at (7.5, 0) {+};
   \draw (8, 0) -- (12, 0);
  \filldraw[fill=lightgray] (9.5, 0) arc[start angle=0, end angle=180, radius=0.5cm] ; 
  \filldraw[fill=lightgray] (11.5, 0) arc[start angle=0, end angle=180, radius=0.5cm] ; 
  \filldraw (8, 0) circle (2pt) node[below=2pt] {$i$};
  \filldraw (8.5, 0) circle (2pt) node[below=2pt] {};
  \filldraw (9.5, 0) circle (2pt) node[below=2pt] {};
  \filldraw (10.5, 0) circle (2pt) node[below=2pt] {};
  \filldraw (11.5, 0) circle (2pt) node[below=2pt] {};
  \filldraw (12, 0) circle (2pt) node[below=2pt] {$j$};
  \node at (9, 0.2) {${\bm\Sigma}$};
  \node at (11, 0.2) {${\bm\Sigma}$};
\end{tikzpicture}\\

where $\Sigma$ is the "self-energy", defined by:
\begin{align}
    {\bf G}(z)=(Z-\Sigma)^{-1},
\end{align}
which implies that 
\begin{align*}
    {\bf G}(z)=Z^{-1}+Z^{-1}\Sigma Z^{-1}+\ldots
\end{align*}
We see that $\Sigma$ represents the generator of irreducible diagrams, that is, non crossings diagrams that cannot be divided in two pieces by just cutting a horizontal line, and without cutting an arc. This implies that if we sandwich ${\bf G}_{ij}$ with the double non crossing arc $\frac{\sigma^2}{N}\delta_{ij}\delta_{kl}$ we get $\Sigma_{kl}$ (since this generates all the non crossing irreducible diagrams):\\

\begin{tikzpicture}
 \filldraw (0, 0) circle (2pt) node[below=2pt] {$k$};
  \filldraw (0.5, 0) circle (2pt) node[below=2pt] {};
  \filldraw (2.5, 0) circle (2pt) node[below=2pt] {};
  \filldraw (3, 0) circle (2pt) node[below=2pt] {$l$};
  \draw (0.5, 0) -- (1, 0);
  \draw (2, 0) -- (2.5, 0);
  \filldraw[fill=lightgray] (1.5, 0) circle (0.5cm);  
  \node at (1.5, 0) {\textbf{G}(z)};
  \draw[thick] (0,0) arc[start angle=180, end angle=0, radius=1.5cm];
  \draw[thick] (0.5,0) arc[start angle=180, end angle=0, radius=1cm];
  \node at (3.5, 0) {=};
  \filldraw[fill=lightgray] (5, 0) arc[start angle=0, end angle=180, radius=0.5cm] ; 
  \node at (4.5, 0.2) {${\bm\Sigma}$};
  \filldraw (4, 0) circle (2pt) node[below=2pt] {k};
  \filldraw (5, 0) circle (2pt) node[below=2pt] {l};
\end{tikzpicture}\\

So that we get:

\begin{align*}
    \Sigma_{kl}=\frac{\sigma^2}{N}\sum G_{ij}\delta_{ij}\delta_{kl}=\sigma^2\mathfrak{g}(z)\delta_{kl}.
\end{align*}
Thus, we finally obtain:
\begin{align*}
    \mathfrak{g}=(z-\sigma^2\mathfrak{g})^{-1}
\end{align*}
which implies that 
\begin{align}
\label{app:eq:g_GOE}
    \mathfrak{g}(z)=\frac{1}{2\sigma^2}\left(z-\text{sign}(\Re(z))\sqrt{z^2-4\sigma^2}\right).
\end{align}
where the choice of sign guarantees that $\mathfrak{g}(z)\to 0$ for $|z|\to \infty$, as it should be from its definition Eq.~\eqref{app:eq:def_stielt}.\\

\noindent \textit{Wick's theorem. }
 We recall Wick's theorem for completeness. If ${\bf X}\sim\mathcal{N}({\bf 0},\Sigma)$, then we have 
\begin{align*}
    \mathbb{E}[X_1X_2\ldots X_{2n+1}]=0
\end{align*}
and 
\begin{align*}
    \mathbb{E}\left[X_1\ldots X_{2n}\right]=\sum_{p\in P_{2n}}\prod_{(i,j)\in p}\mathbb{E}[X_iX_j]
\end{align*}
where $P_{2n}$ is the set of all ways to make distinct pairs of $\{1,\ldots,2n\}$, which has cardinality $n!/(2^{n/2}(n/2)!)$.\\\\

\noindent \textit{Wigner's semicircle.}
 By combining Eq.~\eqref{app:eq:g_GOE} and Eq.~\eqref{app:eq:spectrum_from_stielt}, one can compute
\begin{align}\label{eq:semic}
    \rho(\lambda)=\frac{\sqrt{4\sigma^2-\lambda^2}}{2\pi\sigma^2}\quad\quad x\in[-2\sigma,2\sigma].
\end{align}

\subsubsection{Eigenvalue density of the conditioned Hessian}
\label{app:hessian_spectrum}
\noindent We now study the spectrum of the conditioned matrices $\tilde{\mathcal{M}}^2$ in \eqref{eq:MTilde}.
We begin by performing a change of basis in the subspace $S$, to get a matrix in which all lines and columns are independent from each others. This will facilitate the calculation of the resolvent. We recall that the above matrix has been expressed in the basis $S^\perp\oplus\{{\bf e}_{N-2}^2,{\bf e}_{N-1}^2\}$. In the basis of vectors $\{{\bf e}_{N-2}^2,{\bf e}_{N-1}^2\}$, we have that the matrix $\hat{\Sigma}^{1/2}_{{\bf M}|\tilde{\bf g}}$ reads:
\begin{align}\label{eq:Cov2}
   [ \hat{\Sigma}^{1/2}_{{\bf M}|\tilde{\bf g}}]_{ki, kj}^{22}=\begin{pmatrix}
        \Delta_1({\bf q}) & \Delta_{01}({\bf q})\\
        \Delta_{01}({\bf q}) & \Delta_0({\bf q})
    \end{pmatrix}_{ij} \quad \quad i,j \in \left\{ N-2, N-1\right\}.
\end{align}
We rotate $\tilde{\mathcal{M}}^2/\sqrt{N}$ into the  basis which diagonalizes this covariance matrix. The eigenvalues of  \eqref{eq:Cov2} are given by 
\begin{align}
&\Delta'_{0}({\bf q})=\frac{1}{2}\left(\Delta_0+\Delta_1-\sqrt{\Delta_0^2+\Delta_1^2-2\Delta_0\Delta_1+\Delta_{01}^2}\right)
\end{align}
\begin{align}
    &\Delta'_{1}({\bf q})=\frac{1}{2}\left(\Delta_0+\Delta_1+\sqrt{\Delta_0^2+\Delta_1^2-2\Delta_0\Delta_1+\Delta_{01}^2}\right).
\end{align}
The associated eigenvectors are:
\begin{align}
    &{\bf e}'_{N-1}=\frac{1}{\sqrt{z_{N-1}}}\left({\bf e}_{N-1}^2-\frac{\Delta_0 - \Delta_1 + \sqrt{\Delta_0^2 + 4 \Delta_{01}^2 - 2 \Delta_0 \Delta_1 + \Delta_1^2}}{2 \Delta_{01}}{\bf e}_{N-2}^2\right)
\end{align}
\begin{align}
    &{\bf e}'_{N-2}=\frac{1}{\sqrt{z_{N-2}}}\left({\bf e}_{N-1}^2-\frac{\Delta_0 - \Delta_1 - \sqrt{\Delta_0^2 + 4 \Delta_{01}^2 - 2 \Delta_0 \Delta_1 + \Delta_1^2}}{2 \Delta_{01}}{\bf e}_{N-2}^2\right)
\end{align}
 where
\begin{align}
    &z_{N-1}=2 + \frac{(\Delta_0 - \Delta_1) (\Delta_0 + \sqrt{
    4 \Delta_{01}^2 + (\Delta_0 - \Delta_1)^2} - \Delta_1)}{2\Delta_{01}^2}
\end{align}
\begin{align}
    &z_{N-2}=2 - \frac{(\Delta_0 - \Delta_1) (-\Delta_0 + \sqrt{
    4 \Delta_{01}^2 + (\Delta_0 - \Delta_1)^2} + \Delta_1)}{2\Delta_{01}^2}.
\end{align}
In the new (orthonormal) basis $S^\perp\oplus\{{\bf e}_{N-2}',{\bf e}_{N-1}'\}$ 
the Hessian reads
\begin{align}
\label{app:eq:M2'}
    \frac{(\tilde{\mathcal{M}}^2)'}{\sqrt{N}}=
    \left(
\begin{array}{c c c c c| c c} 
      &  & & &  & v'_1 & u'_1\\
      &  &  & & & \vdots & \vdots\\
      & & {\bf T} & & & \vdots & \vdots\\
      &  &  & & & v'_{N-2} & u'_{N-2}\\
      &  &  & & & v'_{N-3} & u'_{N-3}\\
    \hline
     v'_1 & v'_2 & \ldots & \ldots & v'_{N-3} & x'_1 \mu'_1 & x'_{01}\mu'_{01}\\
     u'_1 & u'_2 & \ldots & \ldots & u'_{N-3} &x'_{01}+ \mu'_{01} & x'_{0}+\mu'_0\\
\end{array}
\right)
\end{align}
where now
\begin{align}
\begin{split}
\label{app:eq:mu'}
&\mu'_0({\bm \epsilon},{\bf q})=\frac{-4\Delta_{01} \mu_{01} + \Delta_1 (\mu_0 - \mu_1) + \Delta_0 (-\mu_0 + \mu_1) + 
 \sqrt{4 \Delta_{01}^2 + (\Delta_0 - \Delta_1)^2} (\mu_0 + \mu_1)}{2 \sqrt{
 4 \Delta_{01}^2 + (\Delta_0 - \Delta_1)^2}}\\
 &\mu'_1({\bm \epsilon},{\bf q})=\frac{4 \Delta_{01} \mu_{01} + \Delta_0 (\mu_0 - \mu_1) + \Delta_1 (-\mu_0 + \mu_1) + 
 \sqrt{4 \Delta_{01}^2 + (\Delta_0 - \Delta_1)^2} (\mu_0 + \mu_1)}{2 \sqrt{
 4 \Delta_{01}^2 + (\Delta_0 - \Delta_1)^2}}\\
&\mu'_{01}({\bm \epsilon},{\bf q})=\frac{\sqrt{\Delta_{01}^2} ((-\Delta_0 + \Delta_1) \mu_{01} + 
   \Delta_{01} (\mu_0 - \mu_1))}{\Delta_{01} (4 \Delta_{01}^2 + (\Delta_0 - \Delta_1)^2)}
\end{split}
\end{align}
and 
\begin{align}
 \mathbb{E}\left[  v_i' v_j'\right]=\delta_{ij}\frac{\Delta_1'({\bf q})}{N}, \quad \quad
 \mathbb{E}\left[  u_i' u_j'\right]=\delta_{ij}\frac{\Delta_0'({\bf q})}{N}, \quad \quad
 \mathbb{E}\left[  v_i u_j\right]=0.
\end{align}
We write Eq.~\eqref{app:eq:M2'} in a block matrix form:
\begin{align}\label{eq:block}
    \frac{(\tilde{M}^2)'}{\sqrt{N}}=
    \begin{pmatrix}
        {\bf T} & {\bf A}_{1/2}\\
         {\bf A}_{1/2}^\top & {\bf A}_1
    \end{pmatrix}
\end{align}
with ${\bf T}\in\mathbb{R}^{(N-3)\times(N-3)}$ and $\mathbf{A}_1\in\mathbb{R}^{2\times 2}$. It is a well-established result in random matrix theory that the continuous part of the eigenvalue distribution of GOE matrices modified by finite-rank perturbations is not affected by the perturbations, and thus coincides with the Wigner's law \eqref{eq:semic}, which in the Hessian case reads 
\begin{align*}
    \rho(x)=\frac{\sqrt{4p (p-1)-x^2}}{2\pi p(p-1)}\quad\quad\quad x\in[-2 \sqrt{p (p-1)},2 \sqrt{p (p-1)}].
\end{align*}

\noindent If the statistics of the components in the blocks ${\bf A}_{1/2}, {\bf A}_1$ was the same as in the block ${\bf T}$, there would be no isolated eigenvalue(s): the latter are generated by the finite-rank perturbations in the directions ${\bf e}'_{N-2}, {\bf e}'_{N-1}$. As recalled in Section \ref{app.GeneralRM}, the isolated eigenvalues are poles of the $1/N$ expansion of the resolvent. Since the eigenvectors of the isolated eigenvalue(s) have an $O(1)$ component along the directions ${\bf e}'_{N-2}, {\bf e}'_{N-1}$,in order to find the   eigenvalues, it suffices to study the  poles of the projection of the resolvent operator on the subspace spanned by ${\bf e}'_{N-2}, {\bf e}'_{N-1}$ \cite{ros2019complex}. Given the block structure \eqref{eq:block}, the block matrix inversion lemma immediately gives that such a projection is the $2 \times 2$ matrix ${\bf A}(z)^{-1}$, with:
\begin{align}
   {\bf A}(z):=z-{\bf A}_1-{\bf A}_{1/2}^\top(z-{\bf T})^{-1}{\bf A}_{1/2}.
\end{align}
We therefore aim at computing the poles  of 
\begin{align*}
    \mathbb{E} \frac{1}{N}\Tr\left[\frac{1}{ {\bf A}(z)}\right] =\frac{1}{N}\Tr\left[\frac{1}{\mathbb{E}{\bf A}(z)} \right]+ O\tonde{\frac{1}{N^2}}.
\end{align*}
The last relation can be easily checked starting from the formal expansion: 
\begin{equation}
    \frac{1}{ {\bf A}(z)}= \sum_{n=0}^\infty \quadre{\frac{1}{z-{\bf A}_1} {\bf A}_{1/2}^\top(z-{\bf T})^{-1}{\bf A}_{1/2} }^n  \frac{1}{ z-{\bf A}_1};
\end{equation}
averaging this expression first over the entries of ${\bf A}_1$, one finds that the leading order contributions in $N$ are the terms in which the indices in the various factors $(z-{\bf T})^{-1}_{\alpha \beta}$ are contracted among themselves, i.e., the number of traces of the matrix $(z-{\bf T})^{-1}$ is maximized (see \cite{paccoros} for further examples of these large-$N$ expansions). 
We are then left with computing $\mathbb{E}{\bf A}(z)$, which to leading order in $N$ reads:
\begin{align}
\mathbb{E}{\bf A}(z)=
\begin{pmatrix}
z-\mu'_0({\bm \epsilon}, {\bf q})-\Delta_0'({\bf q})\mathfrak{g}(z) & -\mu'_{01}({\bm \epsilon}, {\bf q})\\
-\mu'_{01}({\bm \epsilon}, {\bf q}) & z-\mu'_1({\bm \epsilon}, {\bf q})-\Delta_1'({\bf q})\mathfrak{g}(z)
\end{pmatrix}
\end{align}
with $\mathfrak{g}(z)$ the resolvent of a GOE matrix of variance $\sigma^2=p(p-1)$, given in \eqref{app:eq:g_GOE}. 
Hence the poles are solutions to $\det\mathbb{E}{\bf A}(z)=0$,
and therefore the isolated eigenvalues (whenever they exist) are real solutions of the equation
\begin{align}
    \mathcal{F}(z)\equiv [z-\mu_0'({\bm \epsilon}, {\bf q})-\Delta'_0( {\bf q})\mathfrak{g}(z)][z-\mu_1'({\bm \epsilon}, {\bf q})-\Delta'_1( {\bf q})\mathfrak{g}(z)]-[\mu'_{01}({\bm \epsilon}, {\bf q})]^2=0
\end{align}
which do not belong to the support of the continuous density, i.e.  $z\notin[-2 \sqrt{p(p-1)},2\sqrt{p(p-1)}]$. This equation can be solved numerically. If a solution $z^*({\bm \epsilon}, {\bf q})$ is found, the corresponding isolated eigenvalue of the Hessian matrix \eqref{eq:ConDHess} reads:
\begin{equation}
    \tilde \lambda_{\rm iso}({\bm \epsilon}, {\bf q}) =z^*({\bm \epsilon}, {\bf q}) - \sqrt{2} p \epsilon_2.
\end{equation}
Notice that this is the eigenvalue of the rescaled matrices $\nabla^2_\perp h({\bm \sigma}_2)/\sqrt{N}$, and thus differ from the eigenvalue of $\nabla^2_\perp \mathcal{E}({\bf s}_2)/\sqrt{N}$ in \eqref{eq:IsoMain} by a factor of $\sqrt{2}$, consistently with the rescaling \eqref{eq:rescFF}.
We conclude by discussing some limits of this equation. When $q_1,q \to 0$, it can be easily checked that
\begin{align}
\begin{split}
\lim_{q_1,q \to 0}\Delta_0'({\bf q}) =\lim_{q_1,q \to 0}\Delta_0({\bf q}), \quad \quad 
\lim_{q_1,q \to 0}\Delta_1'({\bf q})=\lim_{q_1,q \to 0} \Delta_1({\bf q})
\end{split}
\end{align}
and also that 
\begin{align}
\begin{split}
\lim_{q_1,q \to 0}\mu'_0({\bm \epsilon}, {\bf q})=\lim_{q_1,q \to 0}\mu_0({\bm \epsilon}, {\bf q}), \quad \quad 
\lim_{q_1,q \to 0}\mu'_1({\bm \epsilon}, {\bf q})=\lim_{q_1,q \to 0}\mu_1({\bm \epsilon}, {\bf q}), \quad \quad
&\lim_{q_1,q \to 0}\mu'_{01}({\bm \epsilon}, {\bf q})=0.
\end{split}
\end{align}
We see that in this limit $\mathcal{F}$ reduces to 
\begin{align}
\lim_{q_1,q \to 0}    \mathcal{F}(z)=[z-\lim_{q,q_1\to0}\mu_0({\bm \epsilon}, {\bf q})-\lim_{q,q_1\to0}\Delta_0( {\bf q})\, \mathfrak{g}(z)][z-p(p-1)\mathfrak{g}(z)].
\end{align}
The second factor in this product has no roots, and therefore one is left with solving 
\[
z-\lim_{q,q_1\to0}\mu_0({\bm \epsilon}, {\bf q})-\lim_{q,q_1\to0}\Delta_0( {\bf q})\, \mathfrak{g}(z)=0
\]
which is exactly the equation for the isolated eigenvalue found in \cite{ros2019complexity} in the calculation of the two-point complexity. Once more, the analogous statement holds true for $q, q_0 \to 0$.

\subsubsection{Finite rank perturbations: a comparison with the TAP free energy}
\label{app:hessian_TAP}

To clarify the relation with previous work, in this section we recall some properties of the Hessian matrices associated to the Thouless-Anderson-Palmer (TAP) free energy of the spherical $p$-spin model~\cite{thouless1977solution}. Our goal is to show that the isolated eigenvalues of the Hessians that we discussed in Sec.~\ref{sec:TwoPoint} and Sec.~\ref{sec:stability} are not related to the finite-rank perturbation to the TAP Hessian studied in previous works~\cite{leuzzi2003complexity}. 

The TAP free energy $f_{\rm TAP}({\bf m})$ is a random high-dimensional function defined on the space ${\bf m}=(m_1, \cdots, m_N)$, of local magnetizations $m_i$. The stable local minima of this function (more precisely, those satisfying the so called Plefka’s criterion~\cite{bray1980metastable}) identify the \emph{pure states} of the system, i.e., the states in terms of which the equilibrium Boltzmann  measure at a given inverse temperature $\beta$ can be decomposed, as well as the metastable states. In the pure spherical $p$-spin model, the TAP free energy reads
\begin{equation}
  N f_{TAP}({\bf m})= \sqrt{\frac{p!}{2 N^{p-1}}} \sum_{i_1< \cdots< i_p}  a_{i_1\, \cdots i_p} m_{i_1} \cdots m_{i_p} - \frac{1}{2 \beta}\log [1-q({\bf m})] - \frac{\beta}{4} \quadre{1+ (p-1)q^p({\bf m}) -p \, q^{p-1}({\bf m})}
\end{equation}
where $q({\bf m})= N^{-1}  \sum_i m_i^2$, and the couplings $a_{i_1\, \cdots i_p}$ are the same as those parametrizing the energy landscape $\mathcal{E}({\bf s})$. The states belong to the set of solutions of the TAP equations
\begin{equation}\label{eq:TAPeqs}
\frac{  \partial [N f_{TAP}({\bf m})]}{\partial m_i}= \sqrt{\frac{p!}{2 N^{p-1}}}   \sum_{i_2< \cdots< i_p}  J_{i \, i_2  \cdots i_p} m_{i_2} \cdots m_{i_p}+ \beta^{-1}\, a[q({\bf m})] m_i=0 \quad \quad \forall i,
\end{equation}
with $\beta^{-1}a[q]=[\beta (1-q)]^{-1} + \frac{p (p-1)}{2} \beta (1-q) q^{p-2}$. The Hessian of the TAP free energy has components:
\begin{equation}\label{eq:HessianTAP}
\begin{split}
  \frac{  \partial^2 [N f_{TAP}({\bf m})]}{\partial m_j \partial m_i} = 
  \sqrt{\frac{p!}{2 N^{p-1}}}\sum_{i_3< \cdots< i_p}  a_{i \, j\,  i_3  \cdots i_p} m_{i_3} \cdots m_{i_p}+
  \frac{a[q]}{\beta} \delta_{ij}  +\frac{2}{N}\frac{a'[q]}{\beta} m_i m_j
 \end{split}
\end{equation}
where the last term is a rank-1 projection in the direction ${\bf m}$. A priori, this rank-1 term can contribute to the eigenvalue of \eqref{eq:HessianTAP} whose eigenvector is aligned in the direction ${\bf m}$ of the rank-1 perturbation, often referred to as the “longitudinal" eigenvalue $\lambda_L$~\cite{leuzzi2003complexity}. We now show that this eigenvalue diverges when $\beta \to \infty$, and that this is consistent with the fact that TAP states collapse into single  configurations in the zero-temperature limit. In this limit, the TAP Hessian maps into the Hessian of stationary points of $\mathcal{E}({\bf s})$, which shows no finite-rank perturbation. 

In the pure spherical $p$-spin model, the solutions ${\bf m}^*$ of \eqref{eq:TAPeqs} satisfying the Plefka's criterion can be parametrized as ${\bf m}^*= \sqrt{q} {\bf s}^*$, where ${\bf s}^* \in \mathcal{S}_N(\sqrt{N})$ are stationary points of the energy landscape $\mathcal{E}({\bf s})$, while $q$ satisfies $\beta(1-q)(p-1)q^{p/2-1}=-\epsilon \pm \sqrt{\epsilon^2-\epsilon^2_{\rm th}}$, where $\epsilon \leq \epsilon_{\rm th}$ is the energy density of ${\bf s}^*$. In this model, states are in one-to-one correspondence with the stationary points of the energy landscape, supplemented by the additional parameter $q$ which depends on $\beta$ and measures “the size" of the state itself, i.e., the typical overlap between the configurations belonging to it. Due to this parametrization, the TAP free energy density can be rewritten solely as a function of the parameters $\epsilon$ and $q$. It can be showed that the longitudinal eigenvalue satisfies $\lambda_L \propto \partial^2_q f_{\rm TAP}(\epsilon, q)$.

We now consider the zero-temperature limit $\beta \to \infty$, that is of interest for this work. In this limit, $q \to 1$ (states reduce to single configurations) with $\beta(1-q)$ approaching a constant value. Using  ${\bf m}= \sqrt{q} \, {\bf s}$ and taking this limit, it is simple to show that $N  f_{TAP}({\bf m}) \longrightarrow \mathcal{E}({\bf s})$,  $ \beta^{-1} a[q]\longrightarrow  -p \epsilon$ while
\begin{equation}
\begin{split}
        &\frac{a'[q]}{\beta} \stackrel{\beta \gg 1 }{\sim}  \beta \quadre{ \frac{1}{\beta^2 (1-q)^2} - \frac{p (p-1)}{2}} =O(\beta).
\end{split}
\end{equation}
Therefore, in the $\beta \to \infty$ limit there is a mode of the TAP Hessian \eqref{eq:HessianTAP}, associated to the rank-1 term, which diverges. This is a divergence of $\partial^2_q f_{\rm TAP}(\epsilon, q)$, i.e., of the stiffness of  $f_{\rm TAP}(\epsilon, q)$ in the direction of $q$; this reflects the fact that when $\beta \to \infty$ states collapse into configurations that satisfy a hard spherical constraint $\sum_{i=1}^N s_i^2=N$, and the variable $q$ does not fluctuate but is identical to $q=1$. The TAP Hessian \eqref{eq:HessianTAP}, projected onto the tangent plane to the hypersphere at ${\bf s}$, reduces precisely to the Riemannian Hessian $\nabla^2_\perp \mathcal{E}({\bf s})$, whose statistics we described in Sec.~\ref{sec:stability}.

In summary, the curvature of the function $\mathcal{E}({\bf s})$ restricted to the sphere, measured by the eigenvalues of the Riemannian Hessian, is not at all affected by the rank-1 perturbation present in the TAP Hessian. The isolated eigenvalues computed in \cite{ros2019complexity, ros2020distribution} and discussed in this manuscript are generated by another mechanism, namely, by \emph{enforcing} that the stationary points ${\bf s}_1$ of which one is studying the Hessian are at non-zero overlap from some other stationary point ${\bf s}_0$. This enforcement can be rephrased into a conditioning of the distribution of the Hessian at ${\bf s}_1$, which gives rise to finite-rank perturbations different in nature with respect to the one arising in the TAP context (in particular, they are both additive and multiplicative). To see an analogous mechanism at play within a TAP framework, one should study the subleading contributions of the  Hessian of configurations extracted with a constrained TAP measure, such as that introduced in \cite{barbier2020constrained}. \\\\

\section{Quenched \emph{vs} annealed three-point complexity: a comparison}
\label{app:ann_vs_quench}
\begin{figure}[ht]
\includegraphics[width=0.35
\textwidth, trim=5 5 5 5,clip]{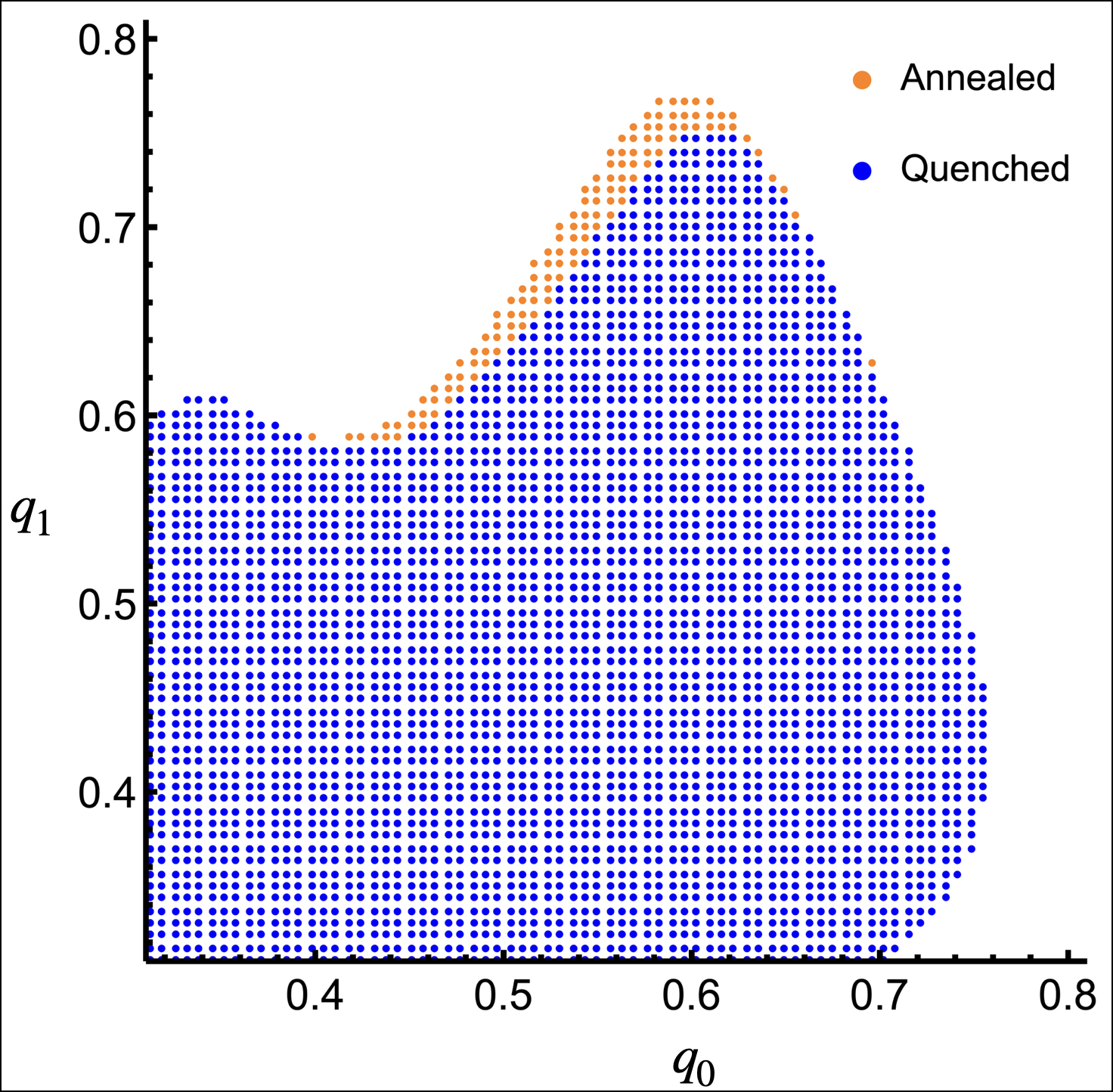}
\caption{The dotted blue region identifies the values of overlaps where $\Sigma^{(3)}\geq 0$, the dotted orange region the values where $\Sigma^{(3)}_{2A}\geq 0$. The plot is made for the same parameters as  Fig. 3 (b):  $\epsilon_0=-1.167$,
$\epsilon_1=\epsilon_2=-1.155$, and $q=0.56$. The largest mismatch is close to regions of clustering (roughly for $q_1>0.6$).}
\label{fig:q_vs_a}
\end{figure}

\noindent In this final Appendix, we illustrate that the annealed three-point complexity (derived in these Appendices) and the quenched three-point complexity (illustrated in the plots given in the main) are quantitatively different, even though for most values of parameters they differ by a very small amount. We find that the region where quenched and annealed three-point complexity differ the most is close to the boundary of the support, in the vicinity of the region where $\Sigma^{(3)}$ vanishes. This is the region associated to the clustering phenomenon, discussed in Sec.~\ref{sec:LandscapeEvolution}.  We illustrate this in Fig.~\ref{fig:q_vs_a}, which compares the domain where the quenched three-point complexity $\Sigma^{(3)}>0$ (blue)  with the domain where $\Sigma^{(3)}_{2A}>0$ (orange).  The plot is given for the same parameters as for the Fig. 3 (b) in the main text. 
We see that the two domains are different. The blue region is enclosed in the orange one, consistently with the bound $\Sigma^{(3)}_{2A} \geq \Sigma^{(3)}$. The strongest 
discrepancy between the domains is exactly in the region where clustering is present. Instead, away from this region, the two computations are practically indistinguishable.

\end{document}